\documentclass[11.75pt,a4paper]{article}

\usepackage[a4paper,top=2cm,bottom=2cm,left=2cm,right=2.5cm,marginparwidth=1.75cm]{geometry}
\usepackage[T1]{fontenc}
\usepackage{bm}
\usepackage{amsmath}
\usepackage{graphicx}
\usepackage{afterpage}
\usepackage{booktabs}
\usepackage{multirow}
\usepackage[bookmarksopen,bookmarksnumbered,colorlinks,linkcolor=blue,citecolor=blue]{hyperref}
\usepackage{apacite} 
\usepackage{natbib}
\usepackage{hyperref}
\usepackage{rotating}
\usepackage{subcaption} 
\usepackage[ruled,vlined]{algorithm2e}

\usepackage{amssymb}
\usepackage{longtable}
\usepackage{pdflscape}

\usepackage[affil-it]{authblk}

\makeatletter

\renewcommand{\section}{\@startsection{section}{1}{\z@}%
  {-2.0ex \@plus -0.5ex \@minus -.2ex} 
  {1.0ex \@plus .1ex} 
  {\large\rmfamily\bfseries}} 

\renewcommand{\subsection}{\@startsection{subsection}{2}{\z@}%
  {-1.8ex \@plus -0.5ex \@minus -.2ex} 
  {0.8ex \@plus .1ex} 
  {\rmfamily\bfseries}} 
\makeatother

\title{\bfseries \normalsize Variational autoencoders for anomaly detection in choice modelling datasets}

\author[1]{Lorenzo Mu\~{n}oz*}
\author[1]{Stephane Hess}
\author[1]{Thomas O. Hancock}
\author[1]{Georges Sfeir}

\affil[1]{Choice Modelling Centre and Institute for Transport Studies, University of Leeds, UK \vspace{-0.1cm}}
\affil[*]{Corresponding author}
\date{\vspace{-5ex}}

\begin{document}
\maketitle

\section*{Abstract}\small

Choice data often contain anomalous observations, which may arise from errors in the data, such as misreported or wrongly computed variables, or from unusual behaviour. Both types can distort parameter estimates and derived outputs such as values of time and elasticities, although unusual observations may also reflect heterogeneity between individuals. While routine checks are generally applied in practice (e.g. dominant alternatives and ranges of trade-offs), a systematic, multivariate check that does not require the type of anomaly to be defined in advance is still missing. Generative machine learning (ML) is well suited for anomaly detection (AD). By learning the underlying data-generating process, these models can identify observations that deviate from the patterns uncovered in the rest of the sample. A variational autoencoder (VAE) compresses each observation into a low-dimensional latent representation and reconstructs it, and observations that are reconstructed poorly receive high anomaly scores. We introduce VAEs for AD in choice modelling (CM) datasets, while accounting for the panel structure of the data through mixed-effects neural networks. We decompose the anomaly score into variable-specific reconstruction probabilities and a KL term, which show which variables make an observation anomalous. In a synthetic case study, the VAE with mixed effects detects 48\% of the induced anomalies, compared to 30\% without mixed effects, 20\% for univariate checks, and 29\% for Mahalanobis distance, and it detects anomalies that are not extreme in any variable, which these methods miss (with an anomaly threshold $\alpha=0.95$). We then apply the framework to a stated preference (SP) and a revealed preference (RP) dataset. To assess how influential the flagged observations are, we exclude the 1--10\% most anomalous observations, which changes values of time by up to 10\% (SP) and 54\% (RP), and elasticities up to 11\% (SP) and 55\% (RP). The proposed approach gives choice modellers a diagnostic tool to identify the observations and the variables that drive their estimates, so that they can decide whether these reflect errors in the data, heterogeneity to be modelled, or a specification to revisit.

\textbf{Keywords}: Anomaly detection, generative machine learning, outliers, variational autoencoders, choice modelling. 

\section{Introduction}
\label{sec:intro}

In choice modelling (CM), the main aim is to understand and predict behaviour \citep{benakiva1985,train2009}. Both depend on the quality of the data, and choice data often contains anomalous observations that differ from the rest of the sample, which may arise either from errors in the data or from unusual behaviour \citep{hess2010outliers, varotto2017}. These anomalies can cause substantial changes in model estimates and mislead applications such as market simulations and demand forecasts \citep{mcfadden1981econometric,ku2017}. Identifying them can therefore help to correct errors in the data, but could also uncover heterogeneity, thus highlighting ways the model specification could be improved.

Several definitions of anomalies coexist in the literature. \cite{hawkins1980} defines anomalies as observations that deviate so much from the other observations as to arouse suspicions that they were generated by a different mechanism. \cite{chandola2009} define them as patterns in data that do not conform to a well-defined notion of normal behaviour. More recently, \cite{ruff2021} define anomalies as an observation that significantly deviates from the expected behaviour in a system, often appearing unusual, inconsistent, or unexpected. 

In the context of trip data, anomalies have also been presented as inconsistencies between observed and externally derived variables. Indeed, significant discrepancies may arise between recorded observations and values obtained from external APIs \citep{bastarianto2025}. Errors are also common in the level-of-service attributes produced by the network models used to forecast transport demand, arising, among other sources, from omissions in coded networks and schedules, the loss of geographical detail caused by zone centroids and connectors, and errors attributed to simplified delay functions \citep{bhatta2011}.

In CM, anomalies (often referred to as outliers) tend to have a behavioural interpretation. \cite{hultkrantz2001} define anomalies as patterns in choice behaviour that are inconsistent with economic theory, i.e., with the assumption that individuals choose the alternative that maximises their utility. Furthermore, \cite{hess2010outliers} identify three main categories in choice behaviour: non-trading behaviour, where individuals always choose the same alternative; lexicographic behaviour, where choices follow a strict ordering of the variables; and inconsistent behaviour, where choices do not follow a predictable pattern. Each type is detected with a different method \citep{mariel2025}, such as frequency counts of the chosen alternatives for non-trading behaviour, choice frequencies by variable level or more flexible parameter distributions for lexicographic behaviour \citep{cazor2024}, and ray diagrams for inconsistent behaviour \citep{saelensminde2002}. Data anomalies, in turn, are usually addressed through data cleaning, for instance by removing observations whose recorded values differ greatly from those obtained from external APIs \citep{bastarianto2025}.

However, each of the diagnostics above targets a single, predefined type of anomaly. Some checks do consider combinations of variables, such as checks for dominated alternatives, for the range of trade-offs implied by an SP design, or for the consistency between recorded values and those obtained from external APIs \citep{bastarianto2025}, but each requires the analyst to decide beforehand which criterion to check. Simpler alternatives, such as flagging unusually long or expensive trips by examining the tails of each variable's distribution, or even cleaning the data manually, look at one variable at a time.  What is missing is a systematic, multivariate check that does not require the type of anomaly to be defined in advance. CM datasets consist of observations with rich multivariate distributions, so anomalies frequently emerge from multi-variable relationships rather than isolated extremes. Furthermore, correlations between variables are complex in CM datasets. For instance, a short trip with a high cost may not be anomalous on either variable separately, but anomalous when both are considered jointly. Accounting for the joint distribution of the data is therefore a central challenge.

In machine learning (ML), the task of finding anomalous observations is known as anomaly detection (AD) \citep{chandola2009,an2015}, and has been applied in domains such as fraud detection \citep{ahmed2016}, cybersecurity \citep{salo2019} and healthcare \citep{schlegl2017}. Among the many AD methods, which we review in \autoref{subsec:ad_methods}, we use reconstruction-based methods. They learn a lower-dimensional representation of the data and use the difference between the original and the reconstructed observations as an anomaly score \citep{an2015}, thereby quantifying departures from the learned joint structure of the data \citep{ruff2021}. They can therefore flag both extreme values and unusual combinations of variables, without the type of anomaly being defined in advance. The representation is typically learned with autoencoders, and in particular with variational autoencoders (VAEs), which learn a probabilistic latent representation and reconstruct the data through a generative network \citep{nguyen2024}.
 
In CM, ML methods have mostly been used to predict choices \citep{wang2020a,wang2021}, to add flexibility to choice models \citep{sifringer2020,han2022}, or to uncover preference heterogeneity \citep{sfeir2021}. Less attention has been given to generative models, which learn the joint distribution of the data rather than the choice conditional on the explanatory variables \citep{ng2002}. In CM, generative models, most of them VAEs, have been used to synthesise populations \citep{borysov2019}, to generate choice sets \citep{yao2022,haghi2026}, and to learn latent representations of travel behaviour \citep{wong2019}. As they learn the data-generating process, generative models can also identify observations that do not fit it, which makes AD one of their main applications \citep{skvara2021}.

However, VAEs have not yet been used for AD in CM datasets, and the way these datasets are typically collected raises an issue for standard VAEs, which assume that observations are independent \citep{bouchacourt2018}. Many CM datasets have a panel structure, with several observations for the same individual, and just as this structure needs to be accounted for when estimating a choice model, it also needs to be accounted for when detecting anomalies. An observation that is unusual for the sample may make sense in the context of a specific individual, and vice versa. Moreover, for AD to be useful to the analyst, it is not enough to know which observations are anomalous, but also which of their variables make them so, as this is what allows the analyst to decide whether a flagged observation reflects an error in the data, heterogeneity to be modelled, or a specification to revisit. It is thus helpful to also identify variable-specific anomaly scores.

This paper proposes a reconstruction-based AD framework for CM data built on VAEs, and makes two main contributions. First, we bring reconstruction-based AD to CM, providing a data-driven approach for identifying atypical observations and unusual combinations of variables, without requiring predefined definitions of anomalous observations. To the best of our knowledge, such approaches have not previously been applied to choice data. Second, we extend the VAE architecture to accommodate panel data through individual-specific random effects, analogous to those used in panel mixed logit models. This allows the framework to account for correlations between observations from the same individual and to capture individual-specific patterns in the data. We first validate the framework on synthetic data, where the true utility specification is known, and anomalies can be deliberately induced, and then apply it to a stated preferences (SP) and a revealed preferences (RP) dataset. The two datasets differ in how anomalies arise: in the SP data, the attributes are set by the experimental design around a reported trip, so anomalies are expected to come from unrealistic attribute combinations and from choices that are inconsistent with them, whereas in the RP data, the attributes of the alternatives are measured for the chosen alternative, and estimated through APIs for the non-chosen alternatives, so anomalies are expected to come from errors in data collection and processing.

This paper is organised as follows: \autoref{sec:theory} introduces the theoretical framework, with a detailed explanation of the use of VAEs for AD; \autoref{sec:cases_synth} presents the synthetic case study, whereas \autoref{sec:cases_real} presents the real-world applications; and \autoref{sec:conclusions} concludes, and presents the limitations of the study and the avenues for future research. We also report the validation checks in \autoref{ap:validation}, and the utility specifications and complementary results in \autoref{ap:comp_results}.

\section{Theoretical framework}
\label{sec:theory}

In this section, we introduce the VAE framework used to reconstruct choice data and to detect anomalous observations. In \autoref{subsec:ad_methods} we review the main families of AD methods and explain why we use a reconstruction-based approach. In \autoref{subsec:vae_arq} we describe the model architecture, i.e., how observations are encoded into a low‑dimensional latent representation and later reconstructed through a generative decoder, together with the training objective and the associated loss function. In \autoref{subsec:menn} we extend the framework to panel data through individual-specific random effects. In \autoref{subsec:rec_probs} we present the anomaly score we use to flag anomalies. Finally, in \autoref{subsec:vae_conv} we present the criteria used to assess convergence and to validate the model, checking that it captures both the marginal distributions and the dependence structure of the data. 

\subsection{Anomaly detection methods}
\label{subsec:ad_methods}
 
In the ML literature, AD techniques can be categorised according to whether the data include labels indicating which observations are anomalous. Supervised approaches rely on labelled data identifying normal and anomalous observations, semi-supervised approaches use limited labelled information, while unsupervised approaches identify anomalies by learning patterns from unlabelled data \citep{huang2025}. Given that anomalous observations are not known in advance in CM datasets, and that we aim to identify unusual patterns without imposing predefined definitions of anomalous behaviour, we focus on unsupervised AD.
 
Within this unsupervised field, methods can be grouped into broad, not mutually exclusive, categories, depending on how they identify observations that deviate from the patterns in the data. Statistical methods, such as copula-based AD \citep{li2020copod} or AD based on empirical cumulative distributions \citep{li2022ecod}, assume a distribution or a probability model for the data, and then identify anomalies with respect to the model using a discordancy test \citep{han2012}. The null hypothesis is that all observations come from the same distribution, and the alternative is that some observations come from a different distribution. Observations for which the null hypothesis is rejected are flagged as anomalies. These methods are well founded when the assumed distribution is appropriate, but the distribution has to be specified in advance, which is difficult for data with many variables of different types.
 
Distance-based methods \citep{knorr2000,hodge2004} flag observations that lie far, under a given metric (e.g., Euclidean or Manhattan distance), from most of the other observations in the dataset, assuming that anomalies are isolated from the majority of the data. They make no distributional assumptions, but are sensitive to parameter choices, making results unstable \citep{chen2010}, and distances become less discriminative as the number of variables grows. They also require a metric that combines variables measured in different units and of different types, such as travel times, costs and trip purposes. Moreover, as variables are typically standardised, these methods assume equal importance across variables, which is rarely the case in practice. 
 
Density-based methods \citep{breunig2000}, such as the local outlier factor \citep{alghushairy2020}, compare the local density around an observation with that around its neighbours, so that observations in regions of low density relative to their neighbours are flagged. Related approaches based on $k$-nearest neighbours have been used to detect errors, noise and abnormal behaviour in road networks, including congestion, incidents, low volumes and abrupt driving behaviour \citep{dang2015}. As they also rely on distances between observations, they share the limitations of distance-based methods in high dimensions. Ensemble methods combine several, complementary methods to improve robustness and reduce the bias of individual methods. For instance, \citet{cao2025} combine deep sequence learning with density-based algorithms to detect anomalies in route trajectory data.
 
Reconstruction-based methods instead learn a lower-dimensional representation of the data and use the difference between the original and the reconstructed observations as an anomaly score \citep{an2015}, thereby quantifying departures from the learned joint structure of the data \citep{ruff2021}. Unlike statistical methods, they do not require the distribution of the data to be specified in advance, and unlike distance-based and density-based methods, they do not rely on a metric in the original space of the variables. Examples of their use for AD include spacecraft data \citep{sakurada2014}, robot-assisted technologies \citep{park2018}, IT systems \citep{su2019} and video surveillance \citep{sun2018jiayu}.
 
The lower-dimensional representation is typically learned with autoencoders \citep{nguyen2024}. Unlike linear dimension-reduction techniques, such as principal component analysis, autoencoders learn non-linear latent representations that capture complex dependencies between variables. Among reconstruction-based approaches, VAEs have gained particular prominence, as they learn compact probabilistic latent representations of high-dimensional data and reconstruct the inputs through deep generative networks \citep{nguyen2024}. We describe the VAE used in this paper in \autoref{subsec:vae_arq}.

We compare the VAE with two simpler methods that are frequently used in CM. The first is a univariate check, where each variable is evaluated separately. In practice, observations above a certain threshold are frequently removed, such as unusually long or expensive trips, or walking trips with a very long distance. These thresholds are either set by the analyst, based on what is considered plausible for each variable, or derived from the distribution of each variable in the sample, e.g. by flagging values above a given percentile. 

The second method is the Mahalanobis distance, which evaluates all continuous variables jointly, as given by:

\begin{equation}
    M_{nt} = \sqrt{(x_{nt} - \bar{x})^\top Q^{-1} (x_{nt} - \bar{x})},
\end{equation}

where $x_{nt}$ is the vector of continuous variables of observation $t$ of individual $n$, and $\bar{x}$ and $Q$ are their sample mean and covariance matrix, respectively. Unlike the univariate check, the Mahalanobis distance accounts for the correlations between variables, so it can flag observations whose values are each plausible on their own but whose combination is unusual. Neither method accounts for the panel structure of the data, so an observation is assessed against the whole sample rather than against the other observations of the same individual.
 
\subsection{VAE architecture}
\label{subsec:vae_arq}

Let the dataset contain $N$ individuals, where individual $n$ has $T_n$ observations, resulting in $T=\sum_{n=1}^{N}T_n$ observations in total. In each observation, individual $n$ chooses one alternative $i$ from a set of $J$ alternatives, indexed by $j=1,\dots,J$. The vector of variables for observation $t$ of individual $n$ is given by ${x}_{nt}=(x_{nt1},\dots,x_{ntK})$, whose $K$ variables may include the attributes of each alternative $j$, individual characteristics, contextual variables, and the choice. Each variable is either continuous ($k \in K_c$) or categorical ($k \in K_d$). Continuous variables are standardised to zero mean and unit variance, and a categorical variable with $C_k$ categories is one-hot encoded through the indicators $x_{ntkc}$, $c=1,\dots,C_k$, which equal one if category $c$ is observed and zero otherwise. When some alternatives are not available for an observation, as is common in RP data, their variables are coded in the dataset as zero, and the availability of each alternative $j$ is included in ${x}_{nt}$ as a binary categorical variable. The VAE can therefore distinguish zeros due to unavailability from observed values, and learn how availability relates to the other variables. The observations for individual $n$ are collected in ${X}_n=\{{x}_{n1},\dots,{x}_{nT_n}\}$, and the dataset is therefore ${X}=\{{X}_1,\dots,{X}_N\}$.

VAEs \citep{kingma2013} are generative models that combine deep learning with probabilistic graphical models to learn complex data distributions, while explicitly inferring structured latent representations from the observed data. A VAE consists of two components: an encoder $q_\phi(z_{nt} \mid x_{nt})$ that compresses each observation $x_{nt}$ into a low‑dimensional latent representation $z_{nt}$, and a decoder $p_\theta(x_{nt} \mid z_{nt})$ that reconstructs the input from this representation, producing the reconstruction $\widehat{x}_{nt}$. The encoder and the decoder are parametrised by $\phi$ and $\theta$, respectively.

By learning a lower-dimensional probabilistic representation, each observation is mapped to a distribution over the latent space by a mean and a standard deviation. VAEs are well suited for AD \citep{ogretir2022}, and this dimensionality reduction lets the modeller focus on the informative structure of the data while filtering out noise \citep{an2015}.

Data used in CM involve specific challenges. First, continuous and categorical variables are often mixed, so a single likelihood cannot be used for all variables. Second, continuous variables such as travel time and cost are often skewed and heavy-tailed, so that long but legitimate trips lie far from the centre of the distribution and may receive high anomaly scores. Third, the variables of unavailable alternatives are often coded as zero, which can produce large likelihood terms that distort the anomaly scores. VAEs have nonetheless shown strong performance in learning latent spaces that support the generation of diverse and realistic tabular samples \citep{nguyen2024}. We address these challenges by log-transforming skewed variables, and by including availabilities in modelling, so that the VAE can learn the structure of these observations. 

In a CM context, VAEs offer at least three advantages relative to existing diagnostic tools (reviewed in \autoref{subsec:ad_methods}). First, they can detect a broad range of irregularities, such as extreme or implausible variable values, misreported variables, or combinations of variables that are individually plausible but jointly atypical, without requiring a prior definition of anomaly. Second, they capture non‑linear and multivariate patterns that are difficult to detect with traditional methods, such as univariate checks and Mahalanobis distance. Third, they scale to large, high‑dimensional datasets, such as GPS-based travel data.

We use a bottleneck architecture, in which the latent dimensionality $D_z$ is lower than the dimensionality of the encoded input vector $D_x$ \citep{hinton2006}. $D_z$ is a hyperparameter that controls the fundamental trade-off in reconstruction-based AD: it must be large enough for the model to capture the main dependencies in the data, but small enough such that the model does not overfit the data by reconstructing every observation well. If $D_z$ exceeds the intrinsic dimensionality of the data, the model can reconstruct anomalies as well as normal observations, and can no longer distinguish between them.

The encoder and the decoder have symmetric architectures, with the decoder mirroring the encoder in the number and size of its hidden layers. We use a fully connected neural network (NN) architecture, with multiple layers of artificial neurons. NNs are highly flexible, non-linear approximations, and are the most common choice for parametrising both the encoder and the decoder \citep{borysov2019}. Each neuron computes the weighted sum of its inputs, followed by a non-linear activation function. Thus, the output of hidden layer $l$ of an NN is computed as $h_l=a(W_l h_{l-1}+b_l)$, where $h_0$ is the input of the network, $W_l$ and $b_l$ are the weight matrix and the bias vector of layer $l$, and $a$ is a non-linear activation function\footnote{The function $a$ is a modelling decision. In this study, we use the Rectified Linear Unit (ReLU) activation in the hidden layers due to its empirical stability and favourable optimisation properties \citep{glorot2011}. Alternative activation functions include hyperbolic tangent (tanh), the exponential linear unit (ELU), and Leaky‑ReLU.}. We define the collection of all weights and biases as $\phi$ and $\theta$ for the encoder ($q_\phi$) and the decoder ($p_\theta$), respectively.

The VAE is trained by minimising the loss function (see \autoref{eq:loss} below) using mini-batch gradient descent. At each step, the gradient of the loss with respect to every weight and bias in the encoder and decoder is estimated on a mini-batch of observations through backpropagation, which is the algorithm used to compute these gradients. The error between the observed and reconstructed data is thus propagated backwards through the network, from the output of the decoder to the input of the encoder, and the chain rule gives how much each weight and bias contributed to that error \citep{rumelhart1986}. All parameters are then adjusted by a small step, controlled by the learning rate, in a direction that is expected to reduce the loss. This is repeated over multiple epochs\footnote{An epoch is one full pass of the training data through the model.} through the training data until the loss on a validation set stops improving (i.e., early stopping).

Following the specification of \cite{borysov2019}, continuous and categorical variables are treated differently within the reconstruction term. The loss combines this reconstruction term with a regularisation term that keeps the latent distribution close to the prior $\mathcal{N}(0,I_{D_z})$, a standard multivariate normal distribution, where $I_{D_z}$ is the $D_z\times D_z$ identity matrix. This formulation balances data fidelity and latent‑space regularisation within a single optimisation problem  \citep{daoud2025}.

\begin{equation}
\mathcal{L}_{\phi,\theta}
= -\sum_{n=1}^{N} \sum_{t=1}^{T_n}
 \mathbb{E}_{q_\phi(z_{nt}\mid x_{nt})}\!\left[\sum_{k=1}^{K}\log p_\theta(x_{ntk} \mid z_{nt})\right]
+ \lambda \sum_{n=1}^{N} \sum_{t=1}^{T_n} D_{KL}\!\left(q_{\phi}(z_{nt} \mid x_{nt}) \,\|\, \mathcal{N}(0,I_{D_z})\right),
\label{eq:loss}
\end{equation}

where the expectation is taken over the latent space $z_{nt}$ given by the encoder. The first term of \autoref{eq:loss} depends on the type of variable $x_{ntk}$. For a continuous variable, we use the Gaussian log-likelihood, given by:

\begin{equation}
    \log p_\theta(x_{ntk} \mid z_{nt}) = -\frac{1}{2} \left( \frac{(x_{ntk} - \mu_{\widehat{x}, ntk})^2}{\sigma_{\widehat{x}, ntk}^2} + \log(2\pi \sigma_{\widehat{x}, ntk}^2) \right), k \in K_c,
\label{eq:loss_num}
\end{equation}

where $\mu_{\widehat{x},ntk}$ and $\sigma_{\widehat{x},ntk}$ are the mean and standard deviation of variable $k$ predicted by the decoder. For a categorical variable with $C_k$ categories, we use the cross-entropy under one-hot encoding:

\begin{equation}
\log p_\theta(x_{ntk} \mid z_{nt})
= \sum_{c=1}^{C_k} x_{ntkc}\log\big(\tau_{ntkc}\big), k \in K_d,
\label{eq:loss_cat}
\end{equation}

where $\tau_{ntkc}$ is the probability of category c predicted by the decoder through a Softmax function. Since $x_{ntkc}$ equals one only for the observed category, only the log-probability of the observed category contributes. The second term of \autoref{eq:loss} is the Kullback--Leibler (KL) divergence \citep{kullback1951} between the encoder distribution and the prior, which has the closed form

\begin{equation}
D_{KL}\!\left(q_{\phi}(z_{nt} \mid x_{nt}) \,\|\, \mathcal{N}(0, I_{D_z})\right) = -\frac{1}{2}\sum_{d=1}^{D_z}\left(1 + \log\sigma_{ntd}^2 - \mu_{ntd}^2 - \sigma_{ntd}^2\right),
\label{eq:kl_z}
\end{equation}

where $\mu_{ntd}$ and $\sigma_{ntd}$ are the mean and standard deviation of latent dimension $d$ produced by the encoder. This term penalises deviations from the prior and regularises the latent space, so that the model cannot simply memorise individual observations. This way, observations that lie far from the learned structure of the data are given high anomaly scores. 

With $\lambda = 1$, \autoref{eq:loss} is the negative evidence lower bound (ELBO) of the log-likelihood of the data. The log-likelihood itself, $\log \int p_\theta(x_{nt}\mid z_{nt})\,\mathcal{N}(z_{nt};0,I_{D_z})\,dz_{nt}$, cannot be maximised directly, as the integral has no closed form when $p_\theta$ depends on $z_{nt}$ through the decoder. The encoder $q_\phi(z_{nt}\mid x_{nt})$ instead approximates the posterior of $z_{nt}$, and the log-likelihood equals the ELBO plus the KL divergence between $q_\phi(z_{nt}\mid x_{nt})$ and the true posterior $p_\theta(z_{nt}\mid x_{nt})$. As this divergence is non-negative, minimising \autoref{eq:loss} maximises a lower bound on the log-likelihood, which is tight when the encoder approximates the posterior well \citep{kingma2013,blei2017}.

To prevent posterior collapse, in which the KL term drives latent dimensions to the prior and the decoder ignores them, $\lambda$ is increased linearly from zero to one over the first epochs of training \citep{bowman2016}. In all case studies, the remaining hyperparameters are selected through a random search, while ensuring that we recover the distributions and correlations present in the data (further explained in \autoref{subsec:vae_conv}).

During training, the encoder maps each observation $x_{nt}$ to two vectors of size $D_z$, $\mu_{nt}$ and $\sigma_{nt}$ (\autoref{eq:encoder}), which define the multivariate normal distribution $q_\phi(z_{nt}\mid x_{nt})=\mathcal{N}\big(\mu_{nt},\mathrm{diag}(\sigma_{nt}^2)\big)$. The latent variable $z_{nt}$ is drawn from this distribution with the reparametrisation trick (\autoref{eq:reparam}), which writes the latent variable as the mean plus a scaled noise term, where $\odot$ denotes element-wise multiplication. This makes the sampling step differentiable and enables gradient-based optimisation:

\begin{equation}
    \mu_{nt} = f_{\mu}(x_{nt};\phi),
    \qquad
    \log \sigma^2_{nt} = f_{\sigma}(x_{nt};\phi),
    \label{eq:encoder}
\end{equation}

\begin{equation}
z_{nt} = \mu_{nt} + \sigma_{nt} \odot \varepsilon_{nt},
\qquad
\varepsilon_{nt} \sim \mathcal{N}(0,I_{D_z}).
\label{eq:reparam}
\end{equation}

\begin{figure}[h!]
\centering
\includegraphics[width=0.9\linewidth]{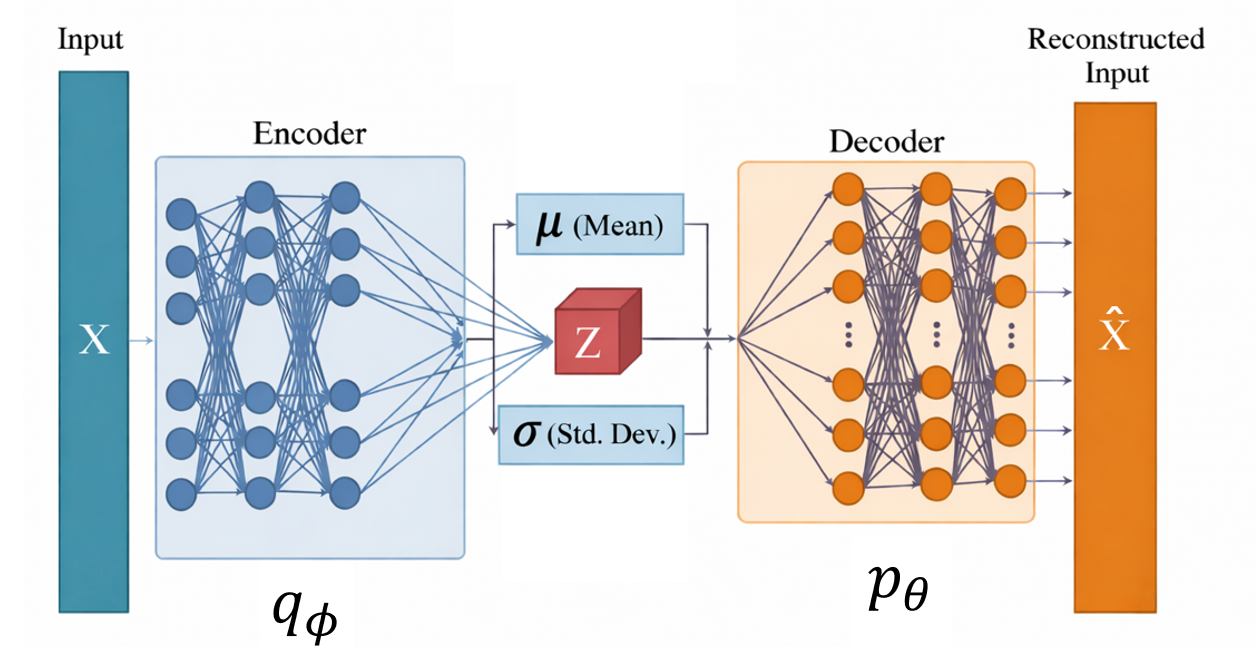}
\caption{VAE structure for AD.}
\label{fig:vae}
\end{figure}

The expectation in \autoref{eq:loss} has no closed form, since $p_\theta$ depends on $z_{nt}$ through the decoder. At each training step, it is estimated with one draw of $z_{nt}$ per observation. The decoder maps the latent variable back into the original feature space and returns the parameters of the reconstruction distribution, i.e.\ $\mu_{\widehat{x},ntk}$ and $\sigma_{\widehat{x},ntk}$ for continuous variables and the category probabilities $\tau_{ntkc}$ for categorical variables:

\begin{equation}
     \left(\mu_{\widehat{x},ntk},\, \sigma_{\widehat{x},ntk},\, \tau_{ntkc}\right) = g(z_{nt};\theta),
    \label{eq:decoder}
\end{equation}

where $g$ is the decoder NN. Although variables are reconstructed independently given $z_{nt}$, they share the same latent space, so the decoder captures dependencies among variables through $z_{nt}$. The VAE can therefore identify anomalies that arise from cross-variable interactions, which are difficult to detect with marginal methods. Note that the VAE estimates the parameters $\phi$ and $\theta$, not the reconstructed variables $\widehat{x}_{ntk}$. \autoref{fig:vae} illustrates the VAE architecture.

\subsection{Mixed-effects VAE for panel data}
\label{subsec:menn}

Standard VAEs typically assume that observations are independently and identically distributed (i.i.d.), and therefore ignore correlations between observations \citep{bouchacourt2018}. However, many CM datasets have a panel structure, in which multiple observations are observed for the same individual, which are correlated rather than independent \citep{louviere1983}\footnote{In ML and other fields, panel data are often referred to as longitudinal or sequential data. We use the panel data terminology throughout, as this work is primarily CM oriented.}. The i.i.d.\ assumption therefore does not hold, and this matters for AD: an observation that appears anomalous when compared with the rest of the dataset can be reasonable in the context of that individual's other observations, and an observation that appears typical in the sample can be atypical for that individual.

To address this limitation, we use mixed-effects neural networks \citep{xiong2019}. These models are designed for grouped data and combine a fixed component, which captures relationships common to all groups, with a random component, which captures group-specific deviations. This is analogous to the panel specification of the mixed logit model, in which random parameters are constant across the choices of the same individual to account for inter-individual heterogeneity \citep{hensher2003mixed}.

In a standard VAE, the encoder $q_\phi$ and the decoder $p_\theta$ play the role of the fixed effects, since both are shared across all individuals $n$ and observations $t$. We extend the decoder in \autoref{eq:decoder} with an individual-specific random-effects latent variable $u_n$ of dimension $D_u$, drawn from a population-level distribution

\begin{equation}
u_n \sim \mathcal{N}(0, \Omega_u),
\label{eq:re_prior}
\end{equation}

where $\Omega_u$ is a full $D_u \times D_u$ covariance matrix, estimated jointly with $\phi$ and $\theta$ rather than fixed in advance \citep{xiong2019} (analogous to the covariance matrix in a mixed logit). $D_u$ is selected through hyperparameter search. 

Each individual's random effect is described by a normal distribution $q_{\psi_n}(u_n) = \mathcal{N}\big(\mu_{u,n}, \mathrm{diag}(\sigma_{u,n}^2)\big)$, whose mean and standard deviation, $\psi_n=(\mu_{u,n},\sigma_{u,n})$, are estimated directly for each individual $n$, rather than computed by the encoder. During training, $u_n$ is drawn from this distribution with the same reparametrisation trick as $z_{nt}$ (\autoref{eq:reparam}). The random effects $u_n$ enter only the decoder: $z_{nt}$ is still inferred from $x_{nt}$ alone, and the two latent variables are concatenated into a single vector $[z_{nt}, u_n]$ of dimension $D_z + D_u$, which is the input of the decoder:

\begin{equation}
\left(\mu_{\widehat{x},ntk},\, \sigma_{\widehat{x},ntk},\, \tau_{ntkc}\right) = g\!\left([z_{nt}, u_n];\ \theta\right).
\label{eq:decoder_me}
\end{equation}

Because $u_n$ is shared by all observations of individual $n$, that individual's observations are conditionally independent given $(z_{nt}, u_n)$ but dependent once $u_n$ is integrated out, as in panel mixed logit models. This relaxes the i.i.d.\ assumption. The loss function changes accordingly: the reconstruction term now conditions on $u_n$, $p_\theta(x_{ntk}\mid z_{nt}, u_n)$, and a second KL term, weighted by $\lambda_u$, keeps each individual's random effect close to the population distribution:

\begin{equation}
\begin{aligned}
\mathcal{L}_{\phi,\theta,\psi,\Omega_u}
= {} & -\sum_{n=1}^{N}\sum_{t=1}^{T_n}
\mathbb{E}_{q_\phi(z_{nt}\mid x_{nt})\,q_{\psi_n}(u_n)}\!\left[\sum_{k=1}^{K} \log p_\theta(x_{ntk}\mid z_{nt},u_n)\right]\\
& + \lambda\sum_{n=1}^{N}\sum_{t=1}^{T_n} D_{KL}\!\left(q_\phi(z_{nt}\mid x_{nt})\,\|\,\mathcal{N}(0,I_{D_z})\right)\\&
+ \lambda_u\sum_{n=1}^{N} D_{KL}\!\left(q_{\psi_n}(u_n)\,\|\,\mathcal{N}(0,\Omega_u)\right),
\end{aligned}
\label{eq:loss_me}
\end{equation}

where $\log p_\theta(x_{ntk}\mid z_{nt},u_n)$ is given by \autoref{eq:loss_num} and \autoref{eq:loss_cat}, for continuous and categorical variables, respectively.

With $\lambda=\lambda_u=1$, \autoref{eq:loss_me} is the negative ELBO of the log-likelihood of the panel,
\[
\sum_{n=1}^{N}\log\int\prod_{t=1}^{T_n}\left(\int p_\theta(x_{nt}\mid z_{nt},u_n)\,\mathcal{N}(z_{nt};0,I_{D_z})\,dz_{nt}\right)\mathcal{N}(u_n;0,\Omega_u)\,du_n,
\]
so minimising it approximately maximises the likelihood with respect to $\theta$ and $\Omega_u$, as in a mixed logit, with the integrals replaced by expectations over $q_\phi$ and $q_{\psi_n}$. The expectation is estimated with one draw of $z_{nt}$ and $u_n$ per observation at each training step. In each mini-batch, the KL term of $u_n$ is divided equally among the $T_n$ observations of individual $n$, so that it is counted once per individual in each epoch.

Since $q_{\psi_n}(u_n)$ and its population distribution are both Gaussian, the added KL term has a closed form:

\begin{equation}
D_{KL}\!\left(q_{\psi_n}(u_n)\,\|\,\mathcal{N}(0,\Omega_u)\right)
= \tfrac{1}{2}\left(\mathrm{tr}\big(\Omega_u^{-1}\mathrm{diag}(\sigma_{u,n}^2)\big) + \mu_{u,n}^\top \Omega_u^{-1} \mu_{u,n} - D_u + \log\frac{\det(\Omega_u)}{\prod_{d=1}^{D_u} \sigma_{u,nd}^2}\right).
\label{eq:kl_re}
\end{equation}

\subsection{Anomaly scores}
\label{subsec:rec_probs}

For our anomaly scores, we use reconstruction probabilities \citep{an2015}, which differ from traditional reconstruction errors in that they not only consider the difference between the observed and reconstructed values but also account for the variability of the reconstruction by considering the variance parameters of the reconstruction distribution \citep{iqbal2023}. Essentially, a point-based metric, such as mean squared error, would only use $\mu_{\widehat{x},ntk}$ by computing the differences between the means. Conversely, reconstruction probabilities use both $\mu_{\widehat{x},ntk}$ and $\sigma_{\widehat{x},ntk}$, so that a given residual counts less for a variable that the decoder reconstructs with high uncertainty. 

The reconstruction probability of variable $k$ in observation $t$ of individual $n$ is averaged over $M$ draws of the latent variables, $z^{(m)}_{nt}\sim q_\phi(z_{nt}\mid x_{nt})$ and $u^{(m)}_n\sim q_{\psi_n}(u_n)$, and expressed as a negative log-probability:

\begin{equation}
\text{Reconstruction probability}_{ntk} = -\frac{1}{M}\sum_{m=1}^{M}\log p_\theta\!\left(x_{ntk}\mid z^{(m)}_{nt},u^{(m)}_n\right),
\label{eq:rec_prob}
\end{equation}

where each term is given by \autoref{eq:loss_num} or \autoref{eq:loss_cat}. Higher values indicate variables that the VAE reconstructs worse.

The anomaly score of observation $t$ of individual $n$ is the sum of the reconstruction probabilities of its variables and the KL term of its latent representation:

\begin{equation}
\text{Anomaly score}_{nt} = \sum_{k=1}^{K}\text{Reconstruction probability}_{ntk} + D_{KL}\!\left(q_\phi(z_{nt}\mid x_{nt})\,\|\,\mathcal{N}(0,I_{D_z})\right),
\label{eq:score1}
\end{equation}

where the KL term is given by \autoref{eq:kl_z}. The first part measures how poorly the variables of the observation are reconstructed, and the second how far its latent representation is from the prior, so that an observation whose variables are reconstructed well can still be anomalous if it can only be encoded in an atypical region of the latent space. 

The anomaly score of individual $n$ adds the KL term of their random effect to the average anomaly score of their observations:

\begin{equation}
\text{Anomaly score}_{n} = \frac{1}{T_n}\sum_{t=1}^{T_n}\text{Anomaly score}_{nt} + D_{KL}\!\left(q_{\psi_n}(u_n)\,\|\,\mathcal{N}(0,\Omega_u)\right),
\label{eq:score_ind}
\end{equation}

where the KL term is given by \autoref{eq:kl_re}, and measures how far the random effect of the individual is from the population distribution. The anomaly scores of the observations are averaged, so that individuals with different numbers of observations can be compared.

High anomaly scores indicate that the observation deviates substantially from the patterns captured by the VAE, and are therefore interpreted as anomalies. An important clarification is that a high anomaly score suggests that the joint configuration of variables of the observation is atypical relative to the rest of the dataset, rather than reflecting an extreme value in a single variable. Such anomalies may arise from data quality issues, such as misreported variables or coding errors, but they may also reflect genuinely unusual choice situations characterised by rare or extreme variable combinations.

\subsection{Model convergence and validation}
\label{subsec:vae_conv}

Ensuring the reliability of a generative model requires both verifying convergence during training and validating the quality of the generated data. Convergence can be checked by inspecting the evolution of the loss function over training epochs, disaggregating it into its different components. The loss function is composed of the reconstruction term and the KL terms (\autoref{eq:loss_me}). We also need to verify that neither component dominates the loss function: if the KL term approaches zero, the latent variables carry no information and the decoder ignores them, whereas if the reconstruction loss dominates, the latent space is barely regularised and the model behaves like a standard autoencoder.

However, convergence is not the only criterion that has to be met. We also need to check that the distributions of the variables generated by the VAE match those of the original data \citep{borysov2019}. For the categorical variables, we use the Total Variation (TV) distance \citep{devroye2018}, and the Jensen-Shannon Divergence (JSD) \citep{menendez1997}. TV measures the maximum discrepancy in probability mass between two distributions and ranges from 0 to 1, with values close to 0 indicating strong similarity. JSD is a symmetric and smoothed version of the KL divergence that quantifies the similarity in information content between two distributions, also bounded between 0 and 1, with lower values indicating better agreement. For continuous variables, we instead compute the Wasserstein (W) distance \citep{panaretos2019} and the Kolmogorov–-Smirnov (KS) statistic \citep{lopes2025}. The Wasserstein distance measures the minimal effort required to transform one distribution into another, making it sensitive to differences in both the shape and the location of the distributions. Lower W values indicate that the reconstructed and original distributions are close in a geometric sense. The KS statistic is the maximum difference between the empirical cumulative distribution functions of the real and reconstructed variables. It ranges from 0 to 1, and small values indicate that the two samples come from similar underlying distributions. Together, these metrics assess whether the VAE reproduces the univariate distributions of the input variables, confirming whether the model has adequately learned to reconstruct each variable from the latent representation \citep{zhang2023vae}.

Nevertheless, this is not sufficient, as a generative model can match each variable separately and still fail to reproduce the dependence structure \citep{borysov2019}. We also need to check that correlations between variables match those in the original data. For example, a model could match the distribution of age and income perfectly and still generate a negative association between them, even if the real data show a positive one. For categorical variables, we can compute Cramér’s V \citep{cramer1999}, which is a statistical measure of association between two categorical variables, ranging from 0 to 1, where 0 indicates no relationship, and 1 indicates perfect association. For continuous variables, there is no gold standard, but we can use Pearson's correlation index \citep{zhangh2023,titar2024, jackson2026}. We therefore compute two matrices, one with Cramér's V between every pair of categorical variables and one with Pearson's correlation between every pair of continuous variables, separately for the real and the synthetic data, and compare each pair of matrices. The synthetic data reproduce the dependence structure well if the matrices look alike, if variables that are strongly related in the real data remain strongly related in the synthetic data, and if the mean absolute difference between corresponding entries is small \citep{zhang2023}.

\section{Synthetic case study}
\label{sec:cases_synth}

We first introduce a synthetic case study where we generate panel choice data with a fully specified data-generating process, so that we know the distributions and correlations of the generated variables, and the utility functions for each individual's choices. As the CM literature lacks a standardised definition for what constitutes an anomaly (or outlier), we induce anomalies of known types, described below, and test whether our approach flags them. We also compare our approach with the two simpler methods described in \autoref{subsec:ad_methods}, univariate checks and the Mahalanobis distance. 

We generate data for 1,000 individuals, with 10 trips each. Each trip has three alternatives, car, bus and rail, which are always available, and each alternative is described by its travel time and travel cost. Each trip also has a recorded trip type with three levels, A, B and C, (which could represent, e.g. trip purposes), and which make up 50\%, 30\% and 20\% of the trips, respectively. 

To make this more consistent with real-world scenarios, travel times and costs depend on the length of each trip, which is not recorded and follows a log-normal distribution with a median of 5, 15 and 40 km for trip types A, B and C, respectively, and a standard deviation of 0.5 on the log scale. The travel time of each alternative is the sum of a fixed component (e.g., access time), and a variable component given the assumed speed of that alternative. Similarly, the travel cost is the sum of a fixed charge and a charge per kilometre. Speeds and charges per kilometre depend on the trip type, and are shown in \autoref{tab:dgp_synth}. Moreover, the fixed time, the speed and the charge per kilometre of each alternative vary across individuals, but little across the trips of the same individual, so that individuals differ consistently from one another.

\begin{table}[h]
\centering
\caption{Parameters used to generate the travel times and costs of the synthetic dataset, disaggregated by alternative and trip type. Fixed times follow a log-normal distribution with a standard deviation of 0.5 on the log scale between individuals and 0.1 within individuals. Speeds and charges per kilometre are multiplied by individual-specific log-normal factors with a median of 1 and a standard deviation of 0.25 on the log scale.}
\label{tab:dgp_synth}
\footnotesize
\setlength{\tabcolsep}{4pt}
\begin{tabular}{lcccccccc}
\toprule
 & \multicolumn{4}{c}{\textbf{Travel time}} & \multicolumn{4}{c}{\textbf{Travel cost}} \\
\cmidrule(lr){2-5}\cmidrule(lr){6-9}
 & \textbf{Median fixed} & \multicolumn{3}{c}{\textbf{Speed (km/h)}} & \textbf{Fixed} & \multicolumn{3}{c}{\textbf{Cost per km (£)}} \\
\cmidrule(lr){3-5}\cmidrule(lr){7-9}
\textbf{Alternative} & \textbf{time (min)} & A & B & C & \textbf{cost (£)} & A & B & C \\
\midrule
Car  & 5  & 30 & 50 & 80  & 3.0 & 0.15 & 0.25 & 0.40 \\
Bus  & 8  & 15 & 25 & 40  & 2.0 & 0.03 & 0.05 & 0.08 \\
Rail & 15 & 40 & 60 & 100 & 3.0 & 0.09 & 0.15 & 0.24 \\
\bottomrule
\end{tabular}
\end{table}

As a result, all travel times and costs are positive and right-skewed, with medians of 21, 39 and 30 minutes and £4.8, £2.4 and £4.1 for car, bus and rail, respectively. They are also strongly correlated (between 0.61 and 0.97), as they all depend on the trip length. However, these relationships are not linear: because of the fixed components, cost does not increase proportionally with time, and because speeds and charges differ by trip type, the data follow a different curve for each trip type, so that a combination of time and cost that is normal for one trip type may be unusual for another. Moreover, as the fixed time, speed, and cost rate are specific to each individual, individuals differ consistently from one another.

Choices are computed from the known utility functions of a panel mixed logit model with random heterogeneity in the travel time component:

\begin{equation}
    V_{jnt} = ASC_j + \beta_{tn}\,\text{time}_{jnt} + \beta_c\,\text{cost}_{jnt},
    \qquad \beta_{tn} = -\exp\left(\mu_{\beta_t} + \sigma_{\beta_t}\,\xi_n\right), \quad \xi_n \sim N(0,1),
\end{equation}

where rail is the reference alternative, the alternative-specific constants are $ASC_{car} = 0.5$ and $ASC_{bus} = -0.5$, $\beta_c = -0.25$, and the time parameter $\beta_{tn}$ follows a negative log-normal distribution across individuals, with $\mu_{\beta_t} = -2.925$ and $\sigma_{\beta_t} = 0.472$. It is important to clarify that $\mu_{\beta_t}$ and $\sigma_{\beta_t}$ follow the standard notation for random parameters in choice models \citep{hess2019apollo}, and are not related to the parameters of the decoder in \autoref{subsec:vae_arq}. The true mean value of travel time (VoTT) is therefore £14.4 per hour, and car, bus and rail are chosen in 50\%, 20\% and 30\% of the trips, respectively.

Because we know the data-generating process, observations that seem unusual under it would not be true anomalies. Hence, we induced anomalies of four types to understand if our approach is able to flag these observations, described below:

\begin{itemize}
    \item Underestimated travel times (250 trips), where the travel time of a non-chosen alternative is recomputed with a trip length two to three times shorter than the true one, while its cost is kept. The alternative is therefore recorded as faster than it was. Each value is plausible on its own, but they do not fit together given the other alternatives of the trip.
    \item Trip type errors (250 trips), where the time and cost of a non-chosen alternative are generated with the speeds and charges of a different trip type. They are normal for the sample as a whole, but not given the trip type.
    \item Individual inconsistent observations (250 trips), where for some individuals we add an 11th trip, which is a normal trip of another individual, as would happen if a record were linked to the wrong respondent. 
    \item Extreme values (250 trips), where the cost of an alternative is multiplied by ten. 
\end{itemize}

Individual inconsistent observations are limited by the number of individuals, as we add at most one to each, and they are normal for the sample, but not at the individual level. Underestimated travel times and trip type errors are induced in a non-chosen alternative, so that the choice was made with the true values. 

We aim to analyse whether excluding the flagged observations implies shifts in the parameter estimates, as the induced anomalies should bias the estimates, and a given method could potentially flag as an anomaly a normal observation, and vice versa. Correctly identifying and removing the anomalies should reduce the bias, whereas removing normal observations should do the opposite. 

To understand the results, we also examine which induced anomalies each method flags. The univariate check follows the thresholds commonly used in practice: it flags the trips in which the travel time or cost of any alternative is above a given percentile of that variable in the sample. For example, flagging 5\% of the observations corresponds to the 99th percentile, i.e., to trips with a travel time above 80, 162 and 82 minutes, or a cost above £43.6, £12.7 and £31.9, for car, bus and rail, respectively. The Mahalanobis distance is computed on the same six variables, i.e., time and cost of the three alternatives, and flags observations that lie far from the centre of the sample, given the correlations between these variables. We expect all methods to detect extreme costs, as these involve a single value far in the tail of its distribution. In contrast, underestimated travel times and trip type errors have values that are plausible on their own, so we expect them to be harder to detect for conventional methods. As the VAE learns how the variables relate to each other and to the trip type, we expect it to detect them more often. Individual inconsistent observations are normal for the sample, so they can only be identified by comparing them with the other observations of the same individual. We therefore expect them to be detected mainly by the VAE with mixed effects, as the other methods assess each observation against the whole sample.

Travel times and costs are log-transformed and standardised, and the trip type and the choice are treated as categorical variables. We estimate two variants of the VAE, with and without mixed effects, each with its own hyperparameters, shown in \autoref{tab:hyper_synth}. Across 5 seeds, the scores have an average Spearman correlation of 0.97; we report the results of the model trained with the first seed.

\begin{table}[h]
\centering
\caption{Hyperparameter search space and selected configuration for each VAE in the synthetic dataset, without (V) and with mixed effects ($\text{V}_{\text{ME}}$).}
\label{tab:hyper_synth}
\footnotesize
\setlength{\tabcolsep}{4pt}
\begin{tabular}{lccc}
\toprule
\textbf{Hyperparameter} & \textbf{Search space} & \textbf{V} & $\textbf{V}_{\textbf{ME}}$ \\
\midrule
Latent dimension ($D_z$)         & 2, 3, 4, 6 & 4 & 3 \\
Random-effect dimension ($D_u$)  & 2, 4, 8 & -- & 8 \\
Hidden layers                    & 1, 2, 3, 4 & 2 & 3 \\
Hidden layer size                & 4, 8, 16, 32, 64, 96, 128 & 128 & 32 \\
Learning rate                    & $10^{-4}$, $5\times10^{-4}$, $10^{-3}$, $2\times10^{-3}$, $5\times10^{-3}$ & $10^{-3}$ & $2\times10^{-3}$ \\
Batch size                       & 8, 16, 32, 64, 128 & 128 & 32 \\
Weight decay                     & 0, $10^{-6}$, $10^{-5}$, $10^{-4}$ & $10^{-6}$ & 0 \\
\bottomrule
\end{tabular}
\end{table}

\autoref{tab:synth_approaches} shows the share of induced anomalies of each type flagged by each method among the 1\%, 5\% and 10\% of observations with the highest scores, and the number of normal observations flagged. As expected, most extreme values are detected by all methods. The Mahalanobis distance and both VAEs recover almost all of them with the top 5\%, whereas the univariate check recovers 76\% with the top 5\% and 87\% with the top 10\%, as a cost multiplied by ten on a short trip can still be lower than the cost of a long trip. Larger differences arise for the other three types. The univariate check flags none of the underestimated travel times and trip type errors, even with the top 10\%, and the Mahalanobis distance flags 11\% and 3\% of them with the top 5\%, and 24\% and 16\% with the top 10\%. In contrast, the VAE with mixed effects flags 17\% of the underestimated travel times and 28\% of the trip type errors with the top 5\%, and 38\% and 66\% with the top 10\%, although underestimated travel times remain the hardest type to detect. It is also the method that detects the most individual inconsistent observations, flagging 20\%, 48\% and 70\% of them with the top 1\%, 5\% and 10\%, compared with at most 11\% for the other methods.

The mixed effects make the largest difference. Without them, the VAE flags only 6\% of the underestimated travel times, 14\% of the trip type errors and 1\% of the individual inconsistent observations with $\alpha=0.95$. Indeed, for the latter, the added trip is a normal trip for the sample as a whole, so it can only be identified by comparing it with the individual's other trips. The mixed effects also reduce the number of normal observations that are flagged, from 210 to 32 with the top 5\%, the fewest of all methods at every share. 

\begin{table}[h]
\centering
\caption{Share of induced anomalies of each type flagged by the univariate check (U), the Mahalanobis distance (M), and the VAE without (V) and with ($\text{V}_{\text{ME}}$) mixed effects among the 1\%, 5\% and 10\% of observations with the highest scores. As there are 250 anomalies of each type and the top 1\% contains 102 observations, at most 41\% of a type can be flagged with the top 1\%. The last row shows the number of normal observations flagged.}
\label{tab:synth_approaches}
\footnotesize
\setlength{\tabcolsep}{4pt}
\resizebox{\textwidth}{!}{%
\begin{tabular}{lcccccccccccc}
\toprule
 & \multicolumn{4}{c}{\textbf{Top 1\%}} & \multicolumn{4}{c}{\textbf{Top 5\%}} & \multicolumn{4}{c}{\textbf{Top 10\%}} \\
\cmidrule(lr){2-5}\cmidrule(lr){6-9}\cmidrule(lr){10-13}
\textbf{Type of anomaly} & U & M & V & $\text{V}_{\text{ME}}$ & U & M & V & $\text{V}_{\text{ME}}$ & U & M & V & $\text{V}_{\text{ME}}$ \\
\midrule
Extreme values (250) & 39\% & 39\% & 35\% & 15\% & 76\% & 100\% & 99\% & 100\% & 87\% & 100\% & 100\% & 100\% \\
Underestimated travel times (250) & 0\% & 0\% & 0\% & 2\% & 0\% & 11\% & 6\% & 17\% & 0\% & 24\% & 15\% & 38\% \\
Trip type errors (250) & 0\% & 0\% & 2\% & 3\% & 0\% & 3\% & 14\% & 28\% & 0\% & 16\% & 28\% & 66\% \\
Individual inconsistent obs. (250) & 0\% & 0\% & 0\% & 20\% & 4\% & 4\% & 1\% & 48\% & 11\% & 8\% & 6\% & 70\% \\
\midrule
All induced anomalies (1,000) & 10\% & 10\% & 9\% & 10\% & 20\% & 29\% & 30\% & 48\% & 24\% & 37\% & 37\% & 68\% \\
\midrule
Normal observations flagged & 4 & 5 & 10 & 2 & 314 & 218 & 210 & 32 & 781 & 657 & 654 & 341 \\
\bottomrule
\end{tabular}}
\end{table}

To examine why observations are flagged, \autoref{tab:anomalies_synth} shows the fifteen observations with the highest scores of the VAE with mixed effects, with the travel time and cost of each alternative, and their percentiles among all normal observations in brackets. All fifteen are induced anomalies. The first ten are individual inconsistent observations, and only two of them are also flagged by another method with $\alpha=0.95$. Most of their values are common in the sample as a whole, but at least one of them lies far from the other trips of the same individual, and in eight of these ten this is the case for the rail time, which is the alternative with the longest fixed time, and therefore the one that differs most between individuals. For example, all values of trip 8326 lie between the 7th and 20th percentiles of the sample, but its rail time is far from those of the other trips of individual 812. The other three types follow closely, with an underestimated travel time, an extreme value and a trip type error at ranks 11, 13 and 14, respectively. Indeed, the VAE with mixed effects does not only detect inconsistencies within individuals: compared with the VAE without mixed effects, it also flags more underestimated travel times and trip type errors, and fewer normal observations (\autoref{tab:synth_approaches}).

The first normal observation appears only at rank 78. As shown in \autoref{tab:synth_approaches}, more normal observations are flagged as $\alpha$ decreases, and the middle block of \autoref{tab:anomalies_synth} shows the five with the highest scores. Two of them, trips 1829 and 2332, are very long trips, with travel times above the 99th percentile of the sample for all alternatives, and the other three have at least two values above the 95th percentile, and some values far from the other trips of the same individual. As their values lie in the tails of the distributions, all five are also flagged by the VAE without mixed effects, and three of them by each of the univariate check and the Mahalanobis distance. They are not errors but rare observations from the tails of the data-generating process, and more of them are flagged with lower values of $\alpha$. Finally, the bottom block shows the five induced anomalies with the highest scores among those not flagged with $\alpha = 0.95$: three trip type errors, one individual inconsistent observation and one extreme value. Their scores are only slightly below the threshold, so all five are flagged with $\alpha = 0.90$. The extreme value is flagged with $\alpha=0.95$ by all three other methods, and one of the trip type errors by the VAE without mixed effects.

\begin{table}[htbp]
\centering
\caption{Fifteen observations with the highest anomaly scores of the VAE with mixed effects (top block), top five normal observations flagged as anomalies (middle block), and the five induced anomalies with the highest scores that were not flagged with $\alpha=0.95$ (bottom block). Travel time (tt) in minutes and travel cost (tc) in £. Percentiles among all normal observations in brackets. Type: individual inconsistent observation (II), underestimated travel time (UT), extreme value (EV), trip type error (TE) and normal observation (N). Also flagged by: other methods flagging the observation with $\alpha=0.95$: univariate check (U), Mahalanobis distance (M) and VAE without mixed effects (V).}
\label{tab:anomalies_synth}
\footnotesize
\setlength{\tabcolsep}{3pt}
\resizebox{\textwidth}{!}{%
\begin{tabular}{cccccccccccccc}
\toprule
 & & & & \multicolumn{2}{c}{\textbf{Car}} & \multicolumn{2}{c}{\textbf{Bus}} & \multicolumn{2}{c}{\textbf{Rail}} & & & & \\
\cmidrule(lr){5-6}\cmidrule(lr){7-8}\cmidrule(lr){9-10}
\textbf{Rank} & \textbf{Trip} & \textbf{Individual} & \textbf{Trip type} & \textbf{tt} & \textbf{tc} & \textbf{tt} & \textbf{tc} & \textbf{tt} & \textbf{tc} & \textbf{Choice} & \textbf{Score} & \textbf{Type} & \textbf{Also flagged by} \\
\midrule
1 & 2022 & 197 & C & 64.0 (96.6) & 25.1 (93.7) & 127.1 (96.8) & 6.0 (92.6) & 89.7 (99.1) & 24.7 (98.4) & Car & 29.4 & II & -- \\
2 & 639 & 62 & A & 26.9 (68.1) & 3.9 (33.3) & 18.6 (7.8) & 2.2 (37.1) & 28.1 (45.7) & 3.4 (23.7) & Car & 24.1 & II & M \\
3 & 4467 & 435 & C & 34.1 (80.5) & 28.0 (95.3) & 108.0 (94.7) & 7.8 (96.6) & 41.5 (77.4) & 14.1 (91.2) & Rail & 23.7 & II & -- \\
4 & 8326 & 812 & A & 11.8 (13.0) & 3.4 (7.6) & 23.1 (16.5) & 2.1 (8.3) & 15.6 (7.3) & 3.4 (20.1) & Car & 23.1 & II & -- \\
5 & 5415 & 528 & A & 17.9 (38.6) & 3.6 (19.8) & 28.4 (28.5) & 2.1 (23.9) & 38.3 (71.6) & 3.4 (22.9) & Car & 22.3 & II & -- \\
6 & 3688 & 359 & A & 14.2 (22.8) & 3.6 (17.4) & 13.2 (1.4) & 2.1 (19.7) & 15.6 (7.3) & 3.3 (12.3) & Car & 21.8 & II & -- \\
7 & 4883 & 476 & A & 18.6 (41.3) & 3.8 (27.4) & 25.5 (21.8) & 2.2 (31.9) & 54.6 (90.8) & 3.5 (25.2) & Car & 21.7 & II & -- \\
8 & 9276 & 905 & C & 92.9 (99.2) & 28.6 (95.5) & 165.9 (98.7) & 7.4 (96.0) & 72.5 (97.2) & 26.0 (98.6) & Bus & 20.8 & II & -- \\
9 & 7733 & 754 & C & 24.0 (61.2) & 13.1 (82.2) & 76.1 (86.2) & 3.7 (80.0) & 20.2 (19.7) & 11.0 (86.1) & Rail & 20.4 & II & V \\
10 & 874 & 85 & B & 26.7 (67.8) & 5.3 (54.5) & 47.3 (63.3) & 2.7 (62.5) & 27.7 (44.5) & 4.2 (51.6) & Car & 20.3 & II & -- \\
11 & 1095 & 107 & B & 55.7 (94.6) & 11.0 (79.1) & 39.7 (51.5) & 3.7 (79.5) & 56.9 (92.2) & 7.4 (77.5) & Car & 19.7 & UT & M, V \\
12 & 41 & 4 & A & 10.4 (8.1) & 3.3 (3.8) & 26.0 (22.8) & 2.1 (5.3) & 7.9 (0.1) & 3.3 (9.6) & Car & 19.3 & II & M \\
13 & 8547 & 834 & C & 96.4 (99.3) & 475.6 ($>$99.9) & 138.6 (97.5) & 12.4 (99.6) & 73.8 (97.4) & 30.7 (99.2) & Rail & 19.3 & EV & U, M, V \\
14 & 1182 & 116 & A & 11.8 (13.0) & 7.9 (70.3) & 29.2 (30.6) & 2.4 (52.6) & 29.8 (51.0) & 4.3 (54.3) & Bus & 19.1 & TE & V \\
15 & 4650 & 453 & B & 38.7 (85.5) & 6.4 (62.4) & 70.5 (83.4) & 3.5 (77.6) & 71.9 (97.1) & 7.0 (76.1) & Car & 19.0 & II & -- \\
\midrule
78 & 1829 & 179 & A & 109.8 (99.6) & 12.5 (81.4) & 236.1 (99.8) & 3.9 (81.0) & 98.4 (99.5) & 9.7 (83.5) & Rail & 13.8 & N & M, V \\
100 & 8346 & 814 & C & 77.2 (98.4) & 38.0 (98.1) & 91.0 (91.2) & 6.6 (94.3) & 44.8 (82.0) & 14.5 (91.7) & Bus & 13.1 & N & V \\
110 & 2332 & 228 & C & 358.2 ($>$99.9) & 96.2 ($>$99.9) & 486.7 ($>$99.9) & 23.9 ($>$99.9) & 175.0 ($>$99.9) & 59.0 ($>$99.9) & Rail & 13.0 & N & U, M, V \\
162 & 8237 & 804 & B & 88.0 (99.0) & 28.9 (95.7) & 212.1 (99.6) & 8.8 (97.8) & 85.6 (98.9) & 20.2 (96.6) & Car & 11.8 & N & U, V \\
166 & 1691 & 165 & C & 129.1 (99.8) & 45.0 (99.1) & 154.9 (98.2) & 14.6 (99.8) & 64.3 (95.2) & 36.4 (99.6) & Bus & 11.8 & N & U, M, V \\
\midrule
514 & 6501 & 634 & A & 20.0 (47.2) & 3.8 (28.5) & 29.8 (32.0) & 2.1 (20.1) & 27.1 (43.0) & 37.4 (99.7) & Car & 8.1 & EV & U, M, V \\
515 & 6328 & 617 & C & 22.6 (56.5) & 10.0 (77.1) & 38.8 (49.7) & 3.0 (71.0) & 22.0 (25.7) & 5.9 (69.5) & Car & 8.1 & II & -- \\
516 & 4200 & 410 & B & 37.6 (84.3) & 7.5 (68.7) & 34.8 (42.3) & 3.2 (74.2) & 40.5 (75.8) & 5.3 (65.6) & Rail & 8.1 & TE & -- \\
517 & 8319 & 812 & B & 20.6 (49.4) & 6.1 (60.5) & 24.2 (19.1) & 3.3 (75.1) & 42.0 (78.1) & 4.7 (58.3) & Car & 8.0 & TE & -- \\
518 & 7119 & 695 & B & 22.2 (55.0) & 5.6 (57.6) & 33.9 (40.3) & 2.5 (55.4) & 15.7 (7.4) & 6.1 (71.4) & Car & 8.0 & TE & V \\
\bottomrule
\end{tabular}}
\end{table}

Finally, we estimate a panel mixed logit model with the true specification on the full sample (including the induced anomalies), after excluding the 1\%, 5\% and 10\% of observations with the highest scores of each method, and after excluding the same shares of observations at random. 

\autoref{tab:synth_exclusion} shows the estimates, their t-ratios based on robust standard errors, and their percentage difference from the true values. All parameters are highly significant in all cases. Excluding the top 1\% of observations does not lead to major changes in the estimates for any method. With the top 5\%, excluding observations at random or by the Mahalanobis distance leaves the estimates almost unchanged, and the univariate check moves the mean VoTT further from its true value (9.9\% below). The VAE with mixed effects brings the mean VoTT closest to its true value (£13.44 per hour, 6.7\% below). With the top 10\%, the VAE with mixed effects brings all estimates close to their true values: $\beta_c$ is 0.7\% smaller in magnitude, $\mu_{\beta_t}$ 2.2\% larger in magnitude, and $\sigma_{\beta_t}$ 0.9\% smaller than their true values, and the mean VoTT is £13.57 per hour, 5.8\% below its true value. The VAE without mixed effects also brings $\beta_c$ and $\mu_{\beta_t}$ close to their true values, but underestimates $\sigma_{\beta_t}$ by 5.8\%, so that the mean VoTT is £13.22 per hour, 8.2\% below its true value. In contrast, excluding the same share at random leaves the estimates almost unchanged, with a mean VoTT 7.6\% below its true value. The Mahalanobis distance brings $\beta_c$ and $\sigma_{\beta_t}$ closer to their true values than random exclusion, but not $\mu_{\beta_t}$, so that the mean VoTT is 10.5\% below its true value. The univariate check brings $\beta_c$ and $\sigma_{\beta_t}$ closer to their true values, but moves $\mu_{\beta_t}$ further from it, so that the mean VoTT is 13.6\% below its true value. Hence, on the mean VoTT, the univariate check and the Mahalanobis distance perform worse than removing observations at random.

\begin{table}[htbp]
\centering
\caption{Estimates of the panel mixed logit model on the full sample and after excluding the 1\%, 5\% and 10\% of observations with the highest scores of each method: random exclusion (R, mean of 20 repetitions), univariate check (U), Mahalanobis distance (M), and the VAE without (V) and with ($\text{V}_{\text{ME}}$) mixed effects. Robust t-ratios in brackets, using the delta method for the mean VoTT, and percentage difference in magnitude from the true value below each estimate. The last column shows the number of induced anomalies removed (out of 1,000).}
\label{tab:synth_exclusion}
\footnotesize
\setlength{\tabcolsep}{4pt}
\resizebox{\textwidth}{!}{%
\begin{tabular}{lccccccc}
\toprule
 & $ASC_{car}$ & $ASC_{bus}$ & $\beta_c$ & $\mu_{\beta_t}$ & $\sigma_{\beta_t}$ & Mean VoTT (£/h) & Induced removed \\
\midrule
True values & 0.500 & $-$0.500 & $-$0.250 & $-$2.925 & 0.472 & 14.40 & -- \\
Full sample & 0.519 (19.7) & $-$0.548 ($-$14.6) & $-$0.231 ($-$33.0) & $-$3.099 ($-$73.4) & 0.505 (13.2) & 13.30 (26.6) & 0 \\
 & $+$3.8\% & $+$9.7\% & $-$7.5\% & $+$5.9\% & $+$6.9\% & $-$7.6\% & \\[2pt]
\midrule
\multicolumn{8}{l}{Top 1\% excluded} \\
R & 0.519 (19.5) & $-$0.548 ($-$14.5) & $-$0.231 ($-$32.9) & $-$3.098 ($-$73.1) & 0.505 (13.2) & 13.32 (26.4) & 10 \\
 & $+$3.7\% & $+$9.5\% & $-$7.5\% & $+$5.9\% & $+$7.0\% & $-$7.5\% & \\[2pt]
U & 0.525 (19.9) & $-$0.550 ($-$14.5) & $-$0.232 ($-$32.8) & $-$3.103 ($-$73.0) & 0.506 (13.1) & 13.22 (26.4) & 98 \\
 & +5.0\% & +10.1\% & $-$7.3\% & +6.1\% & +7.1\% & $-$8.2\% & \\
M & 0.524 (19.9) & $-$0.545 ($-$14.4) & $-$0.232 ($-$32.9) & $-$3.095 ($-$73.3) & 0.503 (13.2) & 13.31 (26.6) & 102 \\
 & $+$4.7\% & $+$9.0\% & $-$7.3\% & $+$5.8\% & $+$6.6\% & $-$7.5\% & \\[2pt]
V & 0.523 (19.8) & $-$0.555 ($-$14.6) & $-$0.233 ($-$33.0) & $-$3.096 ($-$73.0) & 0.498 (13.0) & 13.17 (26.6) & 92 \\
 & $+$4.6\% & $+$10.9\% & $-$6.7\% & $+$5.9\% & $+$5.5\% & $-$8.5\% & \\[2pt]
$\text{V}_{\text{ME}}$ & 0.516 (19.5) & $-$0.547 ($-$14.4) & $-$0.232 ($-$32.9) & $-$3.097 ($-$72.3) & 0.513 (13.4) & 13.34 (26.3) & 100 \\
 & $+$3.2\% & $+$9.5\% & $-$7.3\% & $+$5.9\% & $+$8.5\% & $-$7.4\% & \\[2pt]
\midrule
\multicolumn{8}{l}{Top 5\% excluded} \\
R & 0.519 (19.1) & $-$0.548 ($-$14.2) & $-$0.231 ($-$32.3) & $-$3.097 ($-$72.1) & 0.502 (12.8) & 13.31 (26.0) & 49 \\
 & $+$3.7\% & $+$9.6\% & $-$7.6\% & $+$5.9\% & $+$6.3\% & $-$7.6\% & \\[2pt]
U & 0.529 (19.6) & $-$0.558 ($-$14.4) & $-$0.235 ($-$31.0) & $-$3.093 ($-$69.7) & 0.478 (11.1) & 12.97 (24.9) & 198 \\
 & +5.9\% & +11.5\% & $-$6.0\% & +5.8\% & +1.1\% & $-$9.9\% & \\
M & 0.524 (19.6) & $-$0.544 ($-$13.9) & $-$0.230 ($-$31.2) & $-$3.099 ($-$71.2) & 0.494 (11.7) & 13.31 (24.8) & 294 \\
 & $+$4.9\% & $+$8.7\% & $-$8.1\% & $+$5.9\% & $+$4.5\% & $-$7.6\% & \\[2pt]
V & 0.523 (19.5) & $-$0.553 ($-$14.1) & $-$0.240 ($-$32.5) & $-$3.049 ($-$73.7) & 0.479 (12.1) & 13.29 (25.8) & 302 \\
 & $+$4.5\% & $+$10.5\% & $-$4.0\% & $+$4.3\% & $+$1.4\% & $-$7.7\% & \\[2pt]
$\text{V}_{\text{ME}}$ & 0.523 (19.5) & $-$0.539 ($-$13.7) & $-$0.237 ($-$32.2) & $-$3.062 ($-$71.8) & 0.501 (12.8) & 13.44 (26.0) & 480 \\
 & $+$4.7\% & $+$7.7\% & $-$5.3\% & $+$4.7\% & $+$6.0\% & $-$6.7\% & \\[2pt]
\midrule
\multicolumn{8}{l}{Top 10\% excluded} \\
R & 0.519 (18.7) & $-$0.547 ($-$13.8) & $-$0.231 ($-$31.2) & $-$3.101 ($-$70.1) & 0.505 (12.3) & 13.31 (25.3) & 103 \\
 & $+$3.8\% & $+$9.4\% & $-$7.7\% & $+$6.0\% & $+$7.0\% & $-$7.6\% & \\[2pt]
U & 0.539 (19.0) & $-$0.576 ($-$14.1) & $-$0.241 ($-$26.9) & $-$3.115 ($-$64.9) & 0.484 (9.2) & 12.44 (20.8) & 244 \\
 & +7.9\% & +15.2\% & $-$3.7\% & +6.5\% & +2.4\% & $-$13.6\% & \\
M & 0.532 (19.1) & $-$0.561 ($-$13.8) & $-$0.237 ($-$30.1) & $-$3.099 ($-$64.6) & 0.494 (10.9) & 12.88 (23.3) & 368 \\
 & $+$6.4\% & $+$12.1\% & $-$5.1\% & $+$6.0\% & $+$4.6\% & $-$10.5\% & \\[2pt]
V & 0.531 (19.1) & $-$0.561 ($-$13.9) & $-$0.252 ($-$31.4) & $-$2.990 ($-$70.6) & 0.445 (10.4) & 13.22 (24.9) & 371 \\
 & $+$6.3\% & $+$12.2\% & $+$0.8\% & $+$2.2\% & $-$5.8\% & $-$8.2\% & \\[2pt]
$\text{V}_{\text{ME}}$ & 0.516 (18.4) & $-$0.542 ($-$13.2) & $-$0.248 ($-$31.3) & $-$2.990 ($-$69.6) & 0.468 (10.6) & 13.57 (24.3) & 684 \\
 & $+$3.3\% & $+$8.4\% & $-$0.7\% & $+$2.2\% & $-$0.9\% & $-$5.8\% & \\[2pt]
\bottomrule
\end{tabular}}
\end{table}

Overall, excluding the observations flagged by the VAE with mixed effects brings the estimates closer to their true values than excluding those flagged by the simpler methods. This is because the VAE with mixed effects detects far more individual inconsistent observations than the other methods, while also detecting trip type errors and underestimated travel times more often than the simpler methods, and it flags fewer normal observations of individuals who differ consistently from the rest of the sample. Without mixed effects, the VAE performs little better than the Mahalanobis distance. However, most underestimated travel times remain undetected, so removing the flagged observations corrects only part of the bias in the estimates.

\section{Real world case studies}
\label{sec:cases_real}

This section presents the results of our method applied to two datasets: an SP dataset---the Swiss route choice dataset (\autoref{subsec:swissroute})---and an RP dataset---the Decisions dataset (\autoref{subsec:decisions}). The two differ in how the data are produced. In the SP dataset, variable levels are set by an experimental design pivoted around each individual’s reference trip, and choices are hypothetical. In the RP dataset, choices are observed, and the variables of the alternatives, such as travel times and distances, are computed from external sources, e.g., routing APIs. Analysing both therefore allows us to examine which types of anomalies arise under each data collection process. 

\subsection{Swiss route dataset}
\label{subsec:swissroute}

We analyse the Swiss route dataset \citep{vrtic2002}, available in $Apollo$ \citep{hess2019apollo}. This dataset contains 388 individuals who participated in a stated choice survey regarding the introduction of a tilting train in Switzerland, yielding 3,492 observations. Each individual was first presented with a complete description of their reported trip, in order to verify this, and then faced nine observations between two route alternatives pivoted on their reported trip. The choices are well balanced: route 1 was chosen 1,743 times (49.9\%), and route 2 was chosen 1,749 times (50.1\%). Each alternative is characterised by travel time, cost, headway, and number of interchanges. We summarise the variables reconstructed by the VAE in \autoref{tab:var_swiss}.

\begin{table}[h!]
    \centering
    \caption{Variables reconstructed by the VAE for the Swiss route dataset. Continuous variables are log-transformed and standardised before training. Categorical variables are one-hot encoded.}
    \footnotesize
    \begin{tabular}{lll}
    \hline
     \textbf{Variable} & \textbf{Type} & \textbf{Description (units)} \\
     \hline
     tt1 & Continuous & Travel time (mins) for route 1. \\
     tc1 & Continuous & Travel cost (CHF) for route 1. \\
     hw1 & Continuous & Headway (mins) for route 1. \\
     ch1 & Categorical & Number of interchanges for route 1. \\
     tt2 & Continuous & Travel time (mins) for route 2. \\
     tc2 & Continuous & Travel cost (CHF) for route 2. \\
     hw2 & Continuous & Headway (mins) for route 2. \\
     ch2 & Categorical & Number of interchanges for route 2. \\
     hh\_inc\_abs & Continuous & Household income (CHF/year). \\
     car\_availability & Categorical & Car availability, 1 if yes, 0 otherwise. \\
     purpose & Categorical & Trip purpose: commute, shopping, business or leisure. \\
     choice & Categorical & 1 if choice is route 1, 2 if choice is route 2\\
     \hline
    \end{tabular}
    \label{tab:var_swiss}
\end{table}

The $\text{VAE}_{\text{ME}}$ architecture for the Swiss route dataset, alongside the search space for the hyperparameter tuning, is presented in \autoref{tab:hyperparam_swiss}.

\begin{table}[h]
\centering
\caption{Hyperparameter search space and selected configuration for the Swiss route dataset.}
\label{tab:hyperparam_swiss}
\footnotesize
\setlength{\tabcolsep}{4pt}
\begin{tabular}{lcc}
\toprule
\textbf{Hyperparameter} & \textbf{Search space} & \textbf{Selected} \\
\midrule
Latent dimension ($D_z$)            & 8, 16, 24, 32, 48, 64  & 8 \\
Random-effect dimension ($D_u$)     & 2, 4, 6, 8             & 4 \\
Hidden layers                       & 1, 2, 3, 4        & 1 \\
Hidden layer size                   & 32, 64, 96, 128, 160, 192       & 128 \\
Learning rate  & $10^{-4}$, $5\times10^{-4}$, $10^{-3}$, $2\times10^{-3}$, $5\times10^{-3}$ & $5\times10^{-3}$ \\
Batch size                          & 32, 64, 128, 256, 512      & 128 \\
Weight decay               & 0, $10^{-6}$, $5\times10^{-6}$, $10^{-5}$, $10^{-4}$  & $10^{-6}$ \\
\bottomrule
\end{tabular}
\end{table}

\begin{figure}[h!]
\centering
\includegraphics[width=0.8\linewidth]{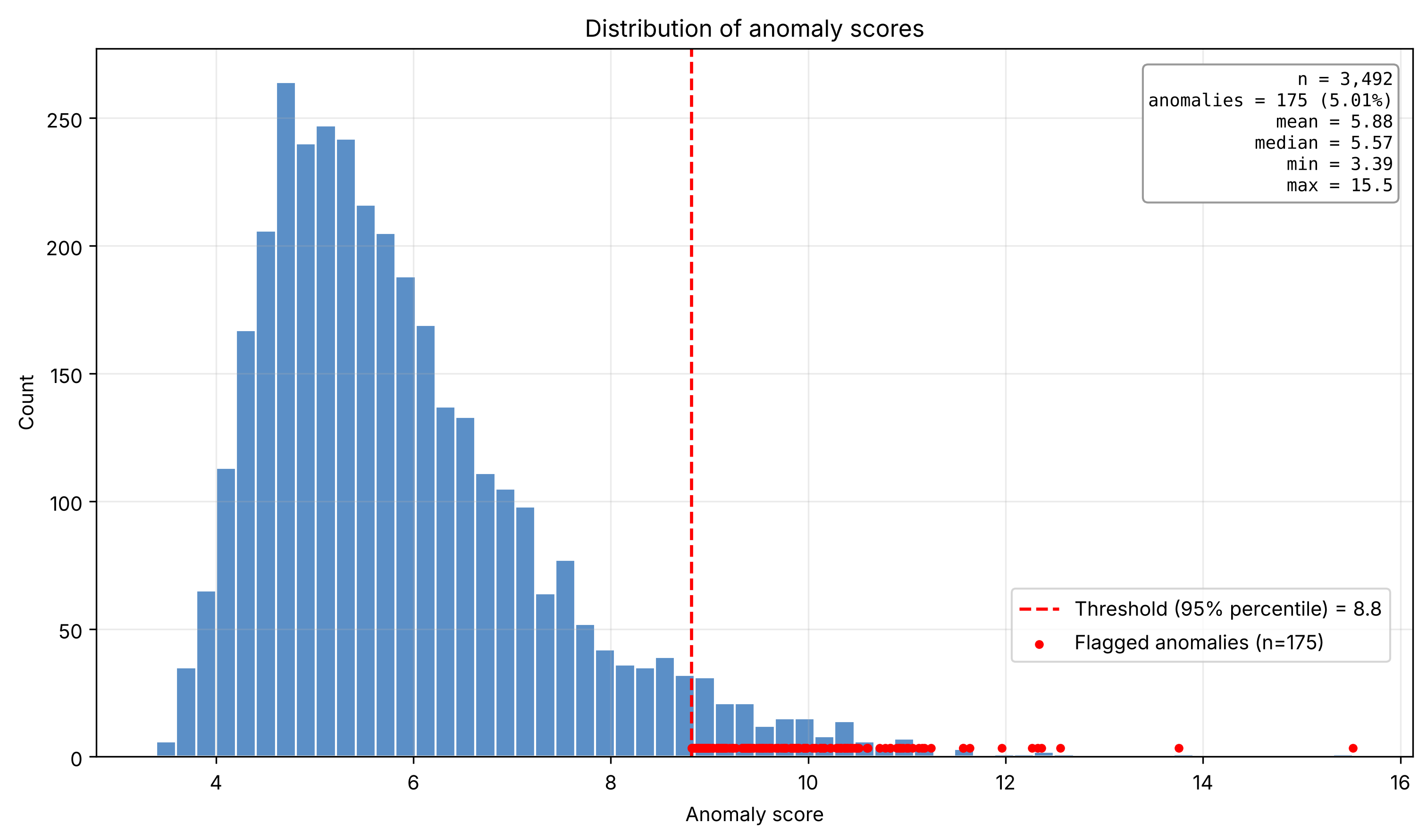}
\caption{Distribution of anomaly scores for the Swiss route dataset. Red dots indicate the 175 observations above the 95th percentile.}
\label{fig:errors_swiss}
\end{figure}

\autoref{fig:errors_swiss} shows the distribution of the anomaly scores. Using the 95th percentile as threshold ($\alpha=0.95$), 175 of the 3,492 observations are flagged as anomalies, shown as red dots. With $\alpha = 0.99$ and $\alpha = 0.90$, 35 and 349 observations are flagged, respectively, and these three thresholds are later used to assess the effect of the flagged observations in CM outputs. The distribution is unimodal and right-skewed: most observations have low scores within a narrow range, and a small number lie in a long upper tail. Most of the flagged observations are just above the threshold, while a few have much larger scores further into the tail. 

The magnitude of the anomaly scores depends on the number and type of variables reconstructed, as the score sums the contributions of all variables and the KL term. The contributions of continuous variables can also be negative, as they are the negative logarithm of a density, which can be larger than one when the variable is reconstructed precisely. The scores are therefore not comparable across datasets, and only their ranking within a dataset is meaningful. This is why the threshold is set as a percentile of the scores rather than as a fixed value, so that the same rule can be applied to any dataset.

An observation is not flagged because of a single value, but because the full set of its variables is unlikely given the rest of the sample. Since the VAE reconstructs the variables of both alternatives together with the socio-demographic variables, the same variable values can be typical for one observation and anomalous for another, depending on, for instance, the trip purpose, income or car availability of the individual. Observations with a high score therefore do not fit the profile of the individual or the trip, which may come from errors in how the data were generated or recorded, or from genuinely unusual trips. This makes the anomaly score a complement to the usual data cleaning checks applied before estimation, which look at each variable separately.

\autoref{tab:anomalies_swiss} presents the ten observations with the highest anomaly scores, with their variable-specific reconstruction probabilities in brackets and their KL terms. We identify two types of observations. Some have variables that do not fit together and are poorly reconstructed. In trip 1526 of individual 17525, which has the highest score, the routes take 25 and 23 minutes but cost only 1 and 2 CHF, and the cost of route 2 is the worst reconstructed variable in the whole sample. In trip 1486, the individual chose route 1, which takes 41 minutes longer and has one more interchange, for 4 CHF less, so the choice is poorly reconstructed. Most of the other observations are reconstructed well, but can only be encoded in an atypical region of the latent space, so that their KL terms are among the highest in the sample. These include very short trips that are expensive for their duration, such as trips 540 and 110, with 4 to 7 minutes for 8 to 12 CHF, and trips 2024 and 2020 of individual 19401, with 3 to 5 minutes for 6 to 9 CHF; very short and cheap trips, such as trips 1573 and 1568 of individual 17669, with 2 to 3 minutes for 1 to 2 CHF; and very long and expensive trips, such as trips 1188 and 1187 of individual 16117, with 101 to 153 minutes for 181 to 268 CHF. Across all flagged observations, the KL term is the largest part of the score, and the flagged observations are much shorter than the rest, with a median travel time of 15 minutes for route 1, compared with 37 minutes in the sample, and more often shopping trips (26\%, compared with 8\%). Among the variables, the headway of route 1 is the most unusual in 26\% of them, followed by the choice (21\%) and the headway of route 2 (19\%).

\begin{sidewaystable}
\centering
\caption{Observations with the ten highest anomaly scores in the Swiss route dataset. Times are in minutes, costs in CHF, headways in minutes, and income in CHF/year. Variable-specific reconstruction probabilities in brackets; higher values indicate variables that the VAE reconstructs worse, and negative values are possible for continuous variables, whose density can exceed one. KL: Kullback--Leibler divergence between the distribution of the latent representation and its prior. The anomaly score is the sum of the variable-specific terms and the KL term. Values discussed in the text are in bold.}
\label{tab:anomalies_swiss}
\small
\setlength{\tabcolsep}{4pt}
\renewcommand{\arraystretch}{1.25}
\resizebox{\textheight}{!}{%
\begin{tabular}{cccccccccccccccc}
\toprule
 & & \multicolumn{4}{c}{\textbf{Route 1}} & \multicolumn{4}{c}{\textbf{Route 2}} & & & & & & \\
\cmidrule(lr){3-6}\cmidrule(lr){7-10}
\textbf{Trip} & \textbf{User} & \textbf{Time} & \textbf{Cost} & \textbf{Headway} & \textbf{Interch.} & \textbf{Time} & \textbf{Cost} & \textbf{Headway} & \textbf{Interch.} & \textbf{Income} & \textbf{Car avail.} & \textbf{Purpose} & \textbf{Choice} & \textbf{KL} & \textbf{Anomaly score} \\
\midrule
1526 & 17525 & 25 ($-$1.3) & \textbf{1 (0.1)} & 15 (0.3) & 0 (0.6) & 23 ($-$0.3) & \textbf{2 (2.0)} & 15 (0.9) & 0 (1.8) & 112,500 (0.1) & 1 (0.0) & Commute (0.1) & 1 (0.5) & 10.6 & 15.5 \\
540 & 14458 & \textbf{6 ($-$1.2)} & \textbf{8 ($-$1.0)} & 30 (0.9) & 2 (1.2) & \textbf{4 ($-$0.7)} & \textbf{11 ($-$0.3)} & 15 (1.1) & 1 (0.6) & 50,000 (0.2) & 0 (0.0) & Shopping (0.0) & 1 (0.8) & 12.1 & 13.8 \\
2024 & 19401 & \textbf{5 ($-$0.8)} & \textbf{6 ($-$0.4)} & 60 (0.3) & 1 (1.0) & \textbf{4 ($-$1.2)} & \textbf{9 (0.5)} & 15 (0.4) & 0 (0.6) & 90,000 (0.1) & 0 (0.0) & Shopping (0.0) & 2 (0.1) & 12.0 & 12.6 \\
110 & 12409 & \textbf{7 ($-$0.6)} & \textbf{9 ($-$1.6)} & 60 (2.1) & 1 (1.0) & \textbf{6 ($-$1.2)} & \textbf{12 ($-$1.3)} & 60 (1.6) & 1 (0.6) & 70,000 (0.1) & 1 (0.0) & Leisure (0.0) & 1 (0.9) & 10.6 & 12.4 \\
1573 & 17669 & \textbf{3 ($-$1.1)} & \textbf{1 ($-$1.3)} & 15 (1.0) & 2 (0.2) & \textbf{2 ($-$0.8)} & \textbf{2 ($-$1.7)} & 60 (1.5) & 0 (0.8) & 70,000 (0.1) & 0 (0.0) & Leisure (0.1) & 2 (0.7) & 12.8 & 12.3 \\
1188 & 16117 & \textbf{145 ($-$1.4)} & \textbf{206 ($-$1.2)} & 30 (0.9) & 2 (0.3) & \textbf{101 ($-$1.0)} & \textbf{268 (0.7)} & 15 (0.4) & 1 (0.6) & 112,500 (0.1) & 0 (0.0) & Leisure (0.1) & 2 (0.1) & 12.7 & 12.3 \\
2020 & 19401 & \textbf{5 ($-$1.2)} & \textbf{7 ($-$1.5)} & 30 (0.9) & 1 (0.7) & \textbf{3 (0.5)} & \textbf{8 ($-$1.7)} & 30 (0.8) & 0 (0.8) & 90,000 (0.1) & 0 (0.0) & Shopping (0.1) & 2 (0.4) & 12.1 & 12.0 \\
1187 & 16117 & \textbf{153 ($-$1.4)} & \textbf{181 ($-$0.1)} & 60 (1.2) & 2 (0.3) & \textbf{115 ($-$1.4)} & \textbf{199 ($-$1.4)} & 30 (1.2) & 2 (1.1) & 112,500 (0.1) & 0 (0.0) & Leisure (0.0) & 2 (0.2) & 11.8 & 11.6 \\
1486 & 17187 & \textbf{274 ($-$1.1)} & \textbf{41 ($-$1.4)} & 30 (1.6) & 1 (0.8) & 233 ($-$0.9) & 45 ($-$1.4) & 30 (0.7) & 0 (0.8) & 50,000 (0.1) & 0 (0.1) & Leisure (0.0) & \textbf{1 (2.2)} & 10.0 & 11.6 \\
1568 & 17669 & \textbf{2 ($-$1.0)} & \textbf{2 ($-$1.7)} & 60 (1.0) & 1 (1.9) & \textbf{3 ($-$1.1)} & \textbf{1 ($-$1.3)} & 60 (0.3) & 1 (0.9) & 70,000 (0.1) & 0 (0.0) & Leisure (0.1) & 2 (0.3) & 12.0 & 11.6 \\
\bottomrule
\end{tabular}}
\end{sidewaystable}

We also summarise the anomaly scores at the individual level in \autoref{tab:anomalies_users_swiss}, which presents the ten individuals with the highest anomaly scores, with the components of their score, the share of their observations flagged with $\alpha = 0.95$, and the characteristics of their reported trip, around which their observations were pivoted. Unlike the observation-level scores, no individual stands out: the scores of these ten individuals are between 14.1 and 14.9, compared with 10.7 on average. Most of the difference comes from the KL terms of their observations, i.e., their observations can only be encoded in atypical regions of the latent space, whereas their variables are reconstructed reasonably well. Individual 14167 is the exception: none of their observations is flagged, but their random effect is the furthest from the population distribution in the whole sample. For the others, between 22\% and 89\% of their observations are flagged, compared with 5\% on average, and together the ten individuals account for 53 of the 175 flagged observations, whereas 80\% of the individuals have no flagged observation at all.

Most of these individuals reported either very long trips or trips with unusual costs. Individuals 16117, 16184 and 16617 reported trips of 173 to 391 km, at or above the 92nd percentile. Individuals 16169 and 12409 reported short trips, of 7 and 10 km, that are expensive per kilometre (0.86 and 0.90 CHF/km, at the 99th percentile), and individual 12640 reported a cost of 52 CHF for a 39 km business trip (1.33 CHF/km). Individuals 14041 and 15704, on the other hand, own a GA travel card and reported very cheap trips, of 1 CHF for 6 km and 9 CHF for 110 km.

\begin{table}[htbp]
\centering
\caption{Ten individuals with the highest anomaly scores in the Swiss route dataset, the characteristics of the reported trip on which their observations were pivoted, and the components of their anomaly score: the average over their nine observations of the sum of the reconstruction probabilities and of the KL term of the latent representation, and the KL term of their random effect. The last column is the share of their observations flagged with $\alpha = 0.95$. Distance in kilometres, time in minutes, cost in CHF. Percentiles across the 388 individuals in brackets. HF: half-fare card, GA: GA travel card.}
\label{tab:anomalies_users_swiss}
\small
\resizebox{\textwidth}{!}{%
\begin{tabular}{cccccccccccc}
\toprule
 & & & & & & & \multicolumn{4}{c}{\textbf{Anomaly score}} & \\
\cmidrule(lr){8-11}
\textbf{ID} & \textbf{Purpose} & \textbf{Distance} & \textbf{Time} & \textbf{Cost} & \textbf{Cost/km} & \textbf{Ticket} & \textbf{Rec. prob.} & \textbf{KL ($z$)} & \textbf{KL ($u$)} & \textbf{Total} & \textbf{Flagged (\%)} \\
\midrule
16117 & Leisure & 215 (96) & 161 (96) & 172 (100) & 0.80 (98) & HF & $-$1.6 & 11.4 & 5.1 & 14.9 & 67 \\
12640 & Business & 39 (46) & 38 (48) & 52 (93) & 1.33 (100) & HF & $-$2.0 & 10.9 & 5.7 & 14.6 & 56 \\
16184 & Business & 173 (92) & 112 (86) & 74 (98) & 0.43 (87) & HF & $-$0.7 & 9.0 & 6.2 & 14.5 & 22 \\
16617 & Leisure & 391 (100) & 296 (100) & 66 (96) & 0.17 (25) & GA & $-$1.5 & 10.5 & 5.4 & 14.5 & 67 \\
14167 & Business & 24 (32) & 19 (23) & 9 (46) & 0.38 (80) & HF & $-$1.7 & 8.0 & 8.2 & 14.4 & 0 \\
20352 & Leisure & 29 (37) & 16 (18) & 21 (71) & 0.72 (97) & None & $-$1.3 & 11.1 & 4.7 & 14.4 & 89 \\
14041 & Shopping & 6 (3) & 28 (36) & 1 (2) & 0.17 (19) & GA & $-$2.5 & 12.2 & 4.5 & 14.2 & 89 \\
15704 & Business & 110 (78) & 108 (84) & 9 (46) & 0.08 (1) & GA & $-$2.5 & 10.0 & 6.6 & 14.2 & 22 \\
16169 & Shopping & 7 (4) & 7 (3) & 6 (31) & 0.86 (99) & None & $-$0.7 & 10.1 & 4.7 & 14.1 & 89 \\
12409 & Leisure & 10 (10) & 8 (5) & 9 (46) & 0.90 (99) & None & $-$1.1 & 10.9 & 4.3 & 14.1 & 89 \\
\bottomrule
\end{tabular}}
\end{table}

Individual-level anomalies are partly a consequence of the SP design. Each observation is pivoted around the individual's reported trip, with variable levels set as deviations from it. This makes the observations more realistic, but it also means that individuals with unusually long, expensive or infrequent trips face variable levels far from the rest of the sample in all of their observations. In addition, SP designs aim for orthogonal variables, no dominant alternatives and enough variation to identify the parameters, which can produce variable levels or combinations that may look extreme at the sample level.

To assess the impact of the flagged observations on CM outputs, we estimate an MNL model on the full sample and re-estimate it after excluding the observations with the highest anomaly scores, i.e., the top 1\%, 5\% and 10\%. The utility of each route is linear in parameters, with generic parameters for travel time, travel cost, headway and number of interchanges. The utility of route 1 also includes alternative-specific constants (ASCs) for car availability and trip purpose, using the variables described in \autoref{tab:var_swiss}. From this, we derive the value of travel time (VoTT), the value of headway (VoH), the value of interchanges (VoI), and direct elasticities with respect to travel time, travel cost and headway. We also exclude the same shares of observations flagged by the univariate check and the Mahalanobis distance, computed on the travel times, costs and headways of both routes and on household income. As removing observations will naturally change the parameter estimates, regardless of which observations are removed, we also use a random benchmark. For each subset, we re-estimate the same model after excluding the same number of observations, i.e., 35, 175 and 349, respectively, but selected at random rather than by their anomaly score. We repeat this ten times, each with a different random subset, and report the mean across all runs.

\autoref{tab:mnl_swiss} presents the results of the estimated models. Excluding the observations flagged by the VAE improves the model fit, with a log-likelihood per observation increasing from $-0.475$ to $-0.470$, $-0.458$ and $-0.446$ in the 99\%, 95\% and 90\% subsets, whereas the univariate check improves it only to $-0.469$ in the 90\% subset. Since the anomaly score includes the choice, part of this improvement comes from excluding choices that are poorly explained. The time and cost parameters increase in magnitude (\autoref{tab:mnl_swiss_estimates}), by 9\%, 11\% and 18\%, and by 12\%, 5\% and 17\%, respectively, and the headway and interchange parameters by up to 10\% and 13\% in the 90\% subset. The time and cost parameters also become more precise despite the smaller sample, with robust t-ratios increasing from $-8.96$ to $-9.97$ and from $-5.66$ to $-6.17$, respectively, in the 90\% subset. The ASCs of the trip purposes remain not significant, whereas the ASC for car availability remains significant.

As the time and cost parameters change in similar proportions, the values of time change much less than the parameters themselves. Excluding the top 1\% reduces VoH and VoI by 9\% and 10\%, more than in any of the random subsets, while VoTT changes by $-2$\%. When excluding the top 5\% and 10\%, the values of time change by between $-6$\% and +5\%. The elasticities, on the other hand, increase in magnitude: the travel time elasticity by 7\%, 6\% and 11\%, more than in any of the random subsets, and the travel cost elasticity by 10\% and 7\% when excluding the top 1\% and 10\%. In comparison, the random subsets change values of time and elasticities by at most 1\% on average. The flagged observations therefore mainly affect the scale of the time and cost parameters, and hence the elasticities used in demand forecasting, rather than the trade-offs between time and cost.

The univariate check and the Mahalanobis distance are computed directly on the travel times, costs, and headways, so excluding the observations they flag mainly changes the time and cost parameters, and hence VoTT. Both increase the time parameter by 5\% to 7\% in the 95\% and 90\% subsets, while the univariate check slightly changes the cost parameter. As a result, VoTT increases by 4\% with the univariate check, whereas VoH and VoI present minor shifts. The Mahalanobis distance also increases the cost parameter, so VoTT changes by $-3$\% and 0\%, while VoH and VoI decrease by up to 9\%. The values of time obtained after excluding observations with a univariate check or the Mahalanobis distance therefore depend largely on which travel times and costs are considered extreme and therefore rely on the analysts' criterion. The VAE changes the time and cost parameters more than both methods, but in similar proportions, so its effect on the values of time is of a similar size.

\begin{sidewaystable}
\centering
\caption{Values of time, elasticities and goodness of fit of the MNL model for the Swiss route dataset, for the full sample, after excluding randomly selected observations (R, mean of ten runs), and after excluding the same number of observations with the highest scores of the univariate check (U), the Mahalanobis distance (M) and the VAE (V). Elasticities are computed with a 1\% increase in the respective variable. Percentage changes relative to the full sample in brackets. Values discussed in the text are in bold. Parameter estimates are reported in \autoref{tab:mnl_swiss_estimates}.}
\label{tab:mnl_swiss}
\small
\setlength{\tabcolsep}{4pt}
\renewcommand{\arraystretch}{1.25}
\resizebox{\textheight}{!}{%
\begin{tabular}{lccccccccccccc}
\toprule
 & & \multicolumn{4}{c}{\textbf{99\% subset}} & \multicolumn{4}{c}{\textbf{95\% subset}} & \multicolumn{4}{c}{\textbf{90\% subset}} \\
\cmidrule(lr){3-6}\cmidrule(lr){7-10}\cmidrule(lr){11-14}
 & \textbf{Full sample} & \textbf{R} & \textbf{U} & \textbf{M} & \textbf{V} & \textbf{R} & \textbf{U} & \textbf{M} & \textbf{V} & \textbf{R} & \textbf{U} & \textbf{M} & \textbf{V} \\
\midrule
\multicolumn{14}{l}{\textbf{Values of time (CHF)}} \\
Travel time (CHF/min) & 0.45 & 0.45 ($-$0\%) & 0.45 (+1\%) & 0.45 (+0\%) & \textbf{0.44 ($-$2\%)} & 0.45 (+0\%) & \textbf{0.47 (+4\%)} & \textbf{0.44 ($-$3\%)} & \textbf{0.47 (+5\%)} & 0.45 (+0\%) & \textbf{0.47 (+4\%)} & \textbf{0.45 ($-$0\%)} & \textbf{0.46 (+1\%)} \\
Headway (CHF/min) & 0.28 & 0.28 ($-$1\%) & 0.29 (+2\%) & 0.28 (+0\%) & \textbf{0.26 ($-$9\%)} & 0.28 ($-$0\%) & 0.28 (+1\%) & \textbf{0.26 ($-$9\%)} & \textbf{0.29 (+3\%)} & 0.28 (+0\%) & 0.29 (+2\%) & 0.26 ($-$7\%) & \textbf{0.27 ($-$6\%)} \\
Interchanges (CHF/interchange) & 8.67 & 8.62 ($-$1\%) & 8.87 (+2\%) & 8.70 (+0\%) & \textbf{7.85 ($-$10\%)} & 8.67 (+0\%) & 8.72 (+1\%) & 7.96 ($-$8\%) & \textbf{8.77 (+1\%)} & 8.77 (+1\%) & 8.74 (+1\%) & 8.16 ($-$6\%) & \textbf{8.37 ($-$3\%)} \\
\midrule
\multicolumn{14}{l}{\textbf{Elasticities}} \\
Travel time & $-$0.945 & $-$0.951 (+1\%) & $-$0.932 ($-$1\%) & $-$0.948 (+0\%) & \textbf{$-$1.016 (+7\%)} & $-$0.951 (+1\%) & $-$0.953 (+1\%) & $-$1.017 (+8\%) & \textbf{$-$1.004 (+6\%)} & $-$0.945 ($-$0\%) & $-$0.966 (+2\%) & $-$1.036 (+10\%) & \textbf{$-$1.052 (+11\%)} \\
Travel cost & $-$0.785 & $-$0.791 (+1\%) & $-$0.771 ($-$2\%) & $-$0.787 (+0\%) & \textbf{$-$0.864 (+10\%)} & $-$0.788 (+0\%) & $-$0.766 ($-$3\%) & $-$0.864 (+10\%) & $-$0.788 (+0\%) & $-$0.784 ($-$0\%) & $-$0.781 ($-$1\%) & $-$0.852 (+8\%) & \textbf{$-$0.839 (+7\%)} \\
Headway & $-$0.366 & $-$0.365 ($-$0\%) & $-$0.367 (+0\%) & $-$0.364 ($-$0\%) & $-$0.367 (+0\%) & $-$0.365 ($-$0\%) & $-$0.368 (+1\%) & $-$0.362 ($-$1\%) & $-$0.377 (+3\%) & $-$0.365 ($-$0\%) & $-$0.376 (+3\%) & $-$0.361 ($-$1\%) & $-$0.372 (+2\%) \\
\midrule
\multicolumn{14}{l}{\textbf{Goodness of fit}} \\
LL(final) per observation & \textbf{$-$0.475} & $-$0.475 & $-$0.475 & $-$0.476 & \textbf{$-$0.470} & $-$0.475 & $-$0.470 & $-$0.477 & \textbf{$-$0.458} & $-$0.475 & \textbf{$-$0.469} & $-$0.475 & \textbf{$-$0.446} \\
$\rho^2$ & 0.314 & 0.315 & 0.315 & 0.313 & 0.322 & 0.314 & 0.322 & 0.312 & 0.339 & 0.315 & 0.324 & 0.314 & 0.356 \\
\bottomrule
\end{tabular}}
\end{sidewaystable}

It is also important to clarify that an observation is flagged because its variables are atypical, not necessarily because the choice is poorly explained by the model. Indeed, the anomaly scores and the observation-specific log-likelihood in the estimated model are weakly correlated (Pearson’s correlation coefficient of $-0.24$), as observations with the highest anomaly score tend to have a lower log-likelihood. However, 64\% of the flagged observations still have a choice probability above 0.5. Since the model is estimated on the full sample (anomalies included) its parameters adjust to these observations, so an observation with atypical variables can still have a reasonably high choice probability. The anomaly score and the model fit therefore measure different things, and should be used together rather than one instead of the other.

For the Swiss route dataset, we present model validation in \autoref{ap:validation}. We also test our VAE models across 5 different seeds, where the scores have a Spearman correlation of 0.98, and we report the results of the model trained with the first seed. We next turn to an RP dataset, where observations are shaped by real constraints, such as infrastructure and mode availability, rather than by an experimental design, and where trip characteristics can be more extreme.

\subsection{Decisions dataset}
\label{subsec:decisions}

The Decisions dataset \citep{decisions} is a large-scale survey collected by the Choice Modelling Centre at the University of Leeds. It combines an online background questionnaire with detailed information on individuals’ life histories, social networks, and travel behaviour recorded over a two‑week period using a smartphone-based tracking application. The online survey gathered extensive socio‑demographic information, travel and vehicle availability, long‑term residential and employment histories, and indicators of social interactions, while the smartphone app recorded trips using GPS and prompted individuals to confirm travel modes, purposes, and costs. The variables to be reconstructed are therefore substantially richer and more complex than in the SP setting. This dataset has 12,004 observations for 415 individuals, and considers six alternatives: Car (48.1\% of choices), Bus (14.4\%), Rail (5.0\%), Taxi (3.3\%), Bicycle (3.3\%), and Walking (26.0\%). We summarise variables reconstructed by the VAE in \autoref{tab:var_decisions}. Note that in RP data, as some alternatives may not be available for some observations, we include availabilities in our modelling. By preserving the distribution and correlation structure of the data, the VAE learns which variables are not available: when an alternative is not available, its availability and all of its variables, such as travel time and cost, are zero. Thus, these are recovered by the VAE.

\begin{table}[h!]
\centering
\caption{Variables reconstructed by the VAE for the Decisions dataset. Continuous variables are transformed as $\log(x+1)$, as some times and costs are zero, and standardised before training. Categorical variables are one-hot encoded.}
\footnotesize
\begin{tabular}{llp{8cm}}
\hline
     \textbf{Variable} & \textbf{Type} & \textbf{Description (units)} \\
    \hline
     car\_travel\_time\_min & Continuous & Car travel time (minutes). \\
     total\_car\_cost & Continuous & Total cost for a car trip including fuel and operating costs (£). \\
     bus\_IVT\_time\_min & Continuous & In-vehicle travel time by bus (minutes). \\
     OVT\_bus\_min & Continuous & Out-of-vehicle time for bus trips (minutes). \\
     bus\_cost\_total\_per\_leg & Continuous & Cost for a bus trip including a distance-based price of a ticket and a discount for season ticket holders (£). \\
     car\_distance\_km & Continuous & API car distance (kilometres). \\
     bus\_distance\_km & Continuous & API bus distance (kilometres). \\
     rail\_IVT\_time\_min & Continuous & In-vehicle travel time by rail (minutes). \\
     OVT\_rail\_min & Continuous & Out-of-vehicle time for rail trips (minutes). \\
     rail\_cost\_total\_per\_leg & Continuous & Cost for a rail trip including a distance-based price of a ticket and a discount for season ticket holders (£). \\
     rail\_distance\_km & Continuous & Travel distance by rail (kilometres). \\
     taxi\_travel\_time\_min & Continuous & Travel time by taxi (minutes). \\
     taxi\_cost & Continuous & Cost of a taxi trip based on average cost rates from local authorities across the region of Yorkshire and the Humber (£) \\
     taxi\_distance\_km & Continuous & Travel distance by taxi (kilometres). \\
     cycling\_distance\_km & Continuous & Travel distance by bicycle (kilometres). \\
     walking\_distance\_km & Continuous & Travel distance on foot (kilometres). \\
     female & Categorical & Gender dummy variable, 1 if female, 0 otherwise. \\
     weekday & Categorical & Type of day dummy variable, 1 if weekday, 0 otherwise. \\
     occ & Categorical & Occupation of the participant. Various categories. \\
     hh\_size & Categorical & Number of people living in the household, excluding the participant.\\
     purpose\_new\_2\_num & Categorical & Stated trip purpose. Various categories. \\
     ampeak & Categorical &  Hour of day dummy variable, 1 if morning peak hour, 0 otherwise. \\
     pmpeak & Categorical & Hour of day dummy variable, 1 if afternoon peak hour, 0 otherwise. \\
     income\_cont & Categorical & Total annual income band of the participant's household.  \\
     bus\_transfers & Categorical & Number of required transfers on a bus trip. \\
     rail\_transfers & Categorical & Number of required transfers on a rail trip. \\
     avail\_car & Categorical &  Availability of the car alternative, 1 if available, 0 otherwise \\
     avail\_bus & Categorical &  Availability of the bus alternative, 1 if available, 0 otherwise \\
     avail\_rail & Categorical &  Availability of the rail alternative, 1 if available, 0 otherwise \\
     avail\_taxi & Categorical &  Availability of the taxi alternative, 1 if available, 0 otherwise \\
     avail\_cycling & Categorical &  Availability of the cycling alternative, 1 if available, 0 otherwise \\
     avail\_walking & Categorical &  Availability of the walking alternative, 1 if available, 0 otherwise \\
     choice & Categorical & Chosen alternative: 1 car, 2 bus, 3 rail, 4 taxi, 5 cycling, 6 walking. \\
    \hline
    \end{tabular}
    \label{tab:var_decisions}
\end{table}

We train the $\text{VAE}_{\text{ME}}$ on the Decisions data using the configuration selected through the hyperparameter search, shown in \autoref{tab:hyperparam_decisions}. 

\begin{table}[h]
\centering
\caption{Hyperparameter search space and selected configuration for the Decisions dataset.}
\label{tab:hyperparam_decisions}
\footnotesize
\setlength{\tabcolsep}{4pt}
\begin{tabular}{lcc}
\toprule
\textbf{Hyperparameter} & \textbf{Search space} & \textbf{Selected} \\
\midrule
Latent dimension ($D_z$)            & 4, 8, 16, 24, 32, 48, 64                  & 16 \\
Random-effect dimension ($D_u$)     & 2, 4, 8                                   & 2 \\
Hidden layers                       & 1, 2, 3, 4                                & 2 \\
Hidden layer size                   & 16, 32, 64, 96, 128, 160, 192, 256        & 96 \\
Learning rate                       & $10^{-4}$, $5\times10^{-4}$, $10^{-3}$, $2\times10^{-3}$, $5\times10^{-3}$ & $10^{-3}$ \\
Batch size                          & 32, 64, 128, 256, 512                     & 32 \\
Weight decay                        & 0, $10^{-6}$, $5\times10^{-6}$, $10^{-5}$, $10^{-4}$ & $10^{-5}$ \\
\bottomrule
\end{tabular}
\end{table}

\begin{figure}[h!]
\centering
\includegraphics[width=0.8\linewidth]{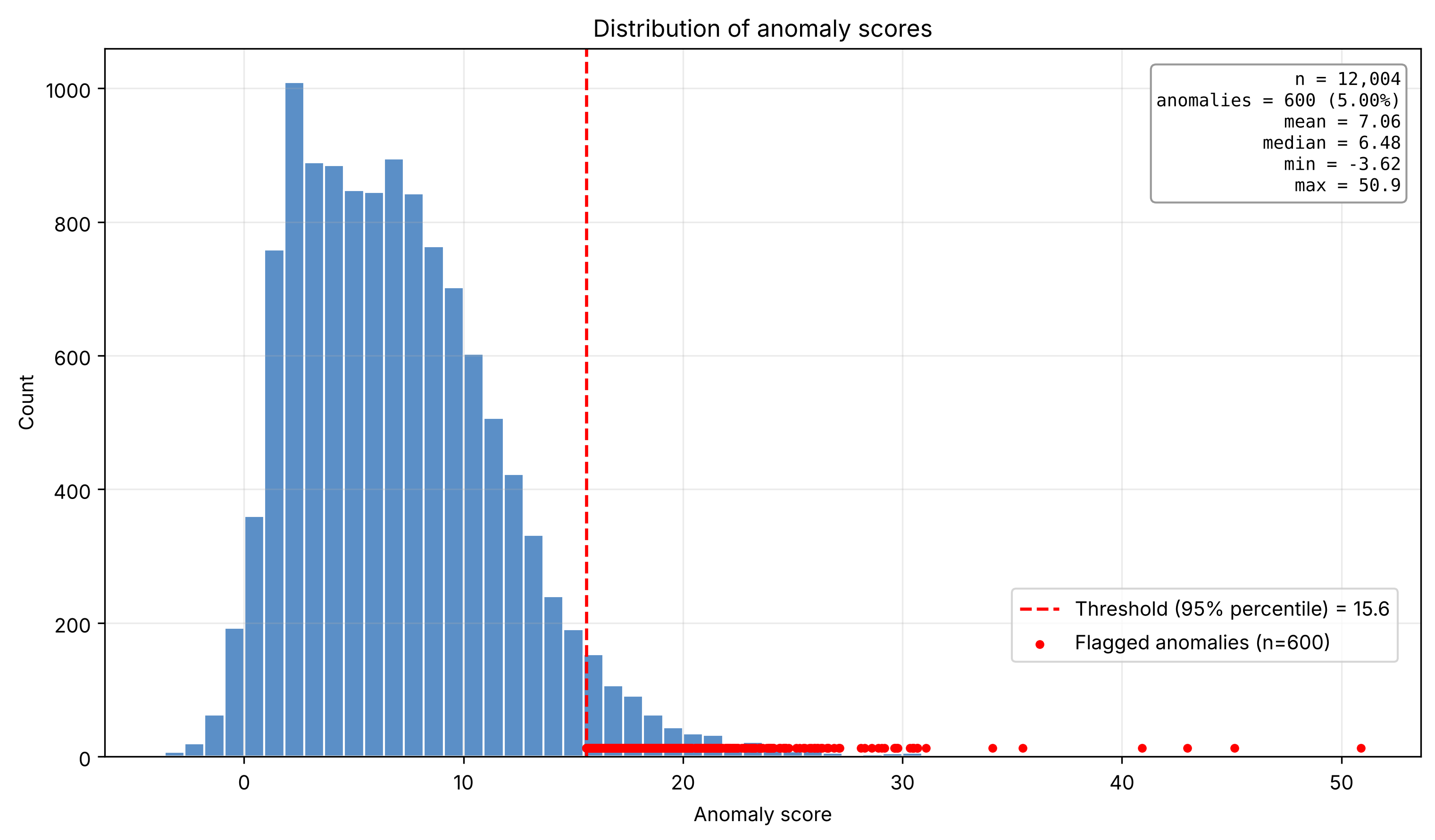}
\caption{Distribution of anomaly scores for the Decisions dataset. Red dots indicate the 600 observations above the 95th percentile.}
\label{fig:errors_decisions}
\end{figure}

\autoref{fig:errors_decisions} shows the distribution of the anomaly scores for the Decisions dataset. Using the 95th percentile as threshold ($\alpha=0.95$), 600 of the 12,004 observations are flagged as anomalies, while with $\alpha = 0.99$ and $\alpha = 0.90$, 120 and 1,200 observations are flagged, respectively. As in the Swiss route case, most observations have low scores, but the scores span a much wider range, and the upper tail is substantially longer. Part of this difference comes from the larger sample size (12,004, compared to 3,492) and number of reconstructed variables, but most of it reflects how the data were collected. Unlike SP data, where variable levels are set by the experimental design within predefined ranges, the variables in RP data come from several sources, each of which can introduce errors. Trips are recorded by GPS, which can miss or misplace parts of a trip; modes, purposes and costs are confirmed by the individuals, who can make mistakes when entering them; and the variables of the non-chosen alternatives, such as travel times and distances, are typically obtained from APIs, which can return unrealistic values when no suitable route or service is found. Most observations have variable values within the usual range for their mode and trip length, and are reconstructed well by the VAE. The long tail, on the other hand, contains both unusual trips, such as very long or rare journeys, and observations affected by these errors.

We present the ten observations with the highest anomaly scores in \autoref{tab:anomalies_decisions}, with their reconstruction probabilities in brackets and their KL terms. Most of them combine times, costs and distances that are not consistent with each other. Three of them, trips 55001, 52302 and 91689, have a bus IVT of only one minute, with a bus fare of £2.1, whereas most such short bus legs cost £1.5 in the data. Trips 56079 and 65096 have car travel times of 18 and 14 minutes for car costs of only £0.8 and £0.6, and trip 57525 is a drop-off trip of 81 km by car, with a car cost of £8.8 and a taxi cost of £47.7. Trip 42616 is a shopping trip, where the API car distance is 13.3 km for a walking distance of only 0.4 km, so that the car distance, the car cost and the taxi cost (£10.5) are all poorly reconstructed. Trip 66481 has a rail OVT of only 4 minutes for a rail IVT of 38 minutes. In the other two observations, the choice itself is unusual: in trip 54982, the individual chose a bus trip of 61 minutes over a rail trip of 48 minutes at the same fare, and in trip 41060, the individual chose a taxi for a 6 km leisure trip, with the highest KL term among the ten.

The trips with very long public transport OVTs are also flagged, mainly through their KL terms, which are among the highest in the sample. Trips 94456, 58672, 59672, 61048 and 54414, with OVTs of between 167 and 424 minutes for both bus and rail, and trips 57990 and 84070, with rail OVTs of 716 and 370 minutes, are encoded to an atypical region of the latent space, so that their OVTs are reconstructed reasonably well given this representation. Trips 59672 and 61048 are the same work trip made by the same individual on a Sunday and a Monday, with almost identical OVTs on both days.

\begin{sidewaystable}
\centering
\caption{Observations with the ten highest anomaly scores in the Decisions dataset. Times in minutes, costs in £, distances in kilometres; transf.: number of interchanges; --: alternative not available, so its variables are not reconstructed. Variable-specific reconstruction probabilities in brackets; for purpose and day, the terms of the trip purpose and of the weekday dummy. Largest other term: largest term among the variables not shown (API distances of car, bus and rail, gender, peak hour dummies, household size and income). KL: Kullback--Leibler divergence between the distribution of the latent representation and its prior. The anomaly score is the sum of the terms of all variables and the KL term. Values discussed in the text are in bold.}
\label{tab:anomalies_decisions}
\footnotesize
\setlength{\tabcolsep}{3pt}
\renewcommand{\arraystretch}{1.25}
\resizebox{\textheight}{!}{%
\begin{tabular}{ccccccccccccccccccccc}
\toprule
 & & & & \multicolumn{2}{c}{\textbf{Car}} & \multicolumn{4}{c}{\textbf{Bus}} & \multicolumn{4}{c}{\textbf{Rail}} & \textbf{Taxi} & \textbf{Cycle} & \textbf{Walk} & & & & \\
\cmidrule(lr){5-6}\cmidrule(lr){7-10}\cmidrule(lr){11-14}\cmidrule(lr){15-15}\cmidrule(lr){16-16}\cmidrule(lr){17-17}
\textbf{Trip} & \textbf{User} & \textbf{Purpose} & \textbf{Day} & \textbf{TT} & \textbf{Cost} & \textbf{IVT} & \textbf{OVT} & \textbf{Transf.} & \textbf{Cost} & \textbf{IVT} & \textbf{OVT} & \textbf{Transf.} & \textbf{Cost} & \textbf{Cost} & \textbf{Dist.} & \textbf{Dist.} & \textbf{Choice} & \textbf{Largest other term} & \textbf{KL} & \textbf{Anomaly score} \\
\midrule
57525 & 1872 & Drop-off (3.3) & Saturday (1.6) & \textbf{56 (2.3)} & \textbf{8.8 (7.8)} & 142 (3.1) & 31 (0.8) & 1 (0.6) & 5.7 (1.0) & -- & -- & -- & -- & \textbf{47.7 (8.0)} & -- & -- & Car (0.0) & \textbf{Car dist. (7.0)} & 8.9 & 50.9 \\
55001 & 1779 & Home (1.1) & Wednesday (0.2) & -- & -- & \textbf{1 (1.5)} & \textbf{4 (0.6)} & 0 (0.0) & \textbf{2.1 (30.7)} & -- & -- & -- & -- & 6.7 (4.4) & -- & 0.5 (1.5) & Walking (0.0) & Income (0.7) & 3.3 & 45.1 \\
52302 & 1618 & Shopping (1.5) & Friday (0.3) & -- & -- & \textbf{1 (1.4)} & \textbf{6 (0.3)} & 0 (0.0) & \textbf{2.1 (31.1)} & -- & -- & -- & -- & 5.9 (2.5) & -- & 0.6 (0.7) & Walking (0.0) & PM peak (0.9) & 3.2 & 43.0 \\
42616 & 1511 & Shopping (0.6) & Saturday (0.6) & 13 (4.2) & \textbf{1.5 (7.3)} & -- & -- & -- & -- & -- & -- & -- & -- & \textbf{10.5 (9.2)} & -- & \textbf{0.4 (2.1)} & Car (0.0) & \textbf{Car dist. (8.8)} & 6.8 & 40.9 \\
91689 & 2167 & Home (1.3) & Saturday (1.6) & -- & -- & \textbf{1 (2.1)} & \textbf{5 (0.6)} & 0 (0.0) & \textbf{2.1 (23.8)} & -- & -- & -- & -- & -- & -- & 0.7 (0.6) & Walking (0.0) & Bus dist. (1.8) & 3.1 & 35.5 \\
56079 & 1696 & Other (1.9) & Thursday (0.2) & \textbf{18 (15.0)} & \textbf{0.8 (4.6)} & 9 (0.2) & 7 (0.6) & 0 (0.0) & 1.5 ($-$1.5) & -- & -- & -- & -- & 6.3 (4.8) & 2.3 ($-$0.4) & 2.1 (0.2) & Car (0.1) & PM peak (3.1) & 5.5 & 34.1 \\
66481 & 1925 & Work (1.2) & Friday (0.2) & -- & -- & 31 (0.5) & 9 (0.8) & 0 (0.1) & 3.4 (0.6) & \textbf{38 (4.2)} & \textbf{4 (11.7)} & 1 (0.4) & 17.0 (1.7) & 12.0 ($-$0.1) & 14.4 (0.0) & -- & Rail (0.4) & Rail dist. (5.4) & 3.9 & 31.1 \\
54982 & 1783 & Home (0.1) & Wednesday (0.0) & -- & -- & \textbf{61 (0.0)} & 22 (0.9) & 1 (0.1) & 2.1 (1.0) & \textbf{48 (3.8)} & 24 (0.7) & 0 (1.8) & 2.1 (1.2) & 33.7 (2.5) & -- & -- & \textbf{Bus (4.4)} & Rail dist. (7.2) & 4.7 & 30.7 \\
41060 & 1527 & Leisure (2.2) & Thursday (0.1) & 15 (0.4) & 3.4 (1.2) & 24 (0.0) & 11 (0.3) & 1 (0.0) & 4.6 ($-$0.9) & 9 (0.1) & 28 (0.1) & 0 (0.1) & 2.9 (5.5) & 7.0 (0.6) & -- & -- & \textbf{Taxi (3.5)} & Rail dist. (4.4) & \textbf{11.1} & 30.7 \\
65096 & 1737 & Leisure (1.1) & Thursday (0.3) & \textbf{14 (8.0)} & \textbf{0.6 (5.1)} & 7 ($-$0.1) & 10 (0.7) & 0 (0.0) & 1.5 ($-$1.7) & -- & -- & -- & -- & \textbf{6.4 (8.2)} & 2.1 ($-$0.5) & 1.9 (0.0) & Car (0.0) & Car dist. (4.7) & 4.8 & 30.5 \\
\bottomrule
\end{tabular}}
\end{sidewaystable}

We also report the ten individuals with the highest anomaly scores in \autoref{tab:anomalies_users_decisions}, with the components of their score, their socio-demographic characteristics and typical trips, and the share of their trips flagged with $\alpha = 0.95$. Unlike the Swiss route case, where all observations of an individual were built around the same reported trip, each individual in the Decisions dataset makes different trips, so a high score can be driven either by unusual trips or by unusual random effects. For these individuals, it is mostly the former. Their scores are between 18.3 and 20.1, compared with 12.7 on average, and all of them are among the 27 individuals with the highest average trip score. Between 5\% and 47\% of their trips are flagged, compared with 5\% on average, and together they account for 60 of the 600 flagged trips, whereas 43\% of the individuals have no flagged trip at all. Most of these individuals make long trips, as seven of them have a mean trip distance above 15 km, compared with a sample average of 8.2 km, and two of them, individuals 1783 and 2167, also made trips in \autoref{tab:anomalies_decisions}.

\begin{table}[h]
\centering
\caption{Ten individuals with the highest anomaly scores in the Decisions dataset, with their socio-demographic characteristics and typical trips, and the components of their anomaly score: the average over their trips of the sum of the reconstruction probabilities and of the KL term of the latent representation, and the KL term of their random effect. The last column is the share of their trips flagged with $\alpha = 0.95$. Income in £/year, distance in kilometres.}
\label{tab:anomalies_users_decisions}
\footnotesize
\setlength{\tabcolsep}{3pt}
\resizebox{\textwidth}{!}{%
\begin{tabular}{ccccccccccccc}
\toprule
 & & & & & & & & \multicolumn{4}{c}{\textbf{Anomaly score}} & \\
\cmidrule(lr){9-12}
\textbf{User} & \textbf{Trips} & \textbf{Age} & \textbf{Gender} & \textbf{Income} & \textbf{Main purpose} & \textbf{Main mode} & \textbf{Mean distance} & \textbf{Rec. prob.} & \textbf{KL ($z$)} & \textbf{KL ($u$)} & \textbf{Total} & \textbf{Flagged (\%)} \\
\midrule
1783 & 13 & 18--24 & M & 35,000 & Work (38\%) & Walking (46\%) & 25.5 & 9.7 & 4.1 & 6.3 & 20.1 & 38 \\
1832 & 33 & 18--24 & F & 15,000 & Home (36\%) & Bus (48\%) & 22.0 & 8.8 & 5.2 & 5.7 & 19.8 & 42 \\
1793 & 15 & 40--49 & F & 15,000 & Work (53\%) & Taxi (60\%) & 18.8 & 8.8 & 5.9 & 4.4 & 19.1 & 47 \\
1487 & 11 & 40--49 & F & 45,000 & Leisure (55\%) & Car (55\%) & 18.1 & 8.3 & 5.8 & 5.0 & 19.1 & 36 \\
2043 & 77 & 40--49 & F & 35,000 & Work (42\%) & Car (94\%) & 11.0 & 7.1 & 5.1 & 6.8 & 19.0 & 12 \\
2168 & 28 & 18--24 & F & 5,000 & Exercise (25\%) & Rail (46\%) & 15.8 & 7.9 & 4.9 & 6.1 & 18.9 & 29 \\
1668 & 15 & 40--49 & M & 62,500 & Work (47\%) & Rail (53\%) & 33.5 & 7.4 & 5.3 & 5.9 & 18.6 & 27 \\
1638 & 19 & 30--39 & M & 25,000 & Work (32\%) & Bus (47\%) & 7.2 & 8.1 & 4.8 & 5.6 & 18.5 & 37 \\
2167 & 15 & 18--24 & F & 5,000 & Home (40\%) & Bus (53\%) & 2.3 & 6.3 & 4.8 & 7.5 & 18.5 & 7 \\
1637 & 19 & 30--39 & M & 45,000 & Other (37\%) & Car (58\%) & 20.0 & 6.4 & 5.6 & 6.4 & 18.3 & 5 \\
\bottomrule
\end{tabular}}
\end{table}

We repeat the sensitivity analysis conducted for the Swiss route dataset. An MNL model is estimated on the full sample and after excluding the top 1\%, 5\% and 10\% of observations by anomaly score, and the results are compared with the mean of ten models, each estimated after excluding the same number of randomly selected observations. We also exclude the same shares of observations flagged by the univariate check and the Mahalanobis distance. As in the Swiss route case, excluding observations with extreme times or costs changes the parameters of these variables by construction, so the random removal remains the benchmark to assess whether the flagged observations themselves influence the estimates. The utility specification is based on \cite{panos2022}, and we present this in \autoref{ap:spec_decisions}. We report values of in-vehicle time (VoIVT), out-of-vehicle time (VoOVT) and travel time (VoTT) by mode, as well as elasticities with respect to these variables and to travel cost. The results are shown in \autoref{tab:mnl_decisions}.

As in the Swiss route case, model fit improves when the flagged observations are excluded, with the log-likelihood per observation increasing from $-0.389$ to $-0.372$, $-0.355$ and $-0.339$ in the 99\%, 95\% and 90\% subsets, and $\rho^2$ from 0.675 to 0.713, whereas the random subsets leave it unchanged. The largest change in the estimates is for bus OVT (\autoref{tab:mnl_decisions_estimates}). In the full sample, its parameter is $-0.116$ with a robust t-ratio of $-3.33$, whereas excluding only the top 1\% of observations changes it to $-0.171$ ($-12.45$), and excluding the top 10\% to $-0.193$ ($-12.90$). The very long OVTs therefore not only bias this parameter but also inflate its standard error. Most other parameters also increase in magnitude as more observations are excluded, by up to 96\% for taxi travel time, 55\% for bus IVT, 46\% for cost and 31\% for cycling distance when excluding the top 10\%. The rail IVT parameter, on the other hand, decreases in magnitude, by 32\%, and is no longer significant when excluding the top 10\%.

The random subsets, used as a baseline, also change values of time and elasticities by at most 4\% on average. When the flagged observations are excluded, the changes are larger and differ across modes and variables. The value of bus OVT increases by 27\%, 19\% and 14\% in the three subsets, as its parameter increases more than the cost parameter, and the value of taxi travel time by 43\% and 34\% when excluding the top 5\% and 10\%. When excluding the top 10\%, the cost parameter increases more than the car travel time, rail IVT and rail OVT parameters, so the value of car travel time decreases by 26\%, rail IVT by 54\% and rail OVT by 19\%. The elasticities follow a similar pattern. The bus OVT elasticity increases by 31\%, 30\% and 31\% in the three subsets, the taxi travel time elasticity by up to 53\%, and the bus and taxi cost elasticities by up to 21\% and 18\%, whereas the rail IVT and car travel time elasticities decrease by 55\% and 28\% when excluding the top 10\%. The flagged observations therefore do not affect all parameters equally, but mainly those related to public transport times and fares, and to taxi travel time.

The univariate check and the Mahalanobis distance also flag the trips with very long OVTs, so excluding them increases the bus OVT parameter by 39\% to 52\% in the 95\% and 90\% subsets, compared with 55\% and 66\% for the VAE. As they act directly on the times, costs and distances, their effect on the values of time depends on which values they consider extreme. The univariate check mainly increases the value of bus OVT, by up to 29\%, and changes the other values of time by at most 10\%, whereas the Mahalanobis distance increases the values of bus OVT and taxi travel time by up to 38\% and 29\%, respectively. Unlike these two methods, the VAE also reduces the values of rail IVT and car travel time, by up to 54\% and 26\%.

\begin{sidewaystable}
\centering
\caption{Values of time, direct elasticities and goodness of fit of the MNL model for the Decisions dataset, for the full sample, after excluding randomly selected observations (R, mean of ten runs), and after excluding the same number of observations with the highest scores of the univariate check (U), the Mahalanobis distance (M) and the VAE (V). Elasticities are computed with a 1\% increase in the respective variable. Percentage changes relative to the full sample in brackets. Values discussed in the text are in bold. Parameter estimates are reported in \autoref{tab:mnl_decisions_estimates}.}
\label{tab:mnl_decisions}
\small
\setlength{\tabcolsep}{4pt}
\renewcommand{\arraystretch}{1.25}
\resizebox{\textheight}{!}{%
\begin{tabular}{lccccccccccccc}
\toprule
 & & \multicolumn{4}{c}{\textbf{99\% subset}} & \multicolumn{4}{c}{\textbf{95\% subset}} & \multicolumn{4}{c}{\textbf{90\% subset}} \\
\cmidrule(lr){3-6}\cmidrule(lr){7-10}\cmidrule(lr){11-14}
 & \textbf{Full sample} & \textbf{R} & \textbf{U} & \textbf{M} & \textbf{V} & \textbf{R} & \textbf{U} & \textbf{M} & \textbf{V} & \textbf{R} & \textbf{U} & \textbf{M} & \textbf{V} \\
\midrule
\multicolumn{14}{l}{\textbf{Values of time (£/hour)}} \\
Car travel time & 38.8 & 38.9 (+0\%) & 37.1 ($-$4\%) & 37.5 ($-$3\%) & 37.5 ($-$3\%) & 38.8 (+0\%) & 38.7 ($-$0\%) & 39.6 (+2\%) & 35.4 ($-$9\%) & 38.7 ($-$0\%) & 34.9 ($-$10\%) & 38.0 ($-$2\%) & \textbf{28.9 ($-$26\%)} \\
Bus IVT & 14.2 & 14.3 (+0\%) & 14.8 (+4\%) & 14.1 ($-$1\%) & 14.8 (+4\%) & 14.3 (+0\%) & 14.9 (+5\%) & 15.9 (+12\%) & 15.6 (+9\%) & 14.3 (+0\%) & 14.0 ($-$2\%) & 14.9 (+5\%) & 15.0 (+6\%) \\
Bus OVT & 39.2 & 39.2 ($-$0\%) & \textbf{50.5 (+29\%)} & 42.0 (+7\%) & \textbf{49.8 (+27\%)} & 39.2 ($-$0\%) & 47.2 (+20\%) & \textbf{54.1 (+38\%)} & \textbf{46.8 (+19\%)} & 38.8 ($-$1\%) & 42.9 (+9\%) & 50.0 (+27\%) & \textbf{44.5 (+14\%)} \\
Rail IVT & 14.3 & 14.4 (+0\%) & 13.5 ($-$6\%) & 14.1 ($-$2\%) & 13.0 ($-$10\%) & 14.6 (+2\%) & 14.4 (+1\%) & 16.2 (+13\%) & 10.8 ($-$25\%) & 14.6 (+2\%) & 13.5 ($-$5\%) & 14.2 ($-$1\%) & \textbf{6.6 ($-$54\%)} \\
Rail OVT & 36.3 & 36.4 (+0\%) & 36.0 ($-$1\%) & 35.5 ($-$2\%) & 32.4 ($-$11\%) & 36.5 (+1\%) & 35.3 ($-$3\%) & 35.2 ($-$3\%) & 30.5 ($-$16\%) & 35.9 ($-$1\%) & 32.6 ($-$10\%) & 34.1 ($-$6\%) & \textbf{29.3 ($-$19\%)} \\
Taxi travel time & 30.9 & 31.0 (+0\%) & 31.7 (+3\%) & 37.2 (+20\%) & 31.4 (+1\%) & 30.7 ($-$1\%) & 31.4 (+2\%) & \textbf{39.8 (+29\%)} & \textbf{44.3 (+43\%)} & 29.8 ($-$4\%) & 31.4 (+1\%) & 37.5 (+21\%) & \textbf{41.5 (+34\%)} \\
\midrule
\multicolumn{14}{l}{\textbf{Elasticities}} \\
Car travel time & $-$0.143 & $-$0.143 (+0\%) & $-$0.145 (+1\%) & $-$0.141 ($-$1\%) & $-$0.149 (+5\%) & $-$0.143 (+0\%) & $-$0.154 (+8\%) & $-$0.143 (+0\%) & $-$0.134 ($-$6\%) & $-$0.143 (+0\%) & $-$0.148 (+3\%) & $-$0.137 ($-$4\%) & \textbf{$-$0.103 ($-$28\%)} \\
Car cost & $-$0.027 & $-$0.027 ($-$0\%) & $-$0.029 (+6\%) & $-$0.028 (+2\%) & $-$0.030 (+9\%) & $-$0.027 (+0\%) & $-$0.028 (+4\%) & $-$0.026 ($-$5\%) & $-$0.029 (+5\%) & $-$0.027 (+0\%) & $-$0.029 (+9\%) & $-$0.025 ($-$9\%) & $-$0.027 ($-$1\%) \\
Bus IVT & $-$0.400 & $-$0.400 (+0\%) & $-$0.430 (+8\%) & $-$0.408 (+2\%) & $-$0.437 (+9\%) & $-$0.403 (+1\%) & $-$0.457 (+14\%) & $-$0.438 (+9\%) & $-$0.467 (+17\%) & $-$0.401 (+0\%) & $-$0.469 (+17\%) & $-$0.419 (+5\%) & $-$0.449 (+12\%) \\
Bus OVT & $-$0.541 & $-$0.539 ($-$0\%) & $-$0.704 (+30\%) & $-$0.590 (+9\%) & \textbf{$-$0.707 (+31\%)} & $-$0.542 (+0\%) & $-$0.702 (+30\%) & $-$0.712 (+32\%) & \textbf{$-$0.704 (+30\%)} & $-$0.535 ($-$1\%) & $-$0.712 (+32\%) & $-$0.695 (+28\%) & \textbf{$-$0.710 (+31\%)} \\
Bus cost & $-$0.159 & $-$0.159 ($-$0\%) & $-$0.168 (+5\%) & $-$0.165 (+3\%) & $-$0.171 (+7\%) & $-$0.160 (+0\%) & $-$0.178 (+12\%) & $-$0.160 (+0\%) & $-$0.182 (+14\%) & $-$0.159 ($-$0\%) & $-$0.198 (+24\%) & $-$0.167 (+5\%) & \textbf{$-$0.194 (+21\%)} \\
Rail IVT & $-$0.308 & $-$0.307 ($-$0\%) & $-$0.306 ($-$1\%) & $-$0.307 ($-$0\%) & $-$0.299 ($-$3\%) & $-$0.314 (+2\%) & $-$0.328 (+7\%) & $-$0.364 (+18\%) & $-$0.235 ($-$23\%) & $-$0.316 (+3\%) & $-$0.324 (+5\%) & $-$0.340 (+10\%) & \textbf{$-$0.137 ($-$55\%)} \\
Rail OVT & $-$1.068 & $-$1.067 ($-$0\%) & $-$1.113 (+4\%) & $-$1.076 (+1\%) & $-$1.030 ($-$4\%) & $-$1.075 (+1\%) & $-$1.143 (+7\%) & $-$1.113 (+4\%) & $-$0.987 ($-$8\%) & $-$1.065 ($-$0\%) & $-$1.163 (+9\%) & $-$1.176 (+10\%) & $-$0.950 ($-$11\%) \\
Rail cost & $-$0.355 & $-$0.354 ($-$0\%) & $-$0.371 (+4\%) & $-$0.358 (+1\%) & $-$0.378 (+6\%) & $-$0.356 (+0\%) & $-$0.373 (+5\%) & $-$0.379 (+7\%) & $-$0.377 (+6\%) & $-$0.357 (+0\%) & $-$0.397 (+12\%) & $-$0.405 (+14\%) & $-$0.359 (+1\%) \\
Taxi travel time & $-$0.839 & $-$0.840 (+0\%) & $-$0.914 (+9\%) & $-$0.990 (+18\%) & $-$0.922 (+10\%) & $-$0.836 ($-$0\%) & $-$0.926 (+10\%) & $-$1.055 (+26\%) & $-$1.275 (+52\%) & $-$0.814 ($-$3\%) & $-$0.985 (+17\%) & $-$1.037 (+24\%) & \textbf{$-$1.286 (+53\%)} \\
Taxi cost & $-$0.915 & $-$0.914 ($-$0\%) & $-$0.974 (+7\%) & $-$0.909 ($-$1\%) & $-$0.994 (+9\%) & $-$0.920 (+1\%) & $-$0.992 (+8\%) & $-$0.906 ($-$1\%) & $-$1.000 (+9\%) & $-$0.920 (+1\%) & $-$1.063 (+16\%) & $-$0.946 (+3\%) & \textbf{$-$1.081 (+18\%)} \\
\midrule
\multicolumn{14}{l}{\textbf{Goodness of fit}} \\
LL(final) per observation & \textbf{$-$0.389} & $-$0.390 & $-$0.377 & $-$0.383 & \textbf{$-$0.372} & $-$0.389 & $-$0.378 & $-$0.372 & \textbf{$-$0.355} & $-$0.390 & $-$0.376 & $-$0.364 & \textbf{$-$0.339} \\
$\rho^2$ & \textbf{0.675} & 0.675 & 0.685 & 0.680 & 0.689 & 0.675 & 0.685 & 0.688 & 0.701 & 0.674 & 0.686 & 0.692 & \textbf{0.713} \\
\bottomrule
\end{tabular}}
\end{sidewaystable}

Finally, for the Decisions dataset, we present model validation in \autoref{ap:validation}. We also test our VAE models across 5 different seeds, where the scores have a Spearman correlation of 0.95, and we report the results of the model trained with the first seed.

\section{Conclusions and next steps}
\label{sec:conclusions}

This paper introduces the use of VAEs for anomaly detection in CM datasets. We extend the VAE to panel data through individual-specific random effects, and propose an anomaly score based on variable-specific reconstruction probabilities and the KL term, which accounts for the correlations between variables and shows which variables make an observation anomalous. The synthetic case study shows that our approach flags not only observations in the tails of each variable's distribution, but also observations whose variables are individually plausible but inconsistent when evaluated jointly. We benchmark the VAE against two traditional methods, a univariate check and the Mahalanobis distance, while also considering random removals. When the anomalies are known, the VAE with mixed effects detects a larger share of them, 48\% of the induced anomalies with the top 5\%, compared with 20\% and 29\%, while flagging far fewer normal observations. With the top 10\%, it also detects most of the trip type errors and individual inconsistent observations (66\% and 70\%), which are not extreme in any single variable, compared with at most 16\% for the traditional methods, and it brings the value of travel time closest to its true value. The two real-world case studies, an SP and an RP dataset, show that the flagged observations have different origins depending on how the data were collected, and that they have a substantial effect on key CM outputs.

While it is widely recognised in the literature that observations with unusual choice behaviour can affect parameter estimates and forecasting applications, we show that this is also the case for observations with unusual variable configurations, which existing diagnostics only identify when the type of anomaly is defined in advance. Removing even a very small share of these observations leads to substantial changes in values of time and elasticities. In the Swiss route dataset, excluding only the 1\% of observations with the highest scores reduces the values of headway and interchanges by 9\% and 10\%, and increases the travel time and cost elasticities by 7\% and 10\%, whereas excluding the top 10\% increases the time and cost parameters by 18\% and 17\%. In the Decisions dataset, excluding the top 1\% increases the value of bus OVT by 27\% and the corresponding elasticity by 31\%, as the very long OVTs in the data mainly affect this parameter, with changes of up to 54\% in values of time and 55\% in elasticities as more observations are excluded, the largest for rail IVT.

The appropriate treatment of anomalies depends heavily on the case study and on the nature of the anomalous observations, which may range from data errors to inconsistent responses. Thus, these results do not imply that the proposed approach should be used as a mechanical procedure for removing outliers. Instead, our findings illustrate how anomaly detection can serve as a diagnostic tool to flag atypical observations or individuals. The proposed framework offers an effective way to identify such cases in large, high-dimensional datasets, not only in transport applications, but also in other domains where complex behavioural or observational data are common, including healthcare, marketing, and finance.

Our results and our framework depend on the share of anomalous observations in the data. As the VAE learns the distribution of the data without knowing which observations are anomalous, it works best when anomalies are rare and differ from each other, so that they cannot be learned as part of the underlying data-generating process. If anomalies were, in contrast, a large share of the sample, the model would learn them and reconstruct them well, giving them low anomaly scores. 

Moreover, some techniques for addressing class imbalance---i.e., when the choices for one or more alternatives are severely under-represented---suggest balancing the data. We advise caution in applying these techniques. In particular, if anomalous observations are present in the original data, such procedures may replicate or amplify noise, inconsistencies, or measurement errors, thereby decreasing model performance and reliability. Therefore, we recommend that practitioners first identify and critically assess anomalous observations prior to any rebalancing, so that any data augmentation is based on consistent and representative data. 

Regarding next steps, a line of research suggests using the anomaly scores to help identify where the model needs to account for heterogeneity. If flagged observations or individuals share common characteristics, such as the short reported trips with high costs per kilometre in the Swiss route data, this suggests segments with different sensitivities, which could then be captured, for instance, through a latent class choice model \citep{hess2010outliers,sfeir2021}. This way, AD can serve as a diagnostic tool to motivate richer behavioural specifications rather than a purely data cleaning procedure. Furthermore, ongoing work is extending this framework with conditional VAEs, i.e., learning the distribution of variable configurations conditional on contextual variables or choices \citep{pol2019}. Ultimately, generative ML can be used to produce synthetic data that reflect the underlying behavioural patterns learned from the training data \citep{borysov2019}.

This work is not exempt from limitations. Some of them relate to the VAE itself. As with other neural networks, its results can be sensitive to the choice of hyperparameters \citep{ding2022}, in particular the size of the latent space, which determines how well both normal and anomalous observations are reconstructed. We select them through a random search and check that the model reproduces the distributions and correlations in the data, but a different configuration could rank observations differently, especially those close to the threshold. Moreover, the decoder may learn to ignore the latent space (i.e., posterior collapse), which we avoid by checking that the KL terms do not approach zero.

Furthermore, variables with ordered levels, such as income bands, household size and the number of interchanges, are reconstructed as categorical, so the decoder treats their levels as unrelated categories, thus ignoring hierarchies. With few such variables and few levels per variable, as in our datasets, this has little effect on modelling and the scores themselves, but datasets with more ordinal variables, or with more levels per variable, would require likelihoods that respect their order, such as the ordered logit models commonly used in CM. 

Finally, there are limits to what the anomaly score can detect. Although the choice is one of the reconstructed variables, it is only one among many, so the framework mainly detects atypical variable configurations rather than behavioural anomalies, and should be used alongside, not instead of, the usual behavioural diagnostics. The VAE also learns what is `normal' from the data itself, so if the sample is not representative, for example due to strong selection bias, the flagged observations will reflect that. Given the lack of a standardised definition of anomaly in CM, the flagged observations cannot be checked against known anomalies in real-world settings, and the share of observations to inspect, set here through the threshold $\alpha$, remains a modelling decision. It is therefore up to the analyst to decide whether flagged observations are data errors or genuine but unusual trips.

\section*{Acknowledgments}
The authors acknowledge the financial support by the European Research Council through the advanced grant 101020940-SYNERGY.

\bibliography{references}

\appendix

\section{Model validation}
\label{ap:validation}

\begin{figure}[h!]
\centering
\begin{subfigure}[b]{0.49\linewidth}
    \centering
    \includegraphics[width=\linewidth]{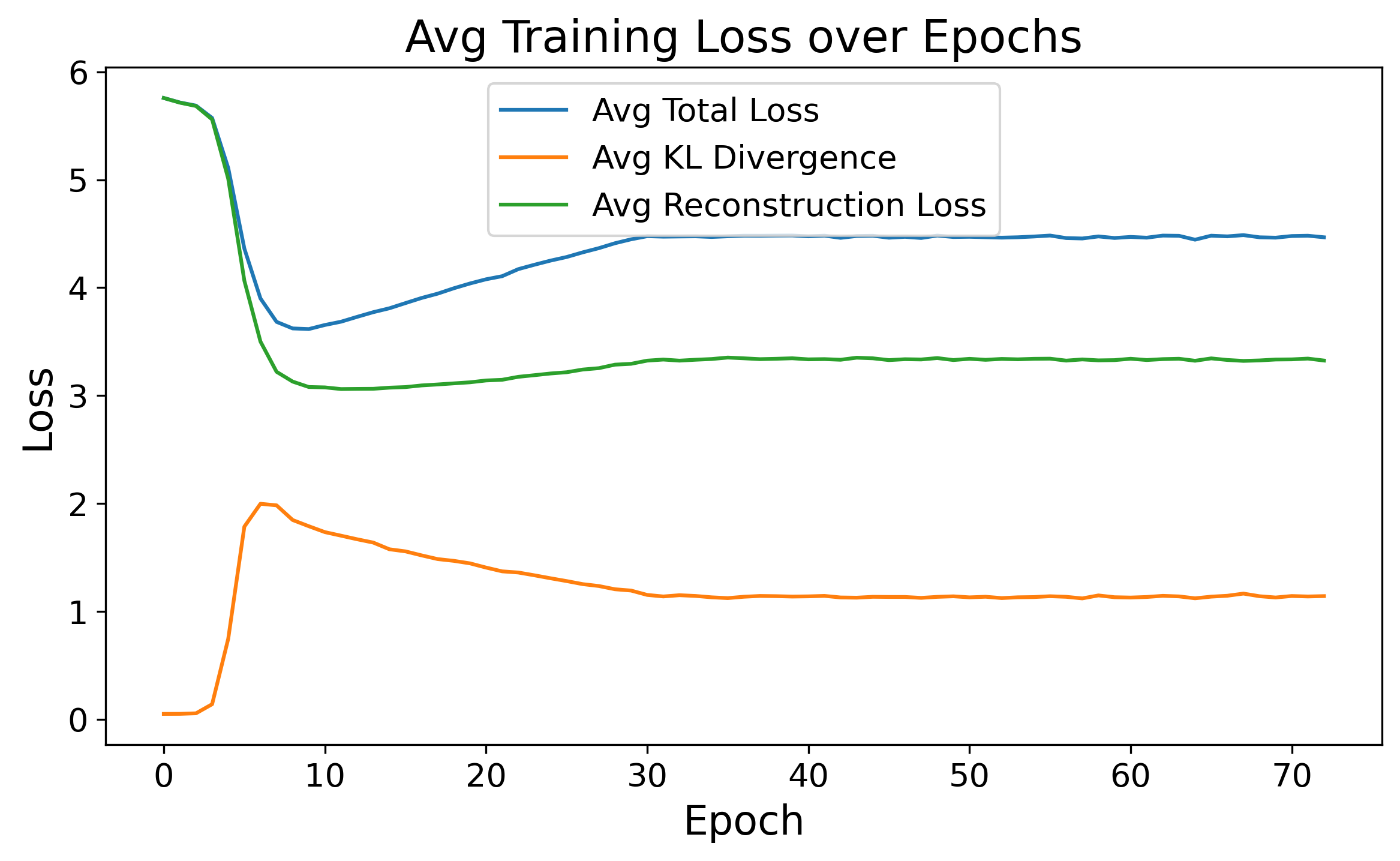}
    \caption{By component (KL and reconstruction loss).}
    \label{fig:loss_synth_components}
\end{subfigure}
\hfill
\begin{subfigure}[b]{0.49\linewidth}
    \centering
    \includegraphics[width=\linewidth]{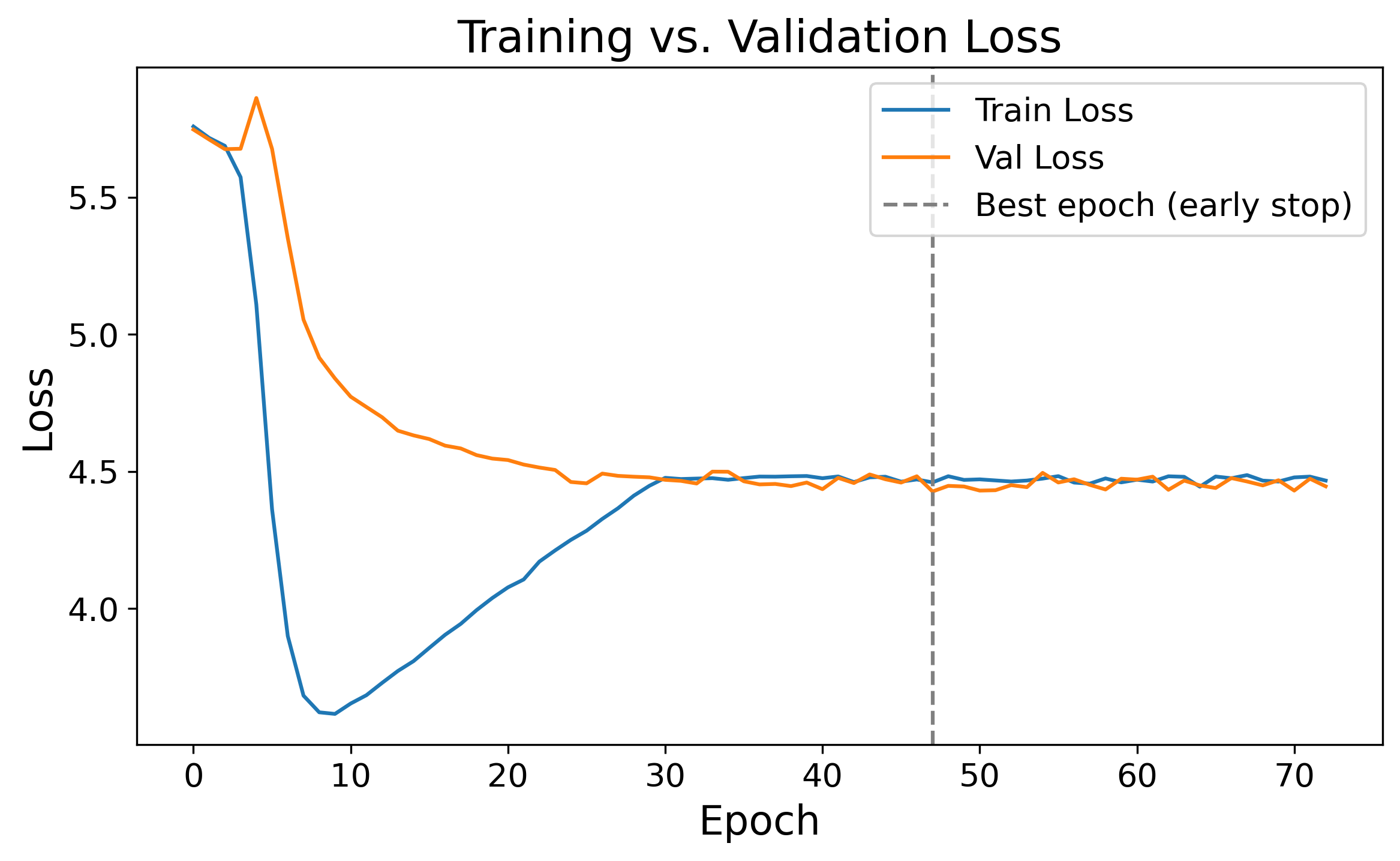}
    \caption{Training and validation loss.}
    \label{fig:loss_synth_val}
\end{subfigure}
\caption{Loss functions for the synthetic dataset.}
\label{fig:loss_synth}
\end{figure}

\begin{figure}[h!]
\centering
\begin{subfigure}[b]{0.49\linewidth}
    \centering
    \includegraphics[width=\linewidth]{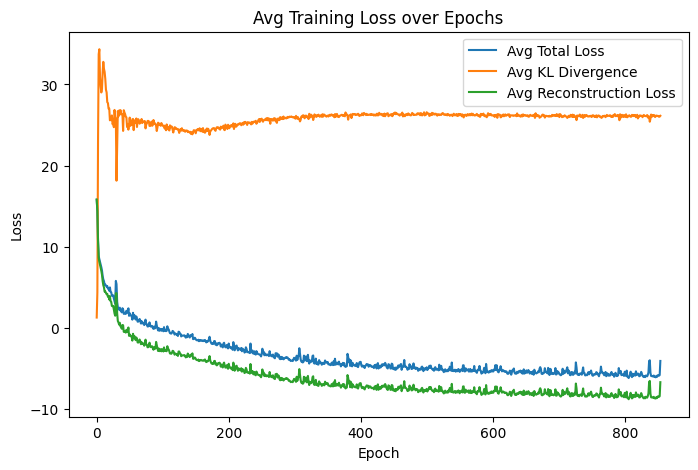}
    \caption{By component (KL and reconstruction loss).}
    \label{fig:loss_swiss_components}
\end{subfigure}
\hfill
\begin{subfigure}[b]{0.49\linewidth}
    \centering
    \includegraphics[width=\linewidth]{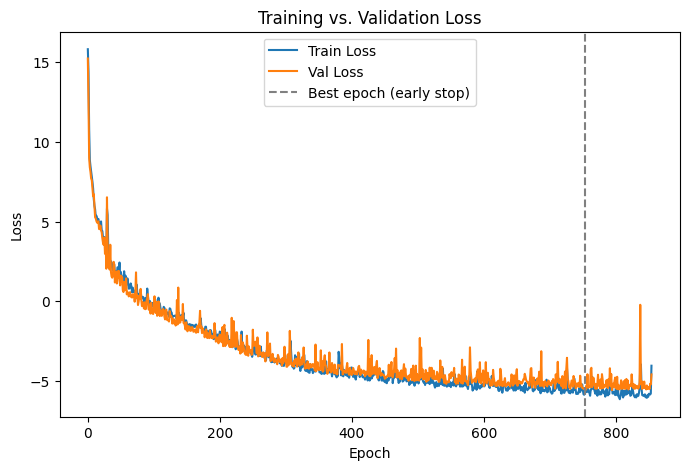}
    \caption{Training and validation loss.}
    \label{fig:loss_swiss_val}
\end{subfigure}
\caption{Loss functions for the Swiss route dataset.}
\label{fig:loss_swiss}
\end{figure}

\begin{figure}[h!]
\centering
\begin{subfigure}[b]{0.49\linewidth}
    \centering
    \includegraphics[width=\linewidth]{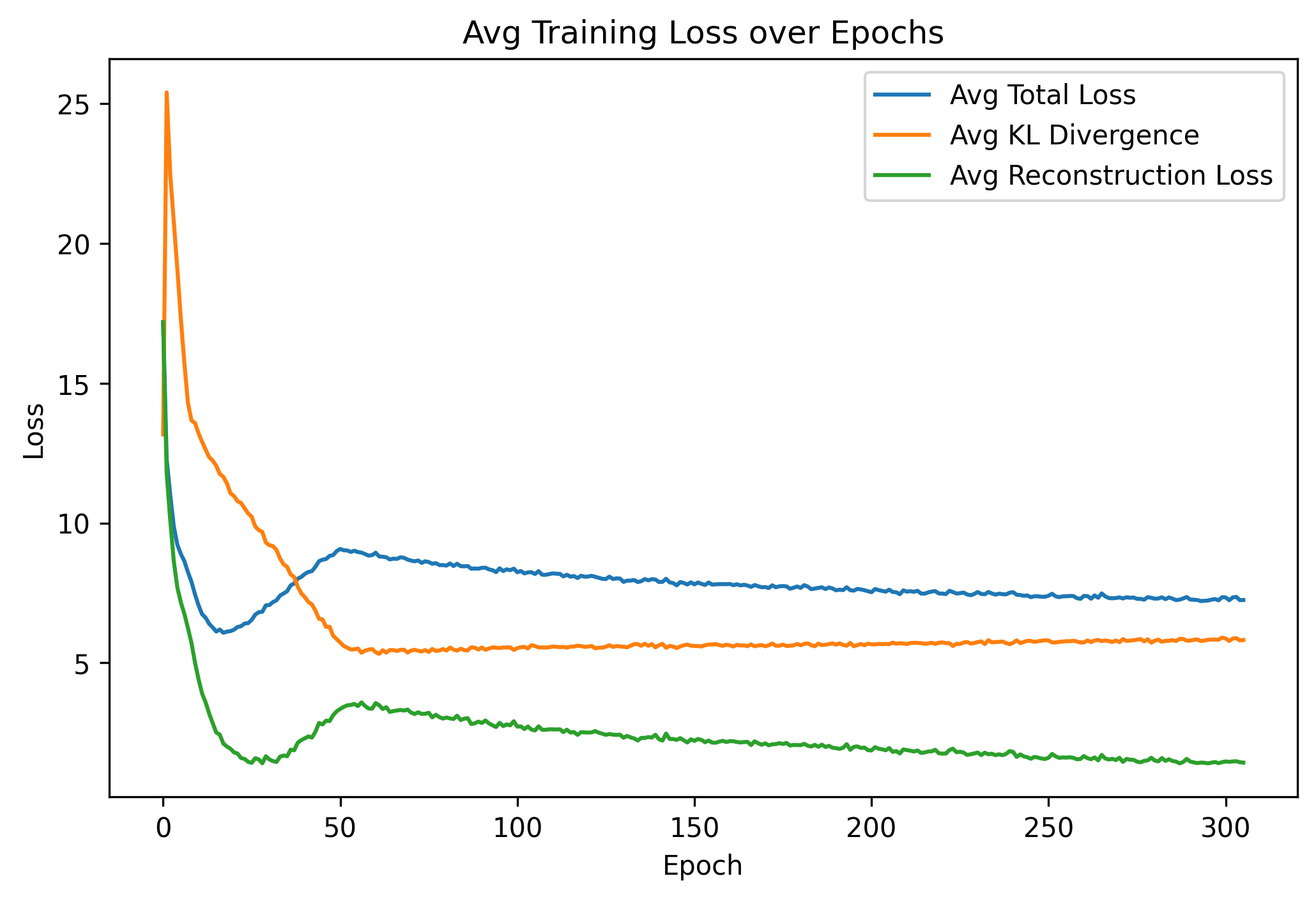}
    \caption{By component (KL and reconstruction loss).}
    \label{fig:loss_decisions_components}
\end{subfigure}
\hfill
\begin{subfigure}[b]{0.49\linewidth}
    \centering
    \includegraphics[width=\linewidth]{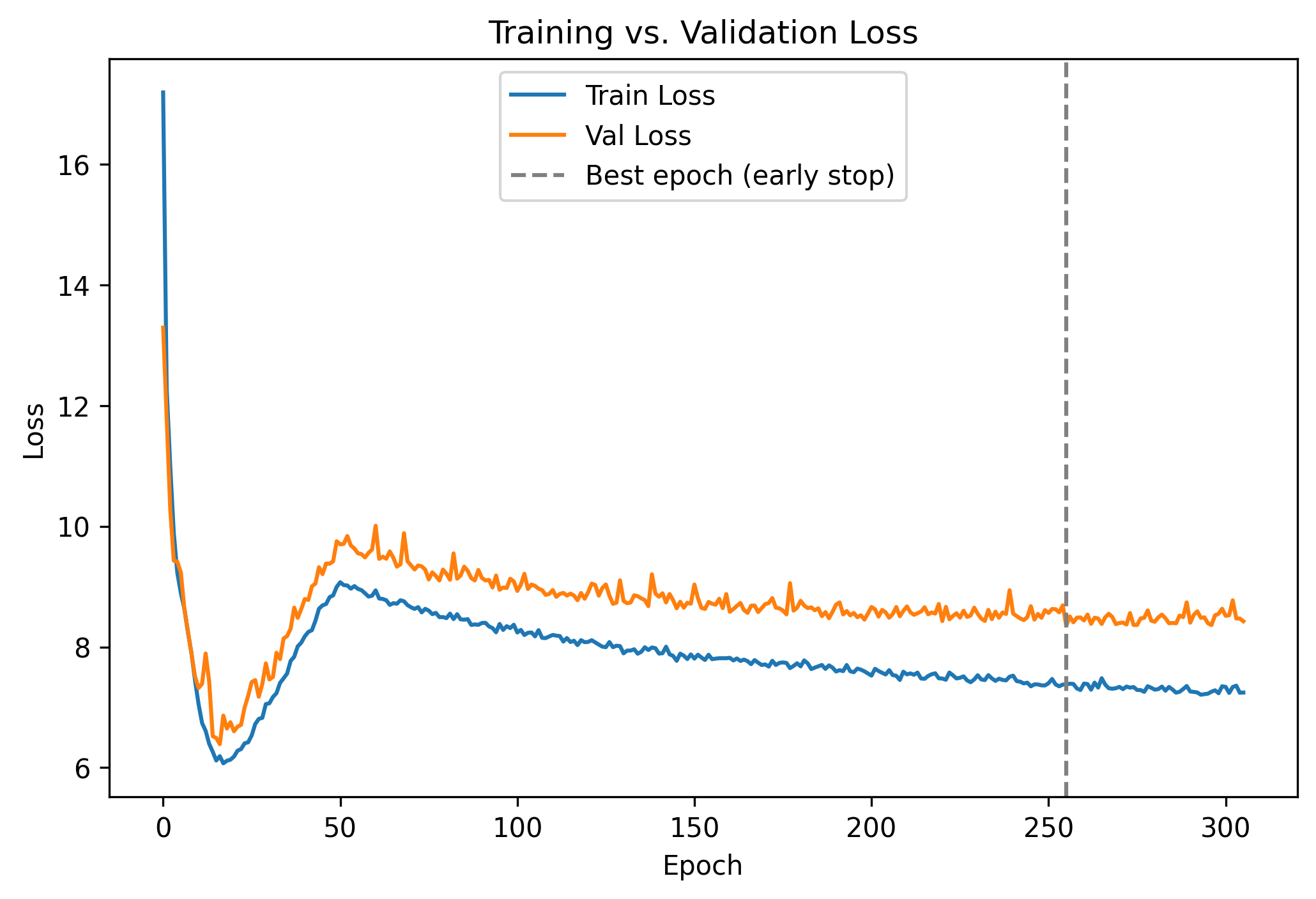}
    \caption{Training and validation loss.}
    \label{fig:loss_decisions_val}
\end{subfigure}
\caption{Loss functions for the Decisions dataset.}
\label{fig:loss_decisions}
\end{figure}

We plot the evolution of the VAE loss components for the synthetic case study in \autoref{fig:loss_synth}, for the Swiss route dataset in \autoref{fig:loss_swiss}, and for the Decisions dataset in \autoref{fig:loss_decisions}. These figures show that all the losses stabilise after a reasonable number of epochs, indicating convergence. Importantly, the KL divergence does not collapse to zero, and the reconstruction and KL terms contribute jointly to the objective. This confirms that the latent space is being actively used and that the model is not collapsing into a standard autoencoder that ignores the latent structure.

\begin{figure}[h!]
\centering
\includegraphics[width=0.7\linewidth]{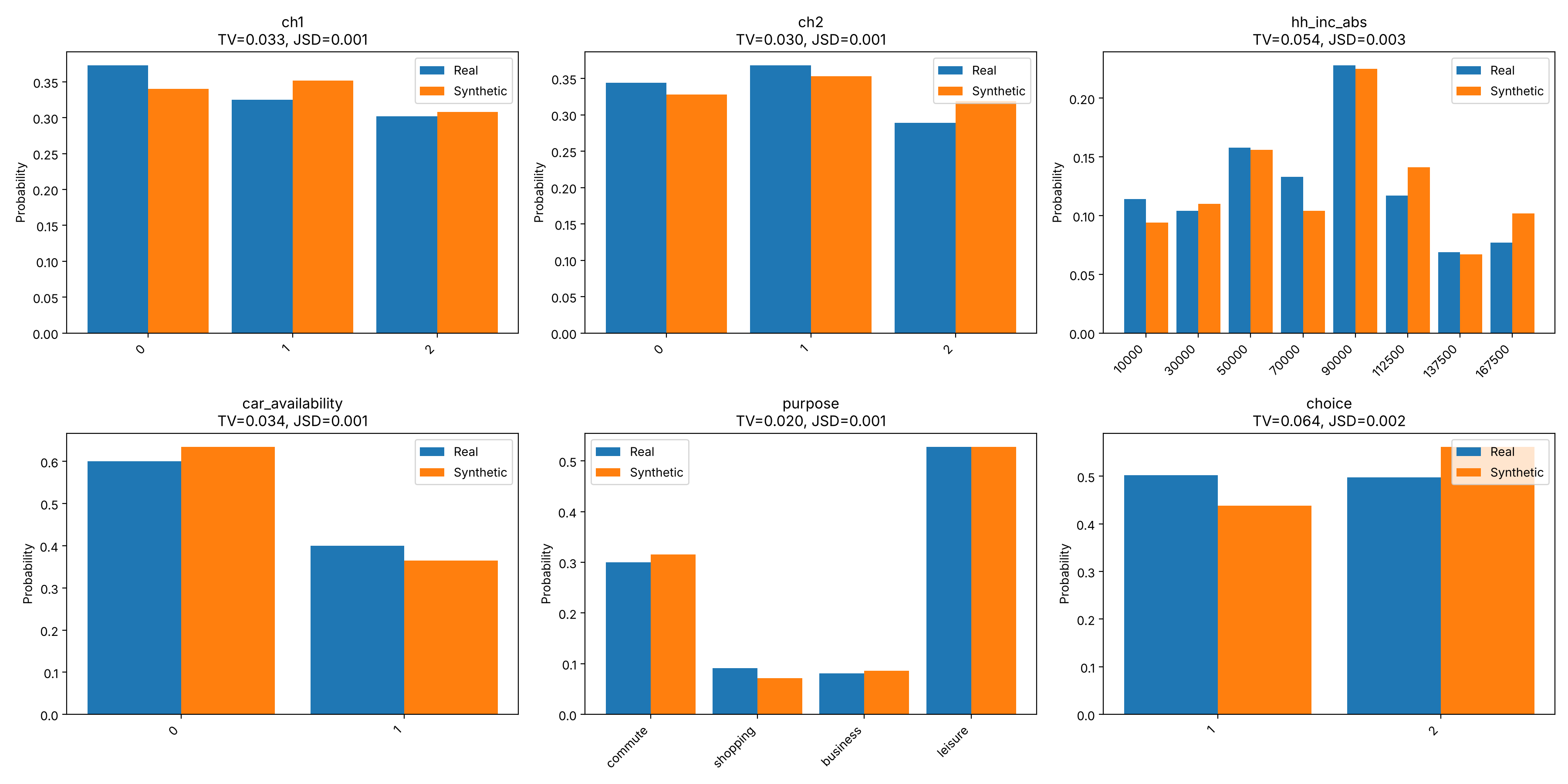}
\caption{Histogram comparison of real and reconstructed categorical variables, Swiss route dataset.}
\label{fig:hist_categ_swiss}
\end{figure}

\begin{figure}[h!]
\centering
\includegraphics[width=0.7\linewidth]{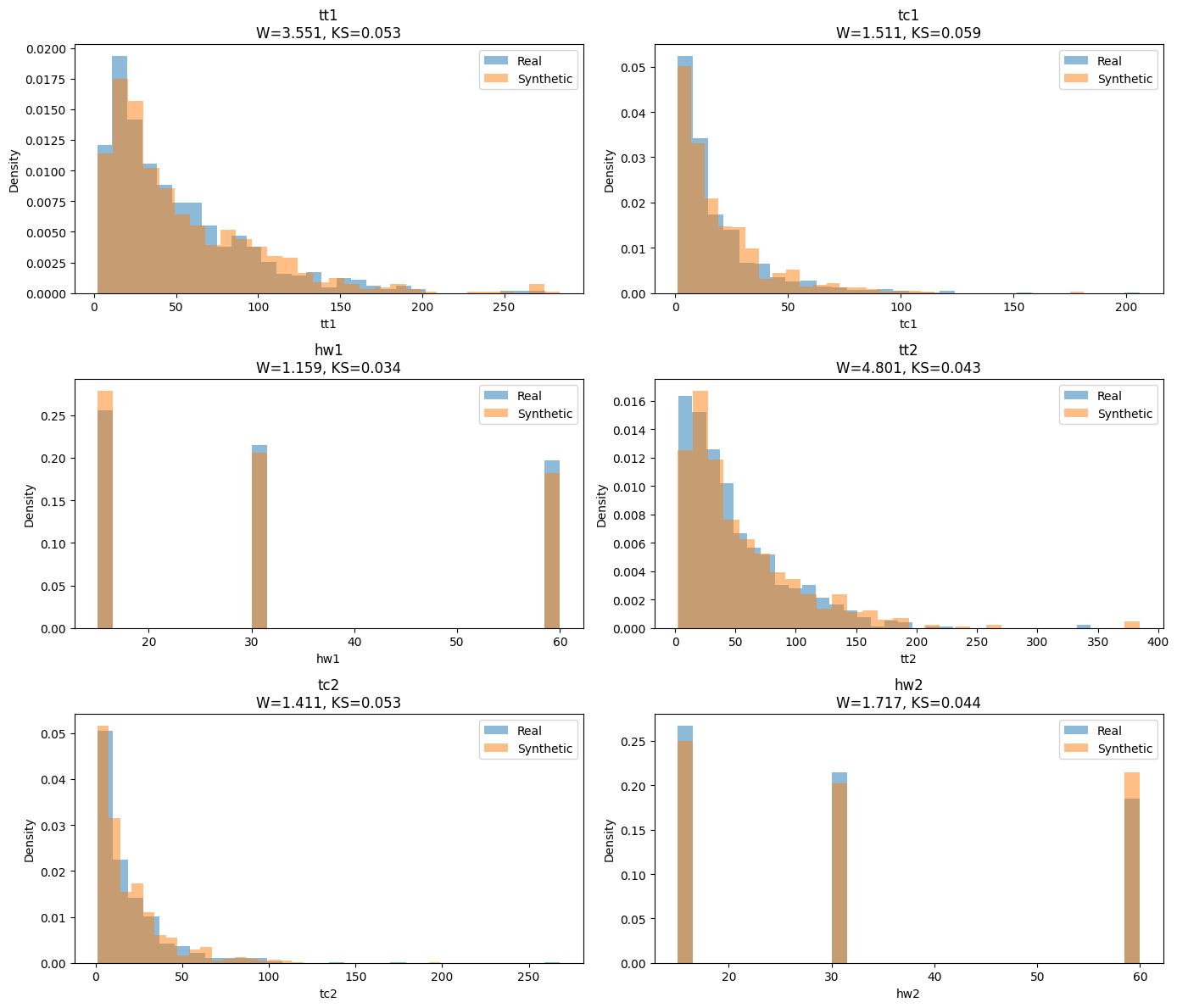}
\caption{Histogram comparison of real and reconstructed continuous variables, Swiss route dataset.}
\label{fig:hist_cont_swiss}
\end{figure}

\begin{figure}[h!]
\centering
\includegraphics[width=0.7\linewidth]{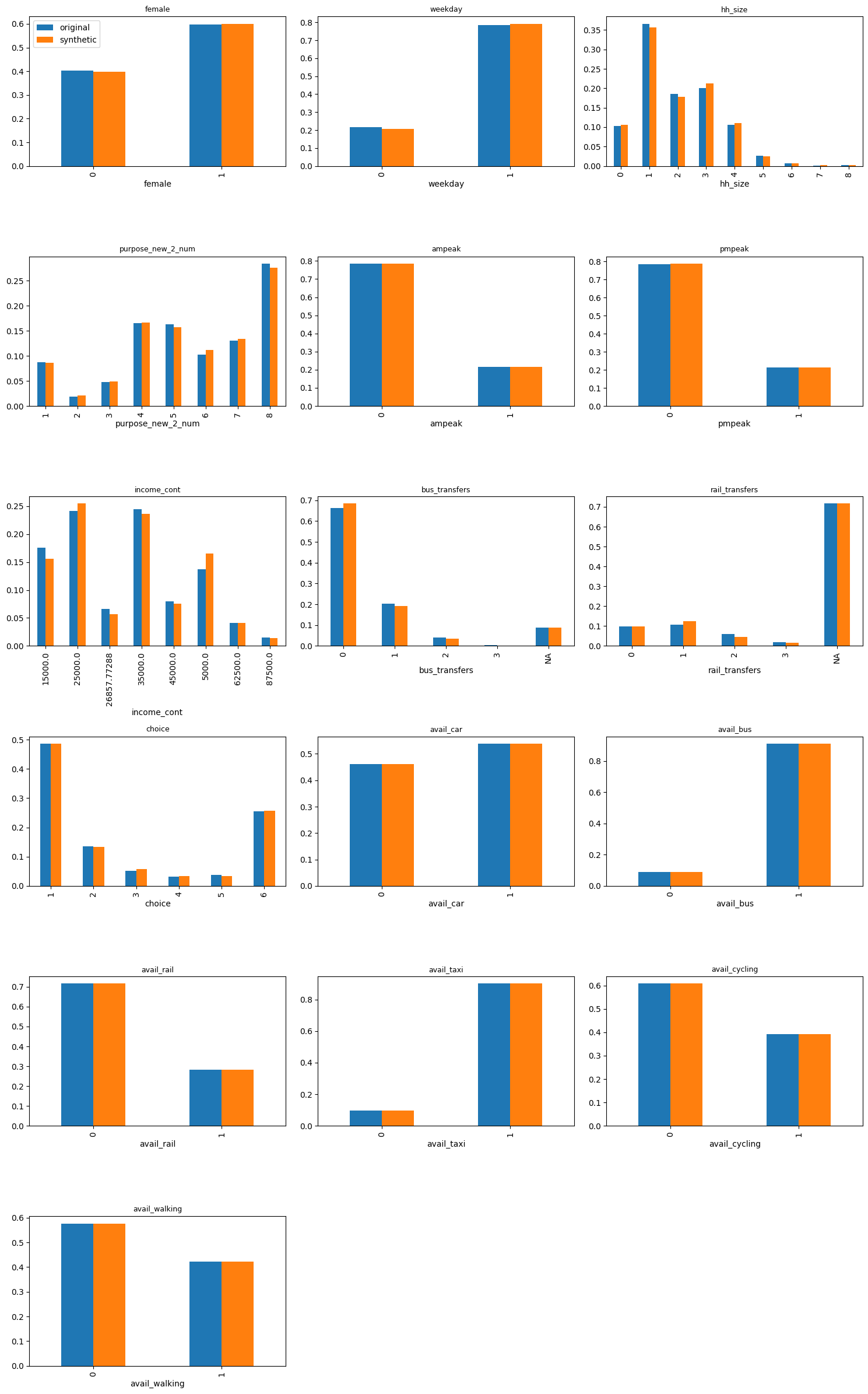}
\caption{Histogram comparison of real and reconstructed categorical variables, Decisions dataset.}
\label{fig:hist_categ_decisions}
\end{figure}

\begin{figure}[h!]
\centering
\includegraphics[width=0.6\linewidth]{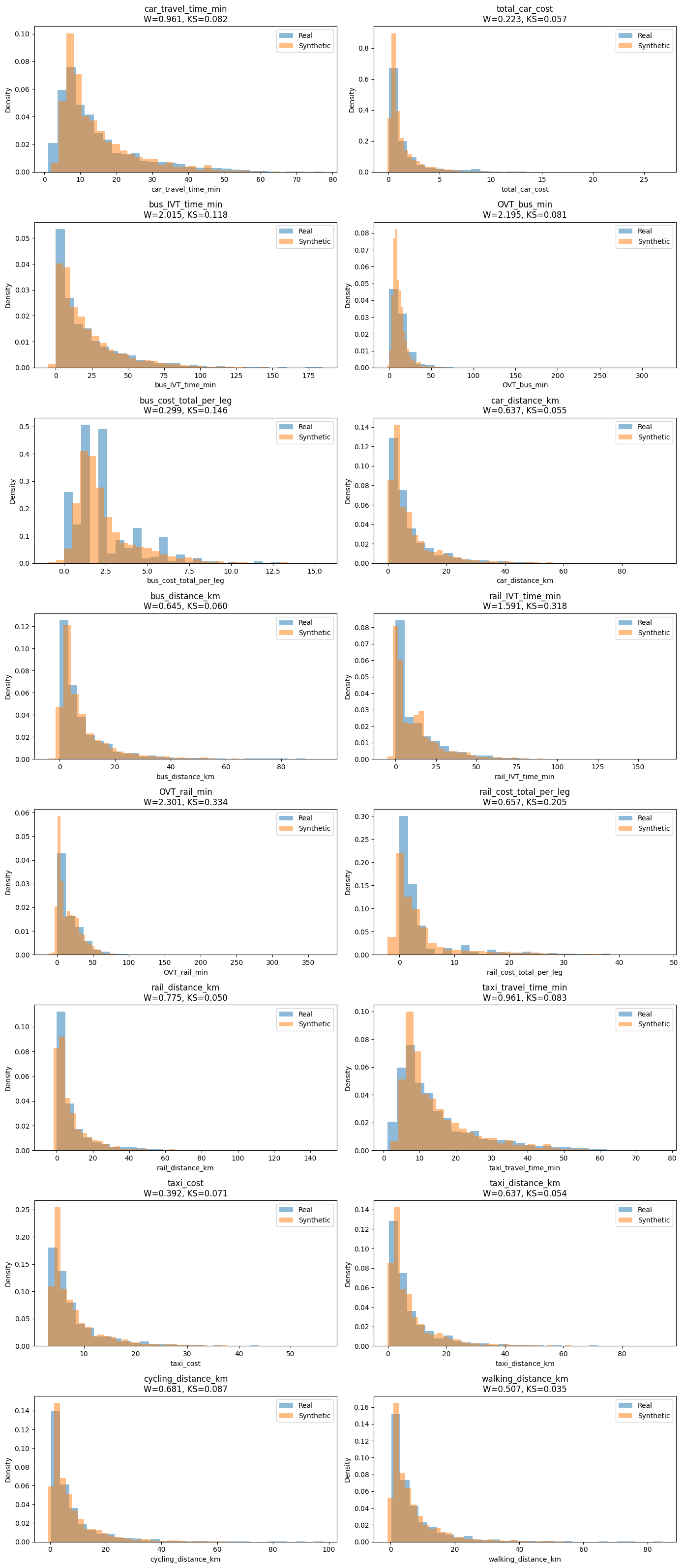}
\caption{Histogram comparison of real and reconstructed continuous variables, Decisions dataset.}
\label{fig:hist_cont_decisions}
\end{figure}

We also compare the empirical distributions of the reconstructed variables with those of the original data. \autoref{fig:hist_categ_swiss} and \autoref{fig:hist_categ_decisions} show the histograms for the categorical variables, while \autoref{fig:hist_cont_swiss} and \autoref{fig:hist_cont_decisions} display the corresponding plots for the continuous variables. In all cases, the reconstructed data closely match the real distributions, both visually through the overlap of the histograms and quantitatively through the distributional indicators discussed in \autoref{sec:theory}. For categorical variables, TV and JSD remain close to zero, indicating that the VAE reproductions have minimal differences with the original data. For continuous variables, the W and KS are also small, showing that both the shape and the values of the reconstructed distributions align well with the empirical data. These results show that the VAE is able to reproduce the marginal structure of the data with a high degree of accuracy. 

\begin{figure}[h!]
\centering
\includegraphics[width=0.9\linewidth]{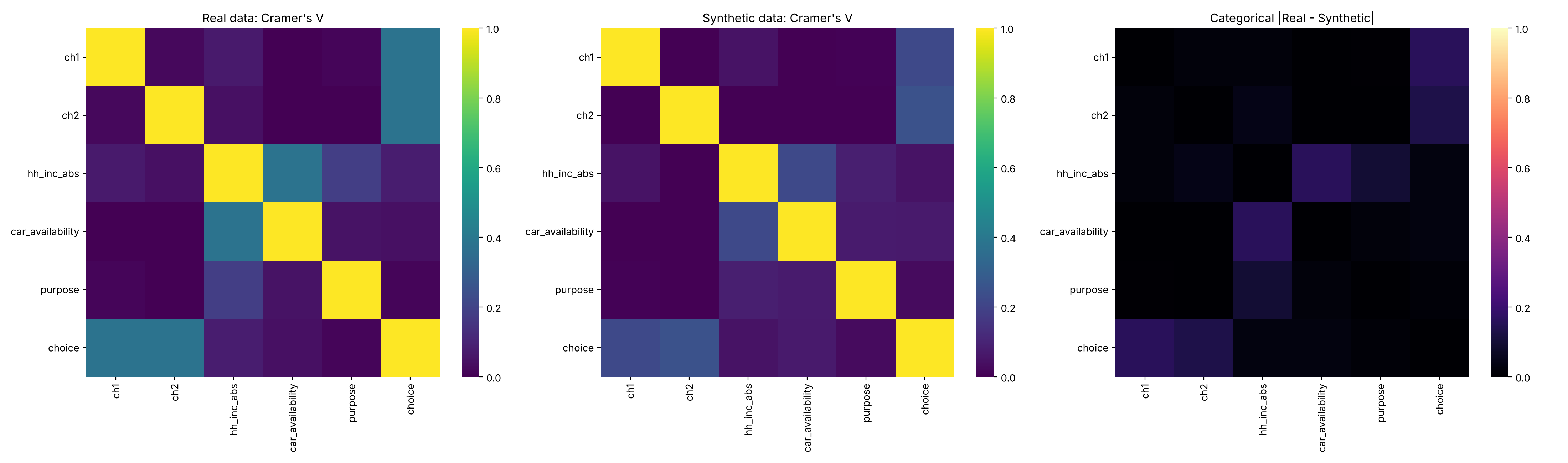}
\caption{Correlation matrices for categorical variables, Swiss route dataset.}
\label{fig:corr_categ_swiss}
\end{figure}

\begin{figure}[h!]
\centering
\includegraphics[width=0.9\linewidth]{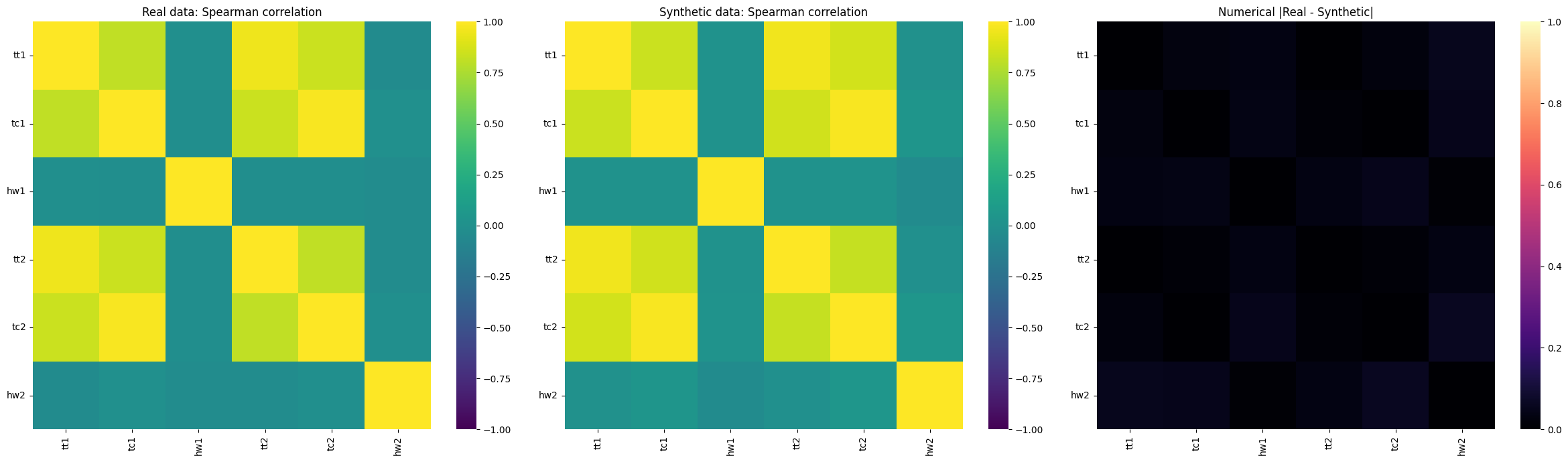}
\caption{Correlation matrices for continuous variables, Swiss route dataset.}
\label{fig:corr_cont_swiss}
\end{figure}

\begin{figure}[h!]
\centering
\includegraphics[width=0.9\linewidth]{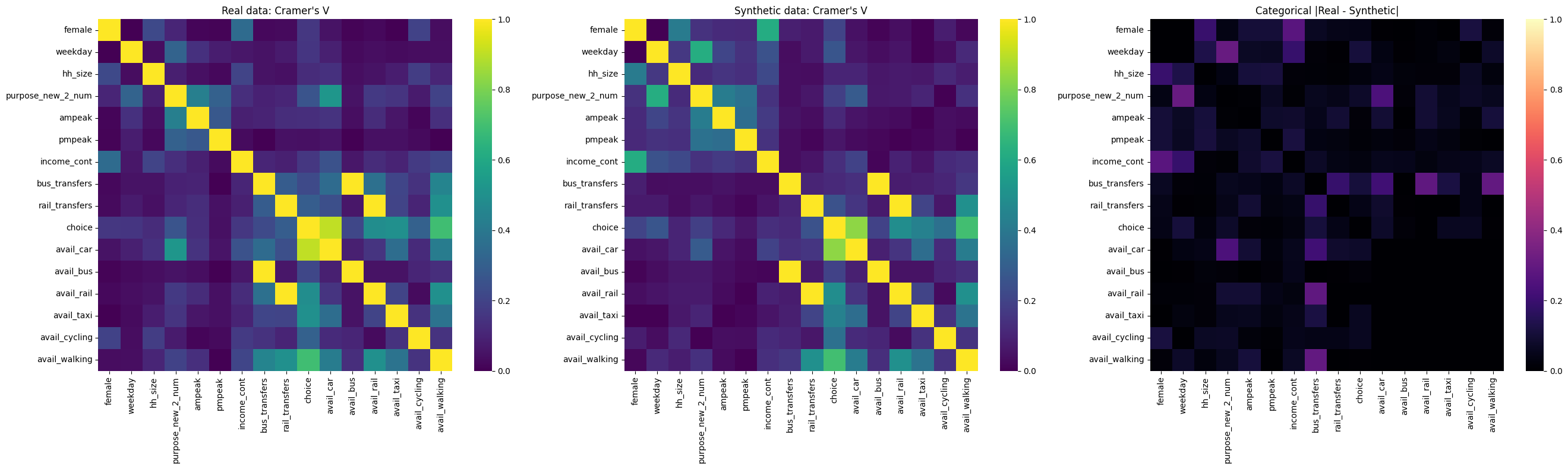}
\caption{Correlation matrices for categorical variables, Decisions dataset.}
\label{fig:corr_categ_decisions}
\end{figure}

\begin{figure}[h!]
\centering
\includegraphics[width=0.9\linewidth]{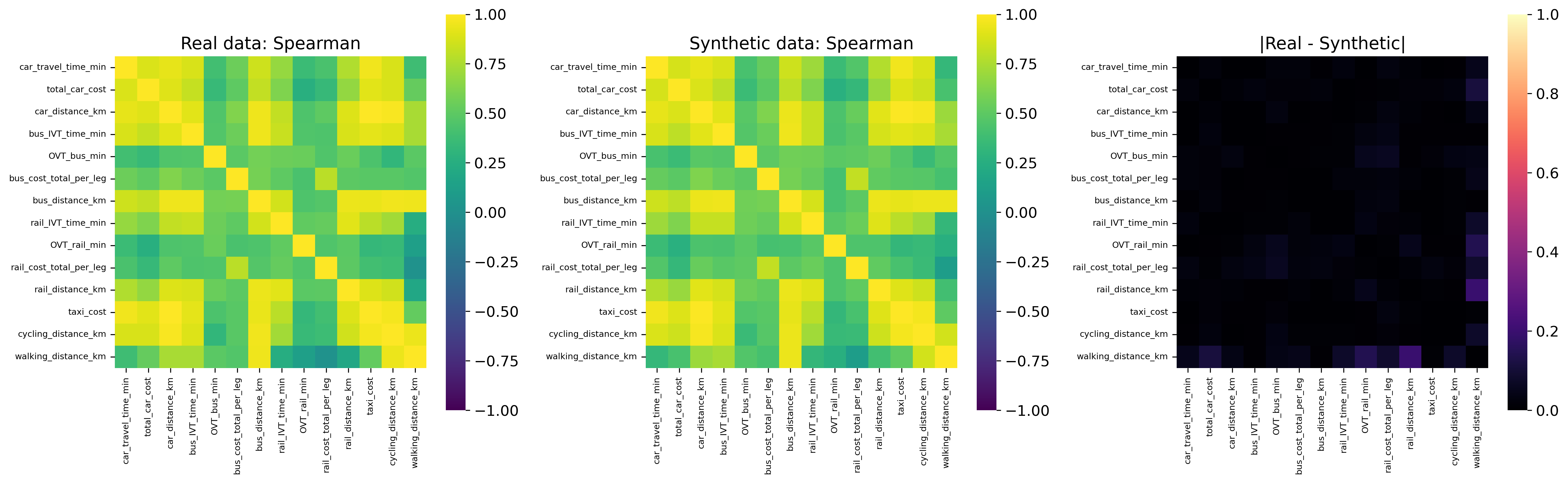}
\caption{Correlation matrices for continuous variables, Decisions dataset.}
\label{fig:corr_cont_decisions}
\end{figure}

We also check for the dependence structure between reconstructed variables. For categorical variables, \autoref{fig:corr_categ_swiss} and \autoref{fig:corr_categ_decisions} present the corresponding matrices based on Cramér’s V, where the reconstructed data closely mirrors the dependence structure observed in the original datasets, with only minor deviations. For continuous variables, \autoref{fig:corr_cont_swiss} and \autoref{fig:corr_cont_decisions} report the Pearson correlation matrices for the original observations and their respective reconstructions, along with their element‑wise absolute difference. The high degree of similarity between the two matrices indicates that the VAE successfully retains the monotonic associations between variables, for both datasets.

\clearpage

\section{Complementary results}
\label{ap:comp_results}

\subsection{Swissroute dataset}
\label{ap:spec_swiss}

We present the parameter estimates and goodness of fit of the Swiss route dataset (\autoref{subsec:swissroute}) in \autoref{tab:mnl_swiss_estimates}.

\begin{sidewaystable}
\centering
\caption{MNL estimates and goodness of fit for the Swiss route dataset, for the full sample, after excluding randomly selected observations (R, mean of ten runs), and after excluding the same number of observations with the highest scores of the univariate check (U), the Mahalanobis distance (M) and the VAE (V). Robust t-ratios in brackets.}
\label{tab:mnl_swiss_estimates}
\small
\setlength{\tabcolsep}{4pt}
\renewcommand{\arraystretch}{1.25}
\resizebox{\textheight}{!}{%
\begin{tabular}{lccccccccccccc}
\toprule
 & & \multicolumn{4}{c}{\textbf{99\% subset}} & \multicolumn{4}{c}{\textbf{95\% subset}} & \multicolumn{4}{c}{\textbf{90\% subset}} \\
\cmidrule(lr){3-6}\cmidrule(lr){7-10}\cmidrule(lr){11-14}
 & \textbf{Full sample} & \textbf{R} & \textbf{U} & \textbf{M} & \textbf{V} & \textbf{R} & \textbf{U} & \textbf{M} & \textbf{V} & \textbf{R} & \textbf{U} & \textbf{M} & \textbf{V} \\
\midrule
\multicolumn{14}{l}{\textbf{Estimates}} \\
$\beta_{tt}$ (travel time) & $-$0.060 ($-$8.96) & $-$0.061 ($-$9.00) & $-$0.059 ($-$8.86) & $-$0.060 ($-$8.94) & $-$0.065 ($-$10.95) & $-$0.060 ($-$9.00) & $-$0.063 ($-$8.82) & $-$0.063 ($-$9.68) & $-$0.066 ($-$10.30) & $-$0.060 ($-$8.65) & $-$0.064 ($-$8.60) & $-$0.064 ($-$9.57) & $-$0.071 ($-$9.97) \\
$\beta_{tc}$ (travel cost) & $-$0.133 ($-$5.66) & $-$0.134 ($-$5.69) & $-$0.131 ($-$5.59) & $-$0.132 ($-$5.65) & $-$0.149 ($-$7.15) & $-$0.133 ($-$5.74) & $-$0.135 ($-$5.34) & $-$0.144 ($-$6.86) & $-$0.140 ($-$6.28) & $-$0.133 ($-$5.59) & $-$0.136 ($-$5.28) & $-$0.143 ($-$6.43) & $-$0.156 ($-$6.17) \\
$\beta_{hw}$ (headway) & $-$0.038 ($-$16.18) & $-$0.038 ($-$16.10) & $-$0.038 ($-$16.08) & $-$0.037 ($-$15.99) & $-$0.038 ($-$16.08) & $-$0.037 ($-$15.88) & $-$0.038 ($-$16.02) & $-$0.037 ($-$15.56) & $-$0.041 ($-$16.40) & $-$0.037 ($-$15.63) & $-$0.039 ($-$15.49) & $-$0.037 ($-$15.47) & $-$0.041 ($-$16.14) \\
$\beta_{ch}$ (interchanges) & $-$1.155 ($-$18.79) & $-$1.158 ($-$18.83) & $-$1.158 ($-$18.66) & $-$1.152 ($-$18.64) & $-$1.168 ($-$18.61) & $-$1.157 ($-$18.55) & $-$1.178 ($-$18.44) & $-$1.149 ($-$17.98) & $-$1.229 ($-$19.07) & $-$1.161 ($-$18.39) & $-$1.184 ($-$17.84) & $-$1.163 ($-$17.86) & $-$1.306 ($-$19.31) \\
ASC car available & $-$0.246 ($-$2.54) & $-$0.247 ($-$2.54) & $-$0.238 ($-$2.45) & $-$0.250 ($-$2.57) & $-$0.235 ($-$2.40) & $-$0.238 ($-$2.41) & $-$0.235 ($-$2.37) & $-$0.252 ($-$2.58) & $-$0.220 ($-$2.18) & $-$0.246 ($-$2.45) & $-$0.229 ($-$2.26) & $-$0.220 ($-$2.17) & $-$0.255 ($-$2.34) \\
ASC commute & 0.115 (1.38) & 0.113 (1.35) & 0.107 (1.28) & 0.114 (1.36) & 0.100 (1.20) & 0.113 (1.33) & 0.123 (1.44) & 0.116 (1.37) & 0.079 (0.95) & 0.121 (1.37) & 0.158 (1.81) & 0.097 (1.11) & 0.033 (0.38) \\
ASC shopping & 0.144 (0.99) & 0.142 (0.97) & 0.143 (0.98) & 0.157 (1.08) & 0.162 (1.04) & 0.135 (0.88) & 0.144 (0.97) & 0.224 (1.49) & 0.182 (1.25) & 0.136 (0.87) & 0.119 (0.79) & 0.187 (1.18) & 0.196 (1.35) \\
ASC business & $-$0.132 ($-$0.86) & $-$0.133 ($-$0.87) & $-$0.136 ($-$0.88) & $-$0.130 ($-$0.84) & $-$0.143 ($-$0.91) & $-$0.136 ($-$0.85) & $-$0.135 ($-$0.86) & $-$0.134 ($-$0.85) & $-$0.126 ($-$0.76) & $-$0.137 ($-$0.84) & $-$0.139 ($-$0.88) & $-$0.141 ($-$0.88) & $-$0.099 ($-$0.58) \\
ASC leisure & 0.084 (1.14) & 0.087 (1.17) & 0.073 (0.98) & 0.089 (1.21) & 0.076 (1.03) & 0.075 (0.99) & 0.083 (1.06) & 0.110 (1.49) & 0.041 (0.54) & 0.078 (1.01) & 0.053 (0.65) & 0.113 (1.44) & 0.057 (0.69) \\
\midrule
\multicolumn{14}{l}{\textbf{Goodness of fit}} \\
Observations & 3,492 & 3,457 & 3,457 & 3,457 & 3,457 & 3,317 & 3,317 & 3,317 & 3,317 & 3,143 & 3,143 & 3,143 & 3,143 \\
LL(0) & $-$2,420.47 & $-$2,396.21 & $-$2,396.21 & $-$2,396.21 & $-$2,396.21 & $-$2,299.17 & $-$2,299.17 & $-$2,299.17 & $-$2,299.17 & $-$2,178.56 & $-$2,178.56 & $-$2,178.56 & $-$2,178.56 \\
LL(final) & $-$1,659.94 & $-$1,640.86 & $-$1,642.27 & $-$1,646.95 & $-$1,625.75 & $-$1,576.37 & $-$1,559.87 & $-$1,582.83 & $-$1,519.45 & $-$1,492.25 & $-$1,473.31 & $-$1,494.03 & $-$1,403.01 \\
Adjusted $\rho^2$ & 0.310 & 0.311 & 0.311 & 0.309 & 0.318 & 0.310 & 0.318 & 0.308 & 0.335 & 0.311 & 0.320 & 0.310 & 0.352 \\
AIC & 3,337.88 & 3,299.72 & 3,302.54 & 3,311.90 & 3,269.50 & 3,170.75 & 3,137.73 & 3,183.66 & 3,056.89 & 3,002.50 & 2,964.63 & 3,006.06 & 2,824.01 \\
BIC & 3,393.30 & 3,355.05 & 3,357.87 & 3,367.23 & 3,324.84 & 3,225.71 & 3,192.69 & 3,238.62 & 3,111.85 & 3,056.97 & 3,019.11 & 3,060.54 & 2,878.49 \\
\bottomrule
\end{tabular}}
\end{sidewaystable}

\subsection{Decisions dataset}
\label{ap:spec_decisions}

We estimate an MNL model in which individual $n$ chooses, for each observation $t$, between up to six modes: car, bus, rail, taxi, cycling and walking. The utilities are

\begin{equation}
\begin{aligned}
V_{car,nt}  = {} & \beta_{car}^{tt}\,tt_{car,nt} + \beta_{cost}\,c_{car,nt},\\
V_{bus,nt}  = {} & ASC_{bus} + \beta_{bus}^{ivt}\,ivt_{bus,nt} + \beta_{bus}^{ovt}\,ovt_{bus,nt} + \beta_{cost}\,c_{bus,nt} +  S_{bus,nt},\\
V_{rail,nt} = {} & ASC_{rail} + \beta_{rail}^{ivt}\,ivt_{rail,nt} + \beta_{rail}^{ovt}\,ovt_{rail,nt} + \beta_{cost}\,c_{rail,nt} +  S_{rail,nt},\\
V_{taxi,nt} = {} & ASC_{taxi} + \beta_{taxi}^{tt}\,tt_{taxi,nt} + \beta_{cost}\,c_{taxi,nt} + S_{taxi,nt},\\
V_{cyc,nt}  = {} & ASC_{cyc} + \beta_{cyc}^{d}\,d_{cyc,nt} + S_{cyc,nt},\\
V_{walk,nt} = {} & ASC_{walk} + \beta_{walk}^{d}\,d_{walk,nt} + S_{walk,nt},
\end{aligned}
\label{eq:v_decisions}
\end{equation}

where $tt$ is the travel time (minutes), $ivt$ and $ovt$ are the in-vehicle and out-of-vehicle times (minutes), with out-of-vehicle time including access, egress and waiting time, $c$ is the travel cost (£), and $d$ is the travel distance (km). The cost parameter $\beta_{cost}$ is generic across modes, while the time and distance parameters are mode-specific. 

The terms $S_{j,nt}$ shift the ASC of each alternative $j$ according to the characteristics of the individual and the trip:

\begin{equation}
\begin{aligned}
S_{j,nt} = {} & ASC_{j}^{male}\,\mathrm{male}_n + ASC_{j}^{stud}\,\mathrm{student}_n + ASC_{j}^{wkd}\,\mathrm{weekend}_{nt} + ASC_{j}^{bsn}\,\mathrm{business}_{nt} \\
& + ASC_{j}^{com}\,\mathrm{commute}_{nt} + ASC_{j}^{am}\,\mathrm{AM}_{nt} + ASC_{j}^{pm}\,\mathrm{PM}_{nt} + ASC_{j}^{short}\,\mathrm{short}_{nt},
\end{aligned}
\label{eq:s_decisions}
\end{equation}

where $\mathrm{male}_n$ and $\mathrm{student}_n$ are dummy variables for male individuals and students, $\mathrm{weekend}_{nt}$, $\mathrm{business}_{nt}$ and $\mathrm{commute}_{nt}$ are dummy variables for weekend, business and commuting trips, $\mathrm{AM}_{nt}$ and $\mathrm{PM}_{nt}$ are dummy variables for trips in the morning and afternoon peaks, and $\mathrm{short}_{nt}$ is a dummy variable for short trips.

\autoref{tab:mnl_decisions_estimates} reports the parameter estimates, alternative-specific constants and their shifts from the sensitivity analysis in \autoref{subsec:decisions}, together with goodness of fit.

\begin{sidewaystable}
\centering
\caption{MNL estimates and goodness of fit for the Decisions dataset (\autoref{eq:v_decisions} and \autoref{eq:s_decisions}), for the full sample, after excluding randomly selected observations (R, mean of ten runs), and after excluding the same number of observations with the highest scores of the univariate check (U), the Mahalanobis distance (M) and the VAE (V). Car is the reference alternative. In the 90\% subset of the univariate check, the short-trip shift of rail is fixed at zero (--), as only one short trip with rail available remains. Robust t-ratios in brackets.}
\label{tab:mnl_decisions_estimates}
\tiny
\setlength{\tabcolsep}{3pt}
\renewcommand{\arraystretch}{1.05}
\resizebox{\textheight}{!}{%
\begin{tabular}{lccccccccccccc}
\toprule
 & & \multicolumn{4}{c}{\textbf{99\% subset}} & \multicolumn{4}{c}{\textbf{95\% subset}} & \multicolumn{4}{c}{\textbf{90\% subset}} \\
\cmidrule(lr){3-6}\cmidrule(lr){7-10}\cmidrule(lr){11-14}
 & \textbf{Full sample} & \textbf{R} & \textbf{U} & \textbf{M} & \textbf{V} & \textbf{R} & \textbf{U} & \textbf{M} & \textbf{V} & \textbf{R} & \textbf{U} & \textbf{M} & \textbf{V} \\
\midrule
\multicolumn{14}{l}{\textbf{Time, distance and cost parameters}} \\
Car travel time & $-$0.115 ($-$7.81) & $-$0.115 ($-$7.83) & $-$0.123 ($-$8.72) & $-$0.119 ($-$7.58) & $-$0.129 ($-$9.16) & $-$0.116 ($-$7.71) & $-$0.133 ($-$9.10) & $-$0.129 ($-$8.65) & $-$0.136 ($-$8.25) & $-$0.115 ($-$7.65) & $-$0.133 ($-$8.86) & $-$0.135 ($-$8.28) & $-$0.125 ($-$6.20) \\
Bus IVT & $-$0.042 ($-$5.77) & $-$0.042 ($-$5.76) & $-$0.049 ($-$7.05) & $-$0.045 ($-$6.01) & $-$0.051 ($-$7.17) & $-$0.043 ($-$5.74) & $-$0.051 ($-$7.07) & $-$0.052 ($-$7.04) & $-$0.060 ($-$7.38) & $-$0.042 ($-$5.67) & $-$0.053 ($-$7.36) & $-$0.053 ($-$6.51) & $-$0.065 ($-$6.78) \\
Bus OVT & $-$0.116 ($-$3.33) & $-$0.116 ($-$3.30) & $-$0.167 ($-$12.57) & $-$0.133 ($-$3.87) & $-$0.171 ($-$12.45) & $-$0.117 ($-$3.25) & $-$0.162 ($-$12.00) & $-$0.176 ($-$13.07) & $-$0.180 ($-$12.40) & $-$0.115 ($-$3.19) & $-$0.164 ($-$11.77) & $-$0.177 ($-$13.00) & $-$0.193 ($-$12.90) \\
Rail IVT & $-$0.043 ($-$2.35) & $-$0.043 ($-$2.35) & $-$0.045 ($-$2.49) & $-$0.044 ($-$2.33) & $-$0.045 ($-$2.50) & $-$0.044 ($-$2.36) & $-$0.049 ($-$2.59) & $-$0.053 ($-$2.83) & $-$0.042 ($-$2.01) & $-$0.043 ($-$2.34) & $-$0.052 ($-$2.71) & $-$0.050 ($-$2.17) & $-$0.029 ($-$1.14) \\
Rail OVT & $-$0.108 ($-$7.05) & $-$0.108 ($-$7.04) & $-$0.119 ($-$7.73) & $-$0.112 ($-$7.04) & $-$0.111 ($-$7.38) & $-$0.109 ($-$7.02) & $-$0.121 ($-$7.33) & $-$0.115 ($-$7.15) & $-$0.117 ($-$7.05) & $-$0.107 ($-$6.81) & $-$0.125 ($-$7.32) & $-$0.121 ($-$7.25) & $-$0.127 ($-$6.56) \\
Taxi travel time & $-$0.092 ($-$2.84) & $-$0.092 ($-$2.83) & $-$0.105 ($-$3.04) & $-$0.117 ($-$3.68) & $-$0.108 ($-$3.09) & $-$0.091 ($-$2.80) & $-$0.108 ($-$2.86) & $-$0.130 ($-$4.09) & $-$0.171 ($-$5.26) & $-$0.089 ($-$2.74) & $-$0.120 ($-$3.58) & $-$0.133 ($-$4.29) & $-$0.180 ($-$5.20) \\
Cycling distance & $-$0.226 ($-$5.30) & $-$0.226 ($-$5.30) & $-$0.247 ($-$5.89) & $-$0.236 ($-$5.42) & $-$0.260 ($-$6.14) & $-$0.227 ($-$5.29) & $-$0.252 ($-$5.97) & $-$0.252 ($-$5.68) & $-$0.303 ($-$6.53) & $-$0.229 ($-$5.27) & $-$0.257 ($-$5.77) & $-$0.253 ($-$5.51) & $-$0.297 ($-$5.87) \\
Walking distance & $-$1.862 ($-$9.35) & $-$1.865 ($-$9.37) & $-$1.952 ($-$10.06) & $-$1.880 ($-$9.44) & $-$1.943 ($-$10.11) & $-$1.839 ($-$9.10) & $-$1.940 ($-$9.70) & $-$2.033 ($-$11.17) & $-$1.985 ($-$10.09) & $-$1.845 ($-$9.05) & $-$1.979 ($-$9.34) & $-$2.048 ($-$11.29) & $-$2.037 ($-$10.19) \\
Cost & $-$0.178 ($-$6.23) & $-$0.178 ($-$6.21) & $-$0.199 ($-$6.75) & $-$0.189 ($-$6.53) & $-$0.206 ($-$6.95) & $-$0.179 ($-$6.17) & $-$0.205 ($-$6.50) & $-$0.195 ($-$6.01) & $-$0.231 ($-$6.75) & $-$0.178 ($-$6.10) & $-$0.229 ($-$6.73) & $-$0.212 ($-$5.51) & $-$0.261 ($-$7.11) \\
\midrule
\multicolumn{14}{l}{\textbf{Alternative-specific constants: bus} ($ASC_{bus}$)} \\
\quad $ASC_{bus}$ & $-$2.718 ($-$6.38) & $-$2.722 ($-$6.37) & $-$2.240 ($-$6.81) & $-$2.595 ($-$6.27) & $-$2.237 ($-$6.63) & $-$2.714 ($-$6.26) & $-$2.384 ($-$7.07) & $-$2.341 ($-$6.85) & $-$2.272 ($-$6.30) & $-$2.712 ($-$6.18) & $-$2.362 ($-$6.92) & $-$2.399 ($-$7.10) & $-$2.100 ($-$5.58) \\
\quad $ASC^{male}$ & 0.257 (1.04) & 0.253 (1.02) & 0.259 (1.02) & 0.279 (1.12) & 0.257 (1.01) & 0.247 (0.99) & 0.251 (0.98) & 0.347 (1.31) & 0.255 (0.95) & 0.256 (1.03) & 0.292 (1.13) & 0.309 (1.14) & 0.238 (0.86) \\
\quad $ASC^{wkd}$ & $-$0.351 ($-$1.68) & $-$0.347 ($-$1.66) & $-$0.384 ($-$1.76) & $-$0.360 ($-$1.69) & $-$0.436 ($-$1.96) & $-$0.354 ($-$1.66) & $-$0.436 ($-$2.03) & $-$0.261 ($-$1.21) & $-$0.523 ($-$2.13) & $-$0.353 ($-$1.63) & $-$0.460 ($-$2.09) & $-$0.243 ($-$1.05) & $-$0.672 ($-$2.45) \\
\quad $ASC^{stud}$ & 0.457 (1.01) & 0.461 (1.02) & 0.447 (0.97) & 0.484 (1.05) & 0.487 (1.05) & 0.457 (1.01) & 0.458 (0.97) & 0.259 (0.55) & 0.515 (1.06) & 0.441 (0.98) & 0.523 (1.13) & 0.318 (0.65) & 0.617 (1.21) \\
\quad $ASC^{bsn}$ & 1.309 (5.76) & 1.306 (5.75) & 1.317 (5.63) & 1.313 (5.71) & 1.299 (5.51) & 1.320 (5.75) & 1.291 (5.54) & 1.408 (5.79) & 1.347 (5.42) & 1.304 (5.66) & 1.336 (5.76) & 1.406 (5.46) & 1.348 (5.08) \\
\quad $ASC^{am}$ & 0.413 (1.97) & 0.413 (1.98) & 0.466 (2.14) & 0.429 (2.00) & 0.453 (2.04) & 0.404 (1.91) & 0.454 (2.04) & 0.528 (2.28) & 0.556 (2.48) & 0.404 (1.90) & 0.457 (2.08) & 0.583 (2.42) & 0.729 (3.10) \\
\quad $ASC^{pm}$ & $-$1.141 ($-$4.40) & $-$1.144 ($-$4.39) & $-$1.152 ($-$4.45) & $-$1.136 ($-$4.34) & $-$1.178 ($-$4.53) & $-$1.106 ($-$4.21) & $-$1.198 ($-$4.42) & $-$0.967 ($-$3.78) & $-$1.200 ($-$4.14) & $-$1.125 ($-$4.23) & $-$1.118 ($-$4.09) & $-$0.874 ($-$3.31) & $-$1.102 ($-$3.63) \\
\quad $ASC^{short}$ & $-$0.738 ($-$2.80) & $-$0.735 ($-$2.80) & $-$0.898 ($-$3.42) & $-$0.770 ($-$2.94) & $-$0.873 ($-$3.30) & $-$0.736 ($-$2.78) & $-$0.912 ($-$3.26) & $-$0.996 ($-$3.72) & $-$0.842 ($-$3.09) & $-$0.766 ($-$2.83) & $-$0.880 ($-$3.08) & $-$0.842 ($-$3.18) & $-$0.873 ($-$3.05) \\
\midrule
\multicolumn{14}{l}{\textbf{Alternative-specific constants: rail} ($ASC_{rail}$)} \\
\quad $ASC_{rail}$ & $-$2.437 ($-$5.21) & $-$2.431 ($-$5.19) & $-$2.316 ($-$4.79) & $-$2.443 ($-$5.18) & $-$2.574 ($-$5.41) & $-$2.441 ($-$5.16) & $-$2.357 ($-$4.53) & $-$2.648 ($-$5.33) & $-$3.091 ($-$5.62) & $-$2.468 ($-$5.12) & $-$2.426 ($-$4.51) & $-$2.876 ($-$5.49) & $-$3.501 ($-$5.37) \\
\quad $ASC^{male}$ & 0.191 (0.50) & 0.183 (0.48) & 0.183 (0.47) & 0.221 (0.56) & 0.199 (0.51) & 0.179 (0.47) & 0.182 (0.46) & 0.200 (0.49) & 0.097 (0.21) & 0.210 (0.54) & 0.343 (0.86) & 0.122 (0.28) & 0.003 (0.01) \\
\quad $ASC^{wkd}$ & $-$0.786 ($-$2.53) & $-$0.795 ($-$2.55) & $-$0.862 ($-$2.72) & $-$0.851 ($-$2.72) & $-$0.825 ($-$2.51) & $-$0.803 ($-$2.52) & $-$0.799 ($-$2.53) & $-$0.718 ($-$2.21) & $-$0.942 ($-$2.72) & $-$0.724 ($-$2.21) & $-$0.874 ($-$2.53) & $-$0.560 ($-$1.61) & $-$1.173 ($-$2.80) \\
\quad $ASC^{bsn}$ & 1.208 (3.34) & 1.208 (3.33) & 1.260 (3.37) & 1.316 (3.56) & 1.137 (3.06) & 1.263 (3.44) & 1.286 (3.41) & 1.538 (4.07) & 1.356 (3.14) & 1.222 (3.31) & 1.377 (3.65) & 1.580 (3.36) & 1.430 (2.95) \\
\quad $ASC^{com}$ & 0.057 (0.26) & 0.056 (0.25) & 0.055 (0.24) & 0.084 (0.36) & 0.017 (0.07) & 0.079 (0.35) & 0.030 (0.12) & 0.040 (0.16) & 0.278 (1.01) & 0.071 (0.31) & 0.028 (0.12) & 0.184 (0.67) & 0.669 (2.01) \\
\quad $ASC^{am}$ & 0.580 (1.81) & 0.565 (1.75) & 0.563 (1.68) & 0.592 (1.75) & 0.495 (1.51) & 0.557 (1.72) & 0.510 (1.50) & 0.447 (1.37) & 0.721 (1.87) & 0.564 (1.73) & 0.595 (1.75) & 0.544 (1.44) & 1.034 (2.24) \\
\quad $ASC^{pm}$ & $-$1.465 ($-$4.41) & $-$1.470 ($-$4.41) & $-$1.515 ($-$4.45) & $-$1.451 ($-$4.20) & $-$1.553 ($-$4.58) & $-$1.401 ($-$4.17) & $-$1.555 ($-$4.38) & $-$1.360 ($-$3.98) & $-$1.625 ($-$4.11) & $-$1.442 ($-$4.22) & $-$1.425 ($-$3.94) & $-$1.257 ($-$3.57) & $-$1.621 ($-$3.63) \\
\quad $ASC^{short}$ & 1.370 (1.90) & 1.378 (1.91) & 1.719 (2.58) & 1.397 (1.92) & 1.578 (2.12) & 1.384 (1.85) & 3.056 (3.70) & 1.626 (1.80) & 1.778 (1.99) & 1.136 (1.43) & -- & 3.329 (3.35) & 1.404 (1.30) \\
\midrule
\multicolumn{14}{l}{\textbf{Alternative-specific constants: taxi} ($ASC_{taxi}$)} \\
\quad $ASC_{taxi}$ & $-$4.392 ($-$9.71) & $-$4.389 ($-$9.69) & $-$4.259 ($-$9.07) & $-$4.092 ($-$9.19) & $-$4.230 ($-$8.55) & $-$4.393 ($-$9.60) & $-$4.273 ($-$8.45) & $-$4.149 ($-$9.11) & $-$3.690 ($-$7.60) & $-$4.474 ($-$9.79) & $-$4.009 ($-$8.64) & $-$4.065 ($-$8.71) & $-$3.495 ($-$7.03) \\
\quad $ASC^{male}$ & $-$0.469 ($-$1.45) & $-$0.481 ($-$1.48) & $-$0.416 ($-$1.27) & $-$0.394 ($-$1.22) & $-$0.422 ($-$1.28) & $-$0.491 ($-$1.50) & $-$0.419 ($-$1.26) & $-$0.304 ($-$0.92) & $-$0.363 ($-$1.09) & $-$0.478 ($-$1.47) & $-$0.366 ($-$1.10) & $-$0.343 ($-$1.03) & $-$0.439 ($-$1.22) \\
\quad $ASC^{wkd}$ & 0.677 (3.06) & 0.678 (3.05) & 0.690 (2.99) & 0.672 (2.96) & 0.651 (2.80) & 0.653 (2.87) & 0.673 (2.84) & 0.777 (3.33) & 0.609 (2.37) & 0.702 (3.03) & 0.674 (2.73) & 0.808 (3.25) & 0.536 (1.88) \\
\quad $ASC^{stud}$ & 1.074 (2.85) & 1.074 (2.85) & 1.040 (2.72) & 1.085 (2.79) & 1.097 (2.81) & 1.071 (2.81) & 1.025 (2.60) & 0.983 (2.44) & 1.087 (2.63) & 1.070 (2.81) & 1.054 (2.68) & 1.004 (2.37) & 1.151 (2.67) \\
\quad $ASC^{bsn}$ & 0.221 (0.61) & 0.225 (0.62) & 0.277 (0.79) & 0.118 (0.30) & 0.236 (0.66) & 0.253 (0.70) & 0.230 (0.65) & 0.177 (0.46) & 0.237 (0.58) & 0.234 (0.65) & 0.209 (0.56) & 0.164 (0.41) & 0.270 (0.65) \\
\quad $ASC^{com}$ & $-$0.090 ($-$0.52) & $-$0.085 ($-$0.49) & $-$0.143 ($-$0.79) & $-$0.106 ($-$0.59) & $-$0.181 ($-$0.98) & $-$0.084 ($-$0.48) & $-$0.149 ($-$0.82) & $-$0.246 ($-$1.32) & $-$0.145 ($-$0.75) & $-$0.054 ($-$0.30) & $-$0.134 ($-$0.72) & $-$0.292 ($-$1.53) & $-$0.111 ($-$0.55) \\
\quad $ASC^{am}$ & 0.597 (2.21) & 0.599 (2.22) & 0.552 (1.97) & 0.532 (1.84) & 0.542 (1.87) & 0.591 (2.16) & 0.571 (2.04) & 0.608 (1.99) & 0.515 (1.58) & 0.626 (2.28) & 0.537 (1.87) & 0.644 (2.05) & 0.659 (1.91) \\
\quad $ASC^{pm}$ & $-$1.126 ($-$3.74) & $-$1.129 ($-$3.73) & $-$1.093 ($-$3.52) & $-$1.071 ($-$3.57) & $-$1.110 ($-$3.51) & $-$1.097 ($-$3.58) & $-$1.154 ($-$3.57) & $-$0.901 ($-$2.98) & $-$1.031 ($-$3.14) & $-$1.089 ($-$3.53) & $-$1.068 ($-$3.35) & $-$0.841 ($-$2.70) & $-$0.948 ($-$2.72) \\
\quad $ASC^{short}$ & $-$0.320 ($-$0.97) & $-$0.317 ($-$0.96) & $-$0.372 ($-$1.12) & $-$0.449 ($-$1.44) & $-$0.315 ($-$0.91) & $-$0.298 ($-$0.90) & $-$0.356 ($-$1.01) & $-$0.557 ($-$1.70) & $-$0.476 ($-$1.48) & $-$0.305 ($-$0.91) & $-$0.350 ($-$1.06) & $-$0.451 ($-$1.39) & $-$0.493 ($-$1.46) \\
\midrule
\multicolumn{14}{l}{\textbf{Alternative-specific constants: cycling} ($ASC_{cyc}$)} \\
\quad $ASC_{cyc}$ & $-$6.025 ($-$11.09) & $-$6.031 ($-$11.08) & $-$6.078 ($-$11.02) & $-$6.057 ($-$11.04) & $-$6.116 ($-$10.94) & $-$6.014 ($-$11.08) & $-$6.194 ($-$10.90) & $-$6.298 ($-$10.95) & $-$6.295 ($-$10.70) & $-$5.995 ($-$11.05) & $-$6.273 ($-$10.88) & $-$6.495 ($-$11.08) & $-$6.402 ($-$10.36) \\
\quad $ASC^{male}$ & 0.999 (2.14) & 0.996 (2.13) & 1.040 (2.20) & 1.021 (2.17) & 1.054 (2.20) & 1.002 (2.14) & 1.083 (2.23) & 1.107 (2.26) & 1.109 (2.21) & 0.992 (2.13) & 1.142 (2.31) & 1.165 (2.40) & 1.122 (2.16) \\
\quad $ASC^{wkd}$ & $-$0.448 ($-$1.42) & $-$0.445 ($-$1.41) & $-$0.406 ($-$1.26) & $-$0.430 ($-$1.36) & $-$0.507 ($-$1.64) & $-$0.475 ($-$1.51) & $-$0.393 ($-$1.23) & $-$0.246 ($-$0.74) & $-$0.455 ($-$1.35) & $-$0.446 ($-$1.39) & $-$0.318 ($-$0.97) & $-$0.163 ($-$0.49) & $-$0.517 ($-$1.44) \\
\quad $ASC^{bsn}$ & 0.782 (2.29) & 0.785 (2.28) & 0.795 (2.31) & 0.776 (2.25) & 0.765 (2.17) & 0.778 (2.25) & 0.821 (2.37) & 0.962 (2.60) & 0.839 (2.29) & 0.765 (2.25) & 0.876 (2.52) & 0.860 (2.22) & 0.841 (2.21) \\
\quad $ASC^{com}$ & $-$0.124 ($-$0.54) & $-$0.132 ($-$0.57) & $-$0.176 ($-$0.75) & $-$0.129 ($-$0.55) & $-$0.178 ($-$0.75) & $-$0.115 ($-$0.49) & $-$0.143 ($-$0.61) & $-$0.067 ($-$0.25) & $-$0.164 ($-$0.66) & $-$0.146 ($-$0.63) & $-$0.140 ($-$0.59) & $-$0.109 ($-$0.42) & $-$0.190 ($-$0.75) \\
\quad $ASC^{am}$ & 1.145 (3.92) & 1.142 (3.91) & 1.119 (3.77) & 1.153 (3.91) & 1.134 (3.70) & 1.160 (3.92) & 1.018 (3.22) & 1.137 (3.45) & 1.272 (4.09) & 1.160 (3.96) & 0.975 (3.01) & 1.246 (3.76) & 1.406 (4.28) \\
\quad $ASC^{pm}$ & $-$0.665 ($-$2.05) & $-$0.660 ($-$2.01) & $-$0.688 ($-$2.08) & $-$0.683 ($-$2.07) & $-$0.687 ($-$2.06) & $-$0.638 ($-$1.94) & $-$0.752 ($-$2.19) & $-$0.544 ($-$1.66) & $-$0.714 ($-$1.95) & $-$0.612 ($-$1.83) & $-$0.637 ($-$1.83) & $-$0.528 ($-$1.51) & $-$0.665 ($-$1.72) \\
\quad $ASC^{short}$ & $-$0.158 ($-$0.45) & $-$0.156 ($-$0.44) & $-$0.162 ($-$0.45) & $-$0.156 ($-$0.44) & $-$0.129 ($-$0.36) & $-$0.151 ($-$0.42) & $-$0.131 ($-$0.36) & $-$0.404 ($-$1.28) & $-$0.091 ($-$0.25) & $-$0.172 ($-$0.48) & 0.034 (0.09) & $-$0.291 ($-$0.91) & $-$0.055 ($-$0.15) \\
\midrule
\multicolumn{14}{l}{\textbf{Alternative-specific constants: walking} ($ASC_{walk}$)} \\
\quad $ASC_{walk}$ & $-$0.181 ($-$0.38) & $-$0.171 ($-$0.36) & $-$0.041 ($-$0.09) & $-$0.181 ($-$0.38) & $-$0.103 ($-$0.22) & $-$0.230 ($-$0.47) & $-$0.147 ($-$0.30) & $-$0.071 ($-$0.15) & $-$0.188 ($-$0.39) & $-$0.233 ($-$0.48) & $-$0.146 ($-$0.28) & $-$0.143 ($-$0.30) & $-$0.100 ($-$0.20) \\
\quad $ASC^{male}$ & 0.178 (0.67) & 0.175 (0.66) & 0.203 (0.75) & 0.209 (0.78) & 0.211 (0.78) & 0.157 (0.59) & 0.235 (0.85) & 0.317 (1.15) & 0.240 (0.87) & 0.174 (0.65) & 0.309 (1.10) & 0.294 (1.04) & 0.236 (0.82) \\
\quad $ASC^{wkd}$ & $-$0.211 ($-$0.97) & $-$0.207 ($-$0.95) & $-$0.222 ($-$0.99) & $-$0.204 ($-$0.93) & $-$0.260 ($-$1.14) & $-$0.223 ($-$1.00) & $-$0.264 ($-$1.13) & $-$0.051 ($-$0.24) & $-$0.258 ($-$1.06) & $-$0.217 ($-$0.95) & $-$0.219 ($-$0.93) & $-$0.035 ($-$0.15) & $-$0.329 ($-$1.22) \\
\quad $ASC^{stud}$ & 1.377 (2.96) & 1.383 (2.96) & 1.357 (2.85) & 1.391 (2.94) & 1.390 (2.92) & 1.373 (2.92) & 1.375 (2.86) & 1.109 (2.30) & 1.428 (2.88) & 1.372 (2.92) & 1.434 (2.97) & 1.105 (2.16) & 1.499 (2.94) \\
\quad $ASC^{bsn}$ & 0.578 (2.36) & 0.576 (2.35) & 0.573 (2.29) & 0.572 (2.31) & 0.552 (2.21) & 0.578 (2.33) & 0.552 (2.21) & 0.689 (2.80) & 0.581 (2.25) & 0.564 (2.26) & 0.604 (2.41) & 0.692 (2.71) & 0.603 (2.24) \\
\quad $ASC^{com}$ & $-$0.619 ($-$3.80) & $-$0.624 ($-$3.82) & $-$0.643 ($-$3.83) & $-$0.612 ($-$3.76) & $-$0.663 ($-$3.98) & $-$0.636 ($-$3.83) & $-$0.639 ($-$3.80) & $-$0.622 ($-$3.59) & $-$0.651 ($-$3.86) & $-$0.622 ($-$3.68) & $-$0.600 ($-$3.48) & $-$0.603 ($-$3.45) & $-$0.671 ($-$3.86) \\
\quad $ASC^{am}$ & 1.068 (4.71) & 1.072 (4.73) & 1.087 (4.74) & 1.071 (4.68) & 1.064 (4.62) & 1.079 (4.72) & 1.051 (4.48) & 1.096 (4.38) & 1.152 (5.02) & 1.069 (4.63) & 1.072 (4.57) & 1.073 (4.22) & 1.286 (5.46) \\
\quad $ASC^{pm}$ & $-$0.602 ($-$2.39) & $-$0.595 ($-$2.35) & $-$0.609 ($-$2.43) & $-$0.598 ($-$2.36) & $-$0.630 ($-$2.48) & $-$0.574 ($-$2.25) & $-$0.712 ($-$2.69) & $-$0.430 ($-$1.75) & $-$0.651 ($-$2.37) & $-$0.593 ($-$2.28) & $-$0.629 ($-$2.35) & $-$0.377 ($-$1.51) & $-$0.557 ($-$1.95) \\
\quad $ASC^{short}$ & $-$0.412 ($-$1.30) & $-$0.417 ($-$1.31) & $-$0.504 ($-$1.57) & $-$0.424 ($-$1.35) & $-$0.460 ($-$1.43) & $-$0.359 ($-$1.12) & $-$0.513 ($-$1.53) & $-$0.723 ($-$2.52) & $-$0.449 ($-$1.41) & $-$0.392 ($-$1.20) & $-$0.473 ($-$1.39) & $-$0.556 ($-$1.92) & $-$0.484 ($-$1.46) \\
\midrule
\multicolumn{14}{l}{\textbf{Goodness of fit}} \\
Observations & 12,004 & 11,884 & 11,884 & 11,884 & 11,884 & 11,404 & 11,404 & 11,404 & 11,404 & 10,804 & 10,804 & 10,804 & 10,804 \\
LL(0) & $-$14,374.87 & $-$14,230.83 & $-$14,241.57 & $-$14,210.35 & $-$14,229.17 & $-$13,659.33 & $-$13,684.47 & $-$13,612.36 & $-$13,614.20 & $-$12,939.14 & $-$12,962.95 & $-$12,793.27 & $-$12,816.54 \\
LL(final) & $-$4,674.77 & $-$4,629.45 & $-$4,479.82 & $-$4,553.08 & $-$4,426.78 & $-$4,437.95 & $-$4,309.87 & $-$4,240.90 & $-$4,053.11 & $-$4,213.55 & $-$4,065.99 & $-$3,935.54 & $-$3,664.62 \\
Adjusted $\rho^2$ & 0.671 & 0.671 & 0.681 & 0.675 & 0.685 & 0.671 & 0.681 & 0.684 & 0.697 & 0.670 & 0.682 & 0.688 & 0.709 \\
AIC & 9,465.54 & 9,374.89 & 9,075.64 & 9,222.17 & 8,969.56 & 8,991.90 & 8,735.73 & 8,597.79 & 8,222.21 & 8,543.10 & 8,245.99 & 7,987.08 & 7,445.24 \\
BIC & 9,894.34 & 9,803.11 & 9,503.85 & 9,650.38 & 9,397.77 & 9,417.72 & 9,161.55 & 9,023.61 & 8,648.03 & 8,965.79 & 8,661.38 & 8,409.76 & 7,867.92 \\
\bottomrule
\end{tabular}}
\end{sidewaystable}

\end{document}